\documentclass[11pt,a4paper]{article}
\pdfoutput=1

\usepackage{amsmath,amssymb}
\usepackage{epsfig,graphicx}
\usepackage{subfigure}
\usepackage{graphicx}
\usepackage{soul}
\usepackage{hyperref}
\usepackage{rotating}
\usepackage{cancel}
\usepackage{bm}
\usepackage{color}
\usepackage[font={small}]{caption}
\usepackage{comment}
\usepackage{cite}
\usepackage{psfrag}

\newcommand{\AC}[1]{\color{red} #1 \color{black}}

\def\beq{\begin{equation}}
\def\eeq{\end{equation}}

\begin{document}
\numberwithin{equation}{section}
\title{{\normalsize  \mbox{}\hfill DESY-22-149}\\
\vspace{2.5cm} 
\huge{\textbf{The Stochastic Relaxion
\vspace{0.5cm}}}}

\author{\Large{Aleksandr Chatrchyan$^{1}$ and G\'{e}raldine Servant$^{1,2}$}\\[2ex]
\small{\em $^{1}$Deutsches Elektronen-Synchrotron DESY, Notkestr.~85, 22607 Hamburg, Germany}\\[0.5ex] 
\small{\em $^{2}$II. Institute of Theoretical Physics, Universit\"{a}t Hamburg D-22761, Germany}\\[0.5ex]
}

\date{}
\maketitle

\begin{abstract}
	\noindent
	
	We revisit the original proposal of cosmological relaxation of the electroweak scale by Graham, Kaplan and Rajendran in which the Higgs mass is scanned during inflation by an axion field, the {\it relaxion}. We investigate the regime where the relaxion is subject to large fluctuations during inflation. The stochastic dynamics of the relaxion is described by means of the Fokker-Planck formalism. We derive a new stopping condition for the relaxion taking into account transitions between the neighboring local minima of its potential. Relaxion fluctuations  have important consequences even in the ``classical-beats-quantum" regime. We determine that for a large Hubble parameter during inflation, the random walk prevents the relaxion from getting trapped at the first minimum. The relaxion stops much further away, where the potential is  less shallow.  Interestingly, this essentially jeopardises the ``runaway relaxion" threat from finite-density effects, restoring most of the relaxion parameter space. We also explore the ``quantum-beats-classical" regime, opening large new regions of parameter space. We investigate the consequences for both the QCD and the non-QCD relaxion. The  misalignment of the relaxion due to fluctuations around its local minimum opens new phenomenological opportunities.

\end{abstract}

\newpage
\tableofcontents
\newpage

\section{Introduction}

Considering that the Standard Model (SM) is an effective field theory valid up the large scale $\Lambda$, the Higgs  mass parameter is surprisingly close to 
the critical value $\mu_h^2=0$  separating the broken and unbroken
phases of the electroweak symmetry, as  $|\mu_h^2/\Lambda^2|\ll1$.
This is the so-called Higgs naturalness or Higgs hierarchy puzzle.
In the last years, a new approach to this puzzle has been proposed, that does not rely on the existence of new symmetries at the TeV scale but instead suggests that a  cosmological dynamics of the Higgs potential parameters drives them towards the critical point.

One major simple concrete proposal in this class is the cosmological relaxation mechanism of Graham, Kaplan and Rajendran (GKR)~\cite{Graham:2015cka}, where the Higgs mass is controlled by the vacuum expectation value (VEV) of an axion-like field $\phi$, the relaxion. 
While the Higgs mass parameter is scanned by a slow-rolling relaxion field  during the early inflationary evolution of the universe, the electroweak phase transition generates a back-reaction able to stop the evolution near the critical point. 
Other dynamical solutions to the hierarchy problem as a result of a selection process in the cosmological evolution include self-organized localization~\cite{Giudice:2021viw}, ``NNaturalness''~\cite{Arkani-Hamed:2016rle}, and many others~\cite{Geller:2018xvz, Giudice:2019iwl, TitoDAgnolo:2021nhd} (see \cite{Dvali:2003br,Dvali:2004tma} for even earlier proposals).

In this paper, we will focus on the original GKR proposal.
The shift symmetry of the relaxion field is broken by non-perturbative effects, which produce a modulation $\propto \Lambda_b^4\cos(\phi/f)$ in the potential, and by the coupling to the Higgs, which lifts the remaining discrete shift symmetry. As a result, the relaxion potential contains a series of local minima, each corresponding to some value of the Higgs mass/VEV. A common assumption in the mechanism is that the relaxion dynamics is governed by classical slow-roll during inflation and not by quantum fluctuations, i.e.~the so-called ``classical beats quantum'' (CbQ) condition. As a consequence of this condition, the relaxion gets trapped near the first local minima of the potential, where the Higgs mass is small. 

However, during inflation, light scalar fields are subject to quantum fluctuations fuelled by the Hubble rate, which correspond, qualitatively, to a random walk.
The aim of this work is to 
drop the CbQ condition and extend the mechanism to the regime where quantum fluctuations during inflation are no longer negligible for the relaxion. 
The dynamics of the relaxion is no longer dominated by slow-roll and we develop a stochastic formalism, based on a Fokker-Planck equation~\cite{Starobinsky:1994bd}, to describe the relaxation process in this ``quantum beats classical'' (QbC) regime of large fluctuations. 
This allows to open up a new parameter region for the relaxion, associated with larger values of the Hubble scale during inflation, which determines the strength of fluctuations.
Our main results include a new stopping condition for the relaxion, as well as the possibility to solve the strong CP problem without the need for extra model-building modifying the relaxion potential at the end of inflation.

We also show that fluctuations play an important role even in the CbQ regime when it comes to the final stages of the dynamics. This is due to the fact that the first local minima of the relaxion potential are usually very shallow~\cite{Graham:2015cka, Banerjee:2018xmn} and sensitive to diffusion effects. The stochastic framework allows one to determine the local minimum in which the relaxion is expected to be trapped for a significantly long time. This is important for analyzing the stability of that minimum in the later stages of cosmic history after inflation, and in particular to derive properly the constraints from dense astrophysical environments such as neutron stars where finite-density effects can destabilise the relaxion.
The role of fluctuations is however much more pronounced in the QbC regime. Here we find that it is possible for the relaxion to get trapped much further away, in the region where the minima are deep. The stopping condition in this case is very different and, in particular, the height of the barriers that can trap the relaxion is controlled by the Hubble scale during inflation, $\Lambda_b \sim H_I$.

The relaxion in the QbC regime was already considered in~\cite{Nelson:2017cfv} and~\cite{Gupta:2018wif}, although neither of these works took into account the modified stopping condition. The main focus in \cite{Nelson:2017cfv} was the QCD relaxion model i.e.~the relaxion being the QCD axion. In fact, the first minima of the relaxion potential have an order-one $\theta$-angle, hence, the QCD relaxion model is compatible with experimental bounds $\theta_{\mathrm{QCD}}<10^{-10}$~\cite{Pendlebury:2015lrz} only at the cost of engineering some modification of the relaxion potential after inflation. Ref.~\cite{Nelson:2017cfv} proposed a way out of this problem by considering relaxation at large Hubble scales, $3 \mathrm{GeV}<H_I<100\mathrm{GeV}$. At these scales, finite-temperature effects suppress the barriers and, as a result, the $\theta$-angle at a given local minimum decreases afterwards as the universe eventually cools down. Ref.~\cite{Gupta:2018wif} focused on the relaxion measure problem associated with eternal inflation in the QbC regime.

Our new stopping condition has important implications for the QCD relaxion model. At large Hubble scales, the relaxion undergoes strong fluctuations and can get trapped near a deep minimum, which indeed allows the mechanism to be reconciled with a small value of $\theta_{\mathrm{QCD}}$. For this, however, we find that the inflationary Hubble scale should be comparable to the topological succeptibility of QCD, $H_I \sim 75\mathrm{MeV}$. At the corresponding de Sitter temperatures $T_I \sim H_I/2\pi$, the QCD barriers are essentially temperature-independent. In constrast, for Hubble scales $H_I>3\mathrm{GeV}$ that were considered in~\cite{Nelson:2017cfv}, our analysis predicts that the relaxion will overshoot the correct local minimum and, therefore, set a wrong value for the Higgs mass.

We also apply our analysis to the non-QCD relaxion model. Compared to the QCD relaxion, the cut-off $\Lambda$ can be pushed to even larger scales. Furthermore, in a large parameter region, a successful relaxation does not require eternal inflation, which allows to avoid many of the measure problems discussed in~\cite{Gupta:2018wif}.

The outline of this paper is the following. In section~\ref{sec:review}, we review the relaxion mechanism from~\cite{Graham:2015cka}. In section~\ref{sec:stochastic}, the stochastic formalism and the Fokker-Planck equation are introduced, and the new stopping condition is derived. 
In Section \ref{sec_implication_CbQ}, we revisit the CbQ regime including stochastic effects. We determine the position of the final relaxion  minimum in the CbQ regime and the resulting astrophysical constraints from finite-density effects.
In section~\ref{sec:QbC} we investigate the QbC regime and describe the resulting extended parameter space for the relaxion. 
In section \ref{sec:properties}, we discuss some phenomenological properties of the relaxion.
 Finally, we summarise our findings in section~\ref{sec:conclusion}. Technical derivations are presented in the appendices.

\section{Review of the relaxion mechanism}
\label{sec:review}

In this section we review the relaxion mechanism, as proposed in the original paper~\cite{Graham:2015cka}. Both the minimal model, where the relaxion is the QCD axion, as well as the non-QCD relaxion model are discussed.

\subsection{The main ingredients}

In the relaxion mechanism, the Higgs mass  $\mu_h^2$ is promoted to a dynamical variable, controlled by the VEV of the relaxion field $\phi$,
\beq
\label{eq_Higgs_mass}
\mu^2_h \rightarrow \mu_h^2(\phi) = \Lambda^2 - g'\Lambda \phi.
\eeq
Here $\Lambda$ is the scale at which the Higgs quadratic divergence gets cutoff and $g'$ is a dimensionless parameter characterizing the new coupling between the relaxion and the Higgs. The smallness of $g'$ is technically natural, due to the approximate shift symmetry of the relaxion.

A second ingredient is a rolling potential for the relaxion, of the following form,
\beq
U_{roll}(\phi) = -g\Lambda^3\phi, \: \: \: \: \:  \: \: \: \: \:  \: \: \: \: \: g \gtrsim g'/4\pi.
\eeq
This term can be radiatively generated by closing the Higgs loop in the term 
$g'\Lambda \phi  H^\dagger{H}$ and thus technical naturalness demands $g \gtrsim g'/4\pi$. 
Higher-order terms are also generated, however suppressed by powers of $g\phi/\Lambda$. The rolling potential allows the relaxion to dynamically minimize the squared mass of the Higgs. We set $g\sim g'$ in most of our expressions, unless stated otherwise.

The third important ingredient for the mechanism are the \textit{Higgs-VEV-dependent barriers} in the relaxion potential,
\beq
U_{br}(\phi)=  \Lambda_b^4(h)[1-\cos(\phi/f)],
\eeq
which allow the relaxion to get trapped in a local minimum of its potential and, thus, select a certain value for $\mu_h^2$. The latter should match the measured value of the Higgs mass, $\mu_h^2=-(88\rm GeV)^2$. Here the negative sign is due to the broken symmetry, which leads to a nonzero Higgs VEV, $\langle h \rangle = v_h = \sqrt{(-\mu_h^2)/{\lambda_h}} = 246 \rm GeV$.

In the minimal model, the relaxion is the QCD axion. The barriers for $\phi$ then originate from the anomalous coupling to gluons, ${\phi}G_{\mu\nu}\tilde{G}^{\mu\nu}$.
The parameter $\Lambda_b$, which is the topological susceptibility of QCD, is computed to be around $\Lambda_b = 75\rm MeV$~\cite{Borsanyi:2016ksw} (for the correct Higgs VEV) at temperatures below the QCD scale, $T\sim \Lambda_{\mathrm{QCD}}\approx 150 \mathrm{MeV}$. The value of $\Lambda_b$ depends on the Higgs VEV at least through the mass of the lightest quark~\cite{Choi:1986zw}. In particular, if the Higgs is in the symmetric phase, the quarks are massless (their mass is proportional to the Yukawa coupling, $m_q = y_q v_h/\sqrt{2}$), and there are no barriers. Once the Higgs develops a symmetry breaking VEV, the barrier height takes the form
\beq
\label{eq:QCDbarrier}
\Lambda_b^4 \approx f_{\pi}^2m_{\pi}^2\frac{m_u m_d}{(m_u+m_d)^2} \approx \Lambda_{\mathrm{QCD}}^3 m_u.
\eeq

In the non-QCD model, the Higgs-dependent barriers originate from an analogous coupling of the relaxion to some hidden gauge group. The dependence on the Higgs VEV in this case is usually of the form $\Lambda_b^4\propto \langle h \rangle^2$.

To summarize, in both models, the dynamics of the relaxion takes place in a potential of the following form,
\beq
V(\phi) =  - g\Lambda^3\phi + \Lambda_b^4(\phi)[1-\cos(\phi/f)].
\eeq
Here it is implicitly assumed that the Higgs adiabatically follows the minimum of its potential, which in turn is determined by the value of $\phi$.

The relaxion gets trapped in one of its local minima, determined by the \textit{stopping mechanism}. The simplest one, as proposed by the authors in~\cite{Graham:2015cka} is realized by assuming that relaxation takes place during inflation and the relaxion is in the slow-roll regime~\footnote{
	For the Hubble friction to be strong enough so that the  relaxion tracks the slow-roll velocity from (\ref{slow_roll}), the rolling time between neighboring  minima $\Delta t = 2\pi f/\dot \phi_{\mathrm{SR}}$ should be larger compared to the Hubble time $\sim H_I^{-1}$. This was explained in~\cite{Fonseca:2019lmc}, where the relaxion scenario was also investigated in both cases.}, governed by
\beq
\dot \phi = \dot \phi_{\mathrm{SR}} =  - \frac{V'(\phi)}{3H_I},
\label{slow_roll}
\eeq
where $H_I$ denotes the inflationary Hubble parameter. Under this assumption, the field should stop near the first local minimum,
\beq
\label{rel_loc_min}
0 = V'(\phi_0) = -g\Lambda^3 + \frac{\Lambda_{b}^4(\phi_0)}{f}\sin\Bigl(\frac{\phi_0}{f}\Bigr).
\eeq
Usually, the relaxion barriers increase by a small amount from one minimum to the next one. This implies that $\sin(\phi_0/f)$ is of order one, hence, the stopping condition can be expressed as
\beq
\label{equal_slopes}
\Lambda_{b}^4(\phi_0) \sim g\Lambda^3{f} .
\eeq

Several conditions must be satisfied for the slow-roll dynamics to be described by Eq.~(\ref{slow_roll}). In particular,

\begin{itemize}
	\item The Hubble parameter during inflation must be large enough so that the change of the potential energy in the relaxion sector, which is of order $\Delta U\sim \Lambda^4(g/g')$ over the typical field range, does not impact the expansion rate,
	\beq
	\label{subdominant}
	H_I^2>\frac{8\pi}{3}\frac{g}{g'} \frac{\Lambda^4}{M^2_{Pl}} \: \: \: \: \: \: \: \: \: \: \: \: \: \: \: \: \: \: \: \: \text{(vacuum energy)}. 
	\eeq
	If this condition is not satisfied, the backreaction of the relaxion on the Hubble expansion must be taken into account (see e.g.~\cite{Geller:2018xvz} which considers similar effects).
	
	\item The classical beats quantum (CbQ) requirement,
	\beq
	\label{cbq}
	H_I^3 < V'  = g\Lambda^3 \: \: \: \: \: \: \: \: \: \: \: \: \: \: \: \: \: \: \: \: \text{(classical beats quantum)}.
	\eeq
	If this condition is not satisfied, inflationary quantum fluctuations, which produce random kicks $\Delta \phi 
	\sim H_I$ per Hubble time $t \sim H_I^{-1}$, cannot be neglected compared to the slow-roll. Later in this work we discuss what happens if this constraint is dropped.
\end{itemize} 

The two above conditions imply that the inflationary Hubble scale should be inside the range
\beq
\label{eq:range_of_H}
\frac{\Lambda^2}{M_{\mathrm{Pl}}} < H_I < g^{1/3} \Lambda.
\eeq
In the above expression, we dropped order-one prefactors for simplicity.

To ensure that the relaxion ends up at the correct Higgs VEV, it must have enough time to scan a typical field range $\Delta \phi \sim \Lambda/g'$. Using (\ref{slow_roll}), one arrives at the required minimum number of e-folds during inflation
\beq
\label{req_num_efolds}
N_I = H_I t_I \gtrsim N_{\mathrm{req}} = \frac{3 H_I^2}{g g'\Lambda^2}.
\eeq
This usually corresponds to a very long period of inflation. The slow-roll makes the dynamics insensitive to the initial conditions, as long as it starts from a positive Higgs mass.

In the next subsections we present the relaxion parameter space in the QCD and the non-QCD models.

\subsection{The QCD model}
\label{ssec:GKR_QCD}

In the model where the relaxion is a QCD axion, its barriers result from the QCD anomaly and $\Lambda_b$ is given by Eq.~(\ref{eq:QCDbarrier}). This model, while minimalistic, leads to the reappearance of the strong CP problem. More specifically, the local minimum of the relaxion potential from (\ref{rel_loc_min}) is displaced from the CP-conserving minimum of the cosine potential at $\sin(\phi_0/f)=0$, due to the rolling term. This generates an order-one $\theta$-angle for QCD,
\beq
\label{theta_angle}
\theta_{\mathrm{QCD}} = \frac{\phi_0}{f} = \arcsin\Bigl(\frac{g\Lambda^3 f}{\Lambda_b^4}\Bigr),
\eeq
in contradiction with the experimental bounds $\theta_{\mathrm{QCD}}<10^{-10}$~\cite{Pendlebury:2015lrz}.

In order to reduce the CP violation, the authors of~\cite{Graham:2015cka} proposed that the slope of the rolling potential changes after inflation, so that $\theta_{\mathrm{QCD}}<10^{-10}$ is satisfied today. As can be understood from (\ref{theta_angle}), the coupling $g_I$ during inflation and its  value today $g$ should then satisfy $$g = \xi g_I  < 10^{-10}g_I.$$
It is argued in~\cite{Graham:2015cka} that such a modification can be achieved by an additional coupling of the relaxion to the inflaton.

The new constraints on the relaxion can be obtained by replacing $g \rightarrow g_I = g/\xi$ in (\ref{equal_slopes}), (\ref{subdominant}) and (\ref{cbq}). One obtains
\beq
\label{QCD_CP_ineq}
\frac{\Lambda^2}{M_{\mathrm{Pl}}} \frac{1}{\sqrt{\xi}} < H_I < \Bigl( \frac{g}{\xi} \Bigr)^{\frac{1}{3}} \Lambda, \: \: \: \: \: \: \: \: \: \text{and} \: \: \: \: \: \: \: \: \:  \Lambda_{b}^4(\phi_0) \sim \frac{g}{\xi}\Lambda^3{f} 
\eeq
Eliminating $H_I$ in the first equation and expressing $g$ from the second equation one arrives at the upper bound on the cut-off scale $\Lambda$ that can be successfully relaxed,
\beq
\Lambda < 3\times 10^{4}\mathrm{GeV} \Bigr(\frac{10^9\mathrm{GeV}}{f}\Bigl)^{1/6} \Bigl(\frac{\xi}{10^{-10}}\Bigr)^{1/4}.
\eeq
Here we used the benchmark value for the axion decay constant $f = 10^9\mathrm{GeV}$ from~\cite{Graham:2015cka}, which is the typical lower bound from astrophysical constraints. We note that this bound is model-dependent.

\begin{figure}[!t]
	\centering 	\includegraphics[width=0.85\textwidth]{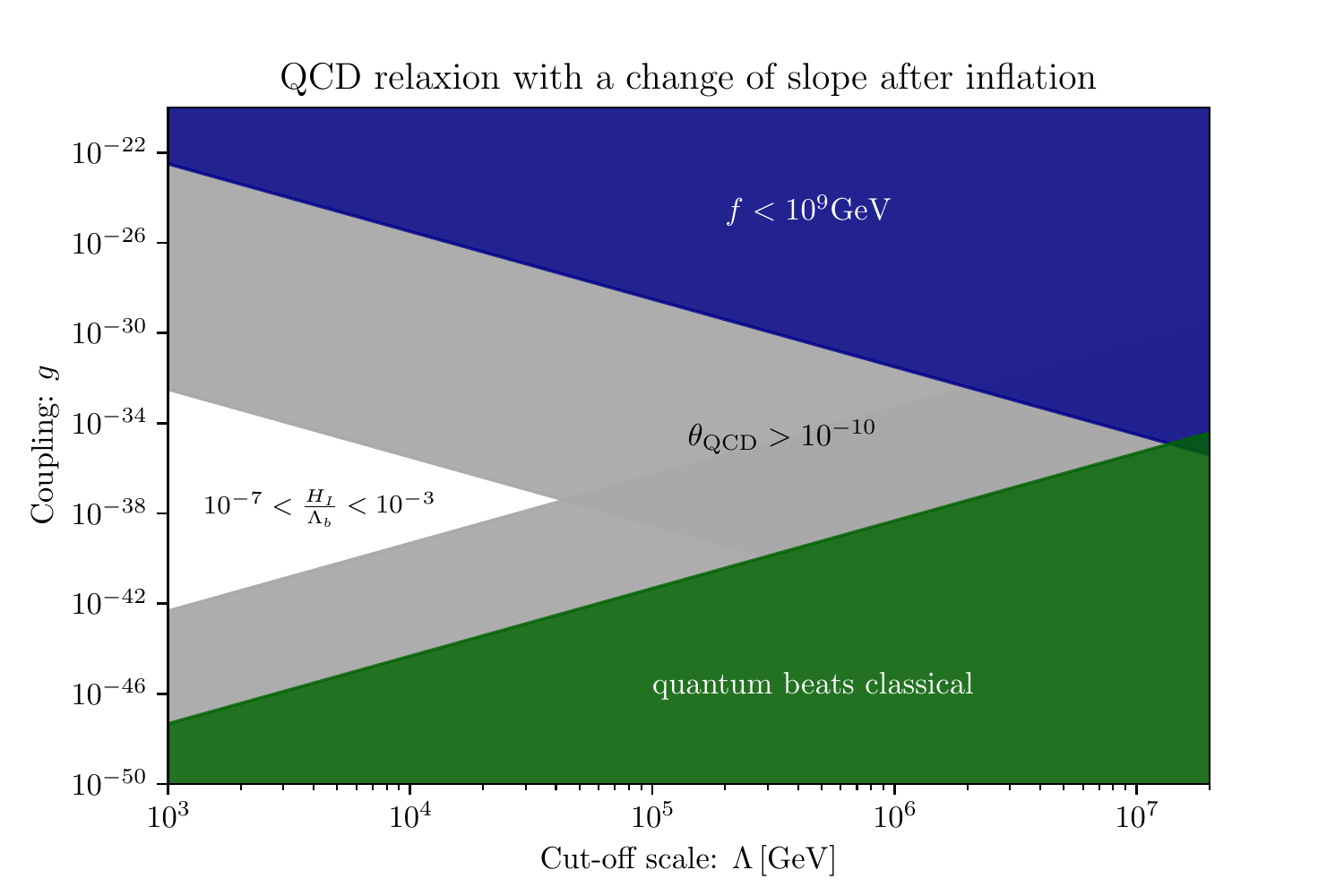}
	\caption{The QCD relaxion parameter space in the $g$ vs $\Lambda$ plane, with the change of the slope after inflation to conserve CP. The allowed region is shown in white. The blue region has an axion decay constant below $10^{9}\rm GeV$, while the green region violates the CbQ constraint. In the grey region $\theta_{\mathrm{QCD}}$ cannot be less than $10^{-10}$.}
	\label{QCD_CbQ}
\end{figure}

The parameter space for this model is shown in Fig.~\ref{QCD_CbQ} in the $g$ vs $\Lambda$ plane. The green region is excluded by the inequality (\ref{eq:range_of_H}) (after eliminating $H_I$), which requires the relaxion to be both subdominant as well as dominated by classical slow-roll. The blue region is excluded by the stopping condition in (\ref{equal_slopes}) combined with requirement $f>10^9 \mathrm{GeV}$. Inside the remaining region the QCD angle can still be large. The inequalities from (\ref{QCD_CP_ineq}) with $\xi = 10^{-10}$ exclude the grey region, leaving the unshaded one with $\Lambda<3\times 10^4 \mathrm{GeV}$ available for the relaxion. Note that the value of $H_I$ is not fixed in the figure. One can check that inside the allowed region it is in the range $10^{-7}\Lambda_b<H_I<10^{-3}\Lambda_b$

\subsection{The non-QCD model}
\label{nonQCD_CbQ}

Larger cut-off scales are possible in the non-QCD relaxion model.  Here the barriers originate from the confinement of some hidden gauge group. The parameter $\Lambda_b$ is therefore an additional free parameter in the non-QCD model and, in particular, can take values larger than $75\rm MeV$. The dependence of the barrier height on the Higgs VEV can usually be parametrized as
\beq
\Lambda_b^4(\langle h \rangle ) = \Lambda_b^4 \Bigl( \frac{ \langle h \rangle  }{v_h}\Bigr)^m, \: \: \: \: \: \: \: \: \: \: \: \: \: \: \: \: \: \: \: \: \: m=2,
\eeq
where $\Lambda_b^4 = \Lambda_b^4(v_h)$ denotes the barrier height at the measured Higgs VEV. Moreover, there is no constraint on the $\theta$ angle anymore and, hence, the trick of changing the slope of the rolling potential is no longer required.

Below we summarize the constraints, that are relevant in the non-QCD model.
\begin{itemize}
	\item The following upper bound is imposed on $\Lambda_b$
	\beq
	\label{Lambda_b_upper}
	\Lambda_b < \sqrt{4\pi}v_{h} \: \: \: \: \: \: \: \: \: \: \: \: \: \: \: \: \: \: \: \: \text{(stability against radiative corrections)},
	\eeq
	which ensures that the barrier potential is stable against radiative corrections and, thus, sensitive to the Higgs VEV~\cite{Fonseca:2019lmc}.
	
	\item Due to the larger barriers, the non-QCD model allows to have larger couplings $g$. Here one has to take care that the local minima of the potential have a separation that is smaller compared to the precision required to scan the Higgs VEV~\cite{Fonseca:2019lmc}. In other words,
	\beq
	\label{eq:precision}
	g'\Lambda (2\pi f) < |\mu_h^2|\: \: \: \: \: \: \: \: \: \: \: \: \: \: \: \: \: \: \: \: \text{(precision of Higgs mass scanning)}.
	\eeq
	
	\item The decay constant is assumed to be in the range
	\beq
	\label{eq:frange}
	\Lambda<f<M_{\mathrm{Pl}},
	\eeq
since $f>M_{\mathrm{Pl}}$ involves trans-Planckian physics, whereas $f>\Lambda$ insures that the relaxion as an effective degree of freedom is present at scales below the cut-off scale $\Lambda$.
\end{itemize}

The upper bound on the cut-off scale can be estimated from (\ref{equal_slopes}), (\ref{subdominant}) and (\ref{cbq}), as it was done in the QCD model. Here we supplement these inequalities with the lower bound on the decay constant from (\ref{eq:frange}) and arrive at
\beq
\label{upper_bound_L}
\Lambda < 4\times 10^9 \mathrm{GeV} \Bigl( \frac{\Lambda_b}{\sqrt{4\pi}v_h}\Bigr)^{4/7}.
\eeq

\begin{figure}[!t]
	\centering
	\includegraphics[width=0.85\textwidth]{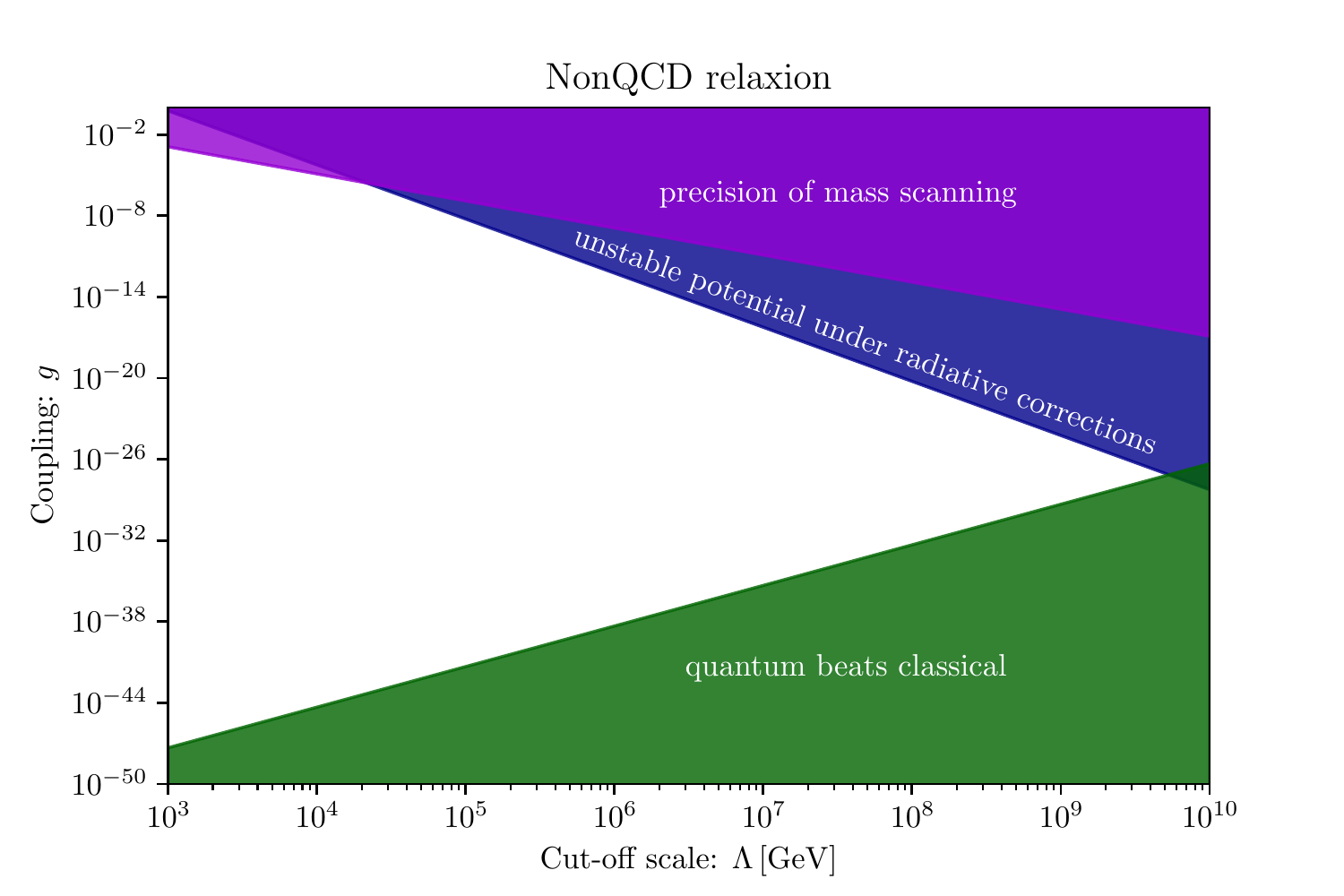}
	\caption{The non-QCD relaxion parameter space in the $g$ vs $\Lambda$ plane, with the allowed region shown in white. The violet, blue and green regions are excluded by Higgs mass scanning precision~(\ref{eq:precision}), problematic radiative corrections for the barriers~(\ref{upper_bound_L_nonQCD}) and the CbQ constraint, respectively.}
	\label{nonQCD_CbQ}
\end{figure}

The parameter region in the $g$ vs $\Lambda$ plane for the non-QCD model is shown in Fig.~\ref{nonQCD_CbQ}. In the white part relaxation can take place. In the violet region, the Higgs mass scanning is too unprecise according to (\ref{eq:precision}) for any allowed value of $f$. In the green region there is no value for the inflationary Hubble parameter, such that the relaxion is both subdominant as well as in the CbQ regime. The blue region is excluded by the stopping condition in (\ref{equal_slopes}) combined with the lower bound on $f$ and the upper bound on $\Lambda_b$.
Further exploration of this parameter space in the CbQ regime was performed in \cite{Fonseca:2019lmc} when including axion fragmentation effects that enable to decrease the duration of inflation. 

We close this review by pointing out that an alternative relaxion mechanism to GKR, using tachyonic production of Standard Model gauge boson as a source of friction was advocated in \cite{Hook:2016mqo,Fonseca:2018xzp}, where the potential barriers are Higgs-independent and relaxation starts in the EW broken phase. A critical investigation of this relaxion mechanism was performed  in the CbQ regime  in \cite{Fonseca:2019lmc} which computed in detail the 
open region of parameter space allowing for a successful relaxion mechanism compatible with all experimental bounds.

\subsection{Going beyond the classical relaxion}

As discussed above, the CbQ condition plays an important role for determining the relaxion parameter space and, in particular, the upper bound on the cut-off scale $\Lambda$. It is thus strongly motivated to explore the relaxion mechanism also in the quantum beats classical (QbC) regime.

One particularly useful application of the QbC relaxion, which was proposed in~\cite{Nelson:2017cfv}, is the possibility to solve the  strong $CP$ problem in the QCD relaxion model. The authors of~\cite{Nelson:2017cfv} proposed to use the temperature-dependence of the axion potential. The later has the form
\beq
\label{QCD_T_dep}
\Lambda_b^4(T, h) \approx \frac{\Lambda_b^4(0, h)}{1+(T/\Lambda_{\mathrm{QCD}})^{m}},
\eeq
with $m\approx8.16$~\cite{Borsanyi:2016ksw}, so that the barriers are strongly suppressed at high temperatures, $T>\Lambda_{\mathrm{QCD}}$, and almost  constant below the QCD scale. The idea is then to consider large enough inflationary Hubble scales $H_I$, so that the corresponding de Sitter temperature is $T_I\sim H_I/2\pi >\Lambda_{\mathrm{QCD}}$. In this case, $\theta_{\mathrm{QCD}}$ can be  of order one at the end of inflation (\ref{theta_angle}), as in the original GKR scenario. It  however shrinks at lower temperatures after inflation as $\Lambda_b$ increases to its zero-temperature value. It is then concluded that for $3\mathrm{ GeV} <H_I< 100\mathrm{GeV}$, the QCD relaxion mechanism can work while the $\theta_{\mathrm{QCD}}$ angle today is consistent with experimental bounds. 
This scenario is more minimal than \cite{Graham:2015cka} as it does not require additional couplings of the relaxion to the inflaton sector in order to reduce the slope of the potential.  As we will explain in Section \ref{QCD_stochastic}, the solution of \cite{Nelson:2017cfv} does not work as they presented, as the QbC regime leads to a different stopping condition which we derive in this paper. Nevertheless, we show that the QbC regime enables to solve the strong CP problem of the QCD relaxion. We will also see that the QbC regime opens new phenomenological possibilities for the relaxion. Besides, we find that even in the CbQ regime, the stochastic behavior is crucial to be taken into account for the proper determination of the final minimum of the relaxion, that controls its behavior in high-density astrophysical environments in the late-time universe.

In the next sections, we introduce the stochastic formalism to describe the relaxion dynamics. We then revisit the mechanism in the CbQ regime when including stochastic effects. We finally explain what happens if the CbQ condition is dropped.

\section{Stochastic formalism for the relaxion}
\label{sec:stochastic}

In this section we set the stage for the discussion of the QbC regime, by introducing the stochastic formalism for the description of relaxion dynamics, using the Fokker-Planck equation which we derive in section~\ref{ssec:FP}. We describe the stochastic dynamics of the relaxion in section~\ref{ssec:stochastic_dynamics} and identify a generalized stopping condition within this framework in section~\ref{ssec:stopcond}. 

\subsection{The Fokker-Planck formalism}
\label{ssec:FP}
If quantum fluctuations of a light scalar field, $m \ll H_I$, are taken into account, its dynamics in de Sitter spacetime can be described by means of a Fokker-Planck (FP) equation of the following form~\cite{Starobinsky:1994bd},
\beq
\label{FP_eq}
\frac{d\rho}{dt} = \frac{1}{3H_I}\frac{\partial(\rho \:  \partial_\varphi V)}{\partial\phi} + \frac{H_I^3}{8\pi^2} \frac{\partial^2\rho}{\partial\phi^2}.
\eeq
The central object in this framework is the function $\rho(\phi, t)$, which gives the probability distribution of the relaxion having its average field value inside the Hubble patch equal to $\phi$ at time $t$. The evolution of this function is governed by the two terms on the right-hand side. The first term is a drift term, describing the slow-roll of the field. The second one is a diffusion term, which arises due to quantum fluctuations. Eq.~(\ref{FP_eq}) is equivalent to a Langevin equation, describing the Brownian motion of a particle.

The origin of the random force in the relaxion dynamics are its low-momentum fluctuations. These are amplified to $\delta \phi \sim H_I$ when they become superhorizon and give a random kick to the field VEV inside the Hubble patch. A derivation of the above equation can be found in~\cite{Starobinsky:1994bd}.

In the absence of a potential, the variance of $\rho(\phi)$ simply grows as $\propto \sqrt{t}$,
\beq
\frac{d\sigma_{\phi}^2}{dt} = \frac{H_I^3}{4\pi^2}, \: \: \: \: \: \: \: \: \: \: \: \: \: \: \: \: \: \: \sigma_{\phi}(t) =\sqrt{ \langle \phi^2 \rangle(t)  - \langle \phi \rangle^2(t) },
\eeq
as one would expect for diffusion.

\subsection{Stochastic dynamics}
\label{ssec:stochastic_dynamics}

The dynamics of the relaxion in the stochastic regime is more complicated than in the deterministic case. We thus split the discussion into several parts.

\bigskip

\textbf{The $\mu_h^2>0$ region:} As usual, it assumed that the relaxion evolution starts in the region with a large and positive Higgs mass squared, $\mu_h^2\sim \Lambda^2$. In this region there are no wiggles and $V'(\phi)$ is approximately constant. The exact solution to the Fokker-Planck equation in this case has the form~\cite{Nelson:2017cfv},
\beq
\label{sol:nowiggles}
\rho(\phi, t) = \sqrt{ \frac{2\pi}{H_I^3 t} } \exp{\Bigl\{- \frac{2\pi^2 }{H^3_It}\Bigl(\phi - \frac{g\Lambda^3t}{3H_I}\Bigr)^2 \Bigr\}},
\eeq
where for simplicity it was  assumed that $\rho(\phi)$ started as a $\delta$-peak  at $t=0$. The solution above is a speading gaussian peak. The center of the peak evolves according to the classical slow-roll equation~(\ref{slow_roll}),
\beq
\label{SR_define}
\frac{d \langle \phi \rangle }{dt} = \dot \phi_{SR} = \frac{g\Lambda^3}{3H_I},
\eeq
and the width of the peak grows as
\beq
\label{width_spread}
\sigma_{\phi}(t) = \sqrt{\frac{H_I^3}{4\pi^2} t} = \frac{H_I}{2\pi} \sqrt{N},
\eeq
due to the diffusion effects. 

Using Eq.~(\ref{sol:nowiggles}) one can estimate the width of the relaxion distribution after traversing the typical distance of $\Delta \phi \sim \Lambda/g'$. Inserting the required number of e-folds from (\ref{req_num_efolds}) into (\ref{width_spread}) one obtains that the spread in the values of the relaxion field and of the Higgs mass squared will be at least
\beq
\label{variance_at_zero_mass}
\sigma_{\phi}=  \sqrt{ \frac{3 g'}{4\pi^2g} } \frac{H_I^2}{g'\Lambda} \sim \frac{H_I^2}{g'\Lambda}, \: \: \: \: \: \: \: \: \: \: \sigma_{\mu_h^2}=  \sqrt{ \frac{3 g'}{4\pi^2g} } H_I^2 \sim H_I^2.
\eeq
The required number of e-folds to scan the field range is again given by~(\ref{req_num_efolds}).

\bigskip

\textbf{The $\mu_h^2<0$ region:} The presence of wiggles in the negative Higgs mass squared region modifies the form~(\ref{FP_eq}) of the distribution function. In particular, $\rho(\phi)$ becomes larger in the regions where the potential $V(\phi)$ is locally minimized and smaller in regions of local maxima (see also~\cite{Nelson:2017cfv}). As the potential wiggles get larger, also the modulation of $\rho(\phi)$ gets more pronounced. There is no exact analytical solution for the FP equation in this regime. A perturbative approach for computing the corrections to (\ref{sol:nowiggles}) was presented in~\cite{Nelson:2017cfv}.

\bigskip

Eventually the distribution reaches the first local minimum where Eq.~(\ref{equal_slopes}) holds. Naively one may expect that the vanishing slope of the potential prevents the mean field value $\langle \phi \rangle$ from evolving further. Recall however, that in the relaxion potential each local minimum is followed by a deeper one, with a barrier separating the two.  As we will now explain, diffusion effects, combined with the asymmetry due to the rolling potential, generate a nonzero flux for the distribution in this regime. 

Let us mention two effects that are relevant for the discussion.
\begin{itemize}
	\item In any bounded potential the probability distribution from (\ref{FP_eq}) eventually reaches an equilibrium of the form~\cite{Starobinsky:1994bd}
	\beq
	\label{eq_eq}
	\rho_{\rm eq}(\phi) \propto \exp\Bigl( - \frac{8\pi^2V(\phi)}{3H_I^4} \Bigr),
	\eeq
	which is maximized at the minimum of $V(\phi)$. In the case of the relaxion the minima are only local and, thus, we can expect the establishment of a local ``meta-equilibrium'' around the minimum after some time. The normalization in (\ref{eq_eq}) is determined by the total population $n_0 = \int\rho(\phi)d\phi$ in the local minimum. 
	
	\item Diffusion effects generate a flux of probability to the next minimum over the barrier. Various approaches to derive the flux can be found in~\cite{Kramers:1940zz, matkowsky1977exit, talkner1987mean}. In appendix~\ref{app:escape} we revisit the so-called ``flux-over-population'' method~\cite{Berera:2019uyp} for computing the escape rate $k_{\rightarrow}$ from a local minimum. The final expression is given by
	\beq
	\label{decay_rate_anal}
	k_{\rightarrow} = \frac{j_{\rightarrow}}{n_{0}} = \frac{\sqrt{V_{0}'' |V_b''|}}{6\pi H_I} e^{  - B },
	\eeq
	where
	\beq
	\label{define_B}
	B = \frac{8\pi^2 \Delta V_b^{\rightarrow}}{3H_I^4}
	\eeq
	Here $n_0$ is the total probability in the local minimum from which the field decays and $j_{\rightarrow}$ is the flux of probability over the barrier. The right arrow corresponds to the direction minimizing the rolling potential. $V_{0}''$ and $V_b''$ are the curvatures of the potential at the decaying local minimum and the local maximum of the barrier, respectively. $\Delta V_b^{\rightarrow} = V_b - V_0$ is the potential difference between these two points. In Appendix ~\ref{app:potential}, we defined $\delta$ as the field separation between the barrier and the local minimum  
		\beq
		\delta \equiv \frac{\phi_b - \phi_0}{2f}.
		\eeq
	As explained in Appendix~\ref{app:potential}, for the relaxion potential we have
	\beq
	V_{0}'' = |V_{b}''| = \frac{\Lambda_b^4(\phi)}{f^2} \times \sin\delta, \:  \:  \:  \:  \:   \:  \:  \:  \:  \Delta V_b^{\rightarrow} = 2\Lambda_b^4(\phi)\times[\sin\delta - \delta \cos\delta],  
	\eeq
	where the $\delta$-dependent multiplicative corrections arise due to the presence of the rolling potential. The parameter $\delta$ can be expressed as (see Appendix~\ref{app:potential})
	\beq
	\label{exp_delta}
	\cos\delta = \frac{g\Lambda^3f}{\Lambda_b^4(\phi)}.
	\eeq
	 Near the first local minimum $\delta \ll 1$ and, as a result, both the curvature/mass as well as the barrier height are suppressed compared to the naive expectation~\cite{Banerjee:2018xmn}.

	As can be seen, the escape rate from the local minimum does not depend on the location of the lower-lying minimum. At the same time, a backwards flux of probability from the lower minimum is generated as well. This flux is however smaller due to the larger barriers in the backwards direction, $$\Delta V_b^{\leftarrow}= \Delta V_b^{\rightarrow}  + g\Lambda^3 (2\pi f),$$which results in a stronger exponential suppression in~(\ref{decay_rate_anal}).
\end{itemize}

Combining all fluxes between the neighboring minima one can compute change of the population of the $i$-th local minimum $n_{0,i}$ in time, 
\beq
\label{rate_eq}
\frac{dn_{0, i}}{dt} = - k_{\rightarrow} n_{0, i} - k_{\leftarrow} n_{0, i}+ k_{\leftarrow} n_{0, i+1} + k_{\rightarrow} n_{0, i-1}.
\eeq 
For simplicity, we assumed a constant height of the barriers (which usually indeed grows very slowly).

From the above rate equation one can compute the mean velocity of the field $\dot{\langle \phi \rangle}$, assuming that most of the distribution is in the region with local minima,
\beq
\label{velo_minima}
\dot{ \langle \phi \rangle } = \int \rho(\phi) \dot{\phi} d\phi \approx \sum_i  \dot{n}_{0, i} \phi_{0, i} = 2\pi f (k_{\rightarrow} - k_{\leftarrow}),
\eeq
where we approximated $\phi_{0, i+1}- \phi_{0, i} \approx 2\pi f$.
The term on the right hand side is proportional to $k_{\rightarrow} - k_{\leftarrow}  \propto e^{-B}$. The mean field velocity is thus exponentially suppressed only after the height of the barriers is comparable to the Hubble parameter, $B\gtrsim 1 $. In fact, Eq.~(\ref{rate_eq}) can be recast into a form of an effective Fokker-Planck equation, which is explained in appendix~\ref{app:stop}.

\bigskip
 
To summarize, Eq.~(\ref{sol:nowiggles}) describes the dynamics in the region with no wiggles, where the relaxion performs most of its displacement. Eq.~(\ref{velo_minima}) is valid at late times, when the potential barriers are large so that $\Delta V_b^{\rightarrow} \gtrsim H_I^4 $. We use~(\ref{velo_minima}) to formulate the stopping condition for the relaxion in the next subsection. Note that an analytical description for the possible intermediate regime of dynamics is unavailable, it is however also not required for our purposes.

\subsection{The new stopping condition}
\label{ssec:stopcond}

As can be understood from the discussion above, the relaxion strictly speaking never stops, if the possibility for it to jump over the barriers due to the random walk is included. At sufficiently late times however its velocity is exponentially suppressed and much smaller compared to the initial velocity. This can be used to formulate the approximate stopping condition: the relaxion stops near the first local minimum, unless the Hubble parameter during inflation is large enough so that the random walk prevents it from getting trapped. In the second case the late-time evolution of the relaxion mean field is governed by Eq.~(\ref{velo_minima}) and one can find the minimum at which the relaxion is expected to be trapped by requiring that the lifetime of that minimum, given by the inverse of the escape rate, is of the order of the duration of inflation, 
\beq
\label{lifetime_of_final_min}
k_{\rightarrow} - k_{\leftarrow} = \frac{\sqrt{V_{0}'' |V_b''|}}{6\pi H_I} \Bigl[ 1-\exp{   \Bigl( - \frac{8\pi^2 g\Lambda^3 (2\pi f)}{3H_I^4} \Bigr)  } \Bigr]   e^{  - B } \sim \frac{1}{t_I} = \frac{H_I}{N_I}.
\eeq
In the most cases the required number of e-folds of inflation is large and, therefore, the final minimum should have a very small escape rate. In this case the rate is dominantly determined by the exponential term $\exp(-B)$ and, ignoring logarithmic corrections, we can use
\beq
\label{late_stop}
B =  \frac{16\pi^2\Lambda_b^4}{3 H_I^4}  (\sin \delta - \delta \cos \delta ) \sim 1.
\eeq
as an approximate relation for the final minimum. In this regard it is convenient to introduce the dimensionless parameter
\beq
\label{parameter_d}
d = \frac{3}{8\pi^2}\frac{H_I^4}{g\Lambda^3f},
\eeq
characterizing the strength of diffusion effects. Using (\ref{exp_delta}) the stopping condition can then be rewritten as
\beq
\label{other_stop}
2[\tan(\delta) - \delta] \sim d.
\eeq
For a given value of $d$ one can use the above expression to find $\delta$ and, using (\ref{exp_delta}), determine also $\Lambda_b$. It is instructive to consider two limiting cases:
\begin{itemize}
    \item When $d\ll 1$, also $\delta\ll 1$ holds, implying that $\cos\delta \approx 1$ and $\Lambda_b^4(\phi) \sim g\Lambda^3 f$. This is the usual stopping condition that was used in e.g.~\cite{Graham:2015cka}.
    \item On the other hand, if $d\gg 1$, it follows from the above expression that $\cos \delta \ll 1$ and, hence, $\Lambda_b^4(\phi) \gg g\Lambda^3 f$. In fact, from (\ref{late_stop}) it follows that $\Lambda_b^4(\phi) \sim 3H_I^4/16\pi^2$, a very different stopping condition compared to (\ref{equal_slopes}). 
\end{itemize} 
We can thus express the final stopping condition approximately as
\beq
\label{eq_new_stop}
\boxed{ \Lambda_b^4 \sim \mathrm{max}\Bigl(g \Lambda^3 f, \frac{3H_I^4}{16\pi^2} \Bigr).}
\eeq

\bigskip

The above observations are verified in the real-time numerical simulations of the Fokker-Planck equation. These were performed using a simple finite difference method. We solved the rescaled version of the FP equation, given by (\ref{FP_rescaled}) as derived in the appendix~\ref{app:stop}. Here, in addition to the initial conditions, for which we use a gaussian distribution with a certain width centered in the $\mu_h^2>0$ region, the two free parameters are $d$ and $b$, the later given by
\beq
b = \frac{(-\mu_h^2)\Lambda^2}{\Lambda_b^4},
\eeq
and characterizing how fast the barriers of the potential increase with the field value. We assume this dependence on $\phi$ to be linear, as is usually the case in the non-QCD relaxion model, although the results are not sensitive to this choice. In our simulations we choose the spatial and the temporal steps small enough so that the results are insensitive to these parameters.

\begin{figure}[!t]
	\centering
	\includegraphics[width=0.84\textwidth]{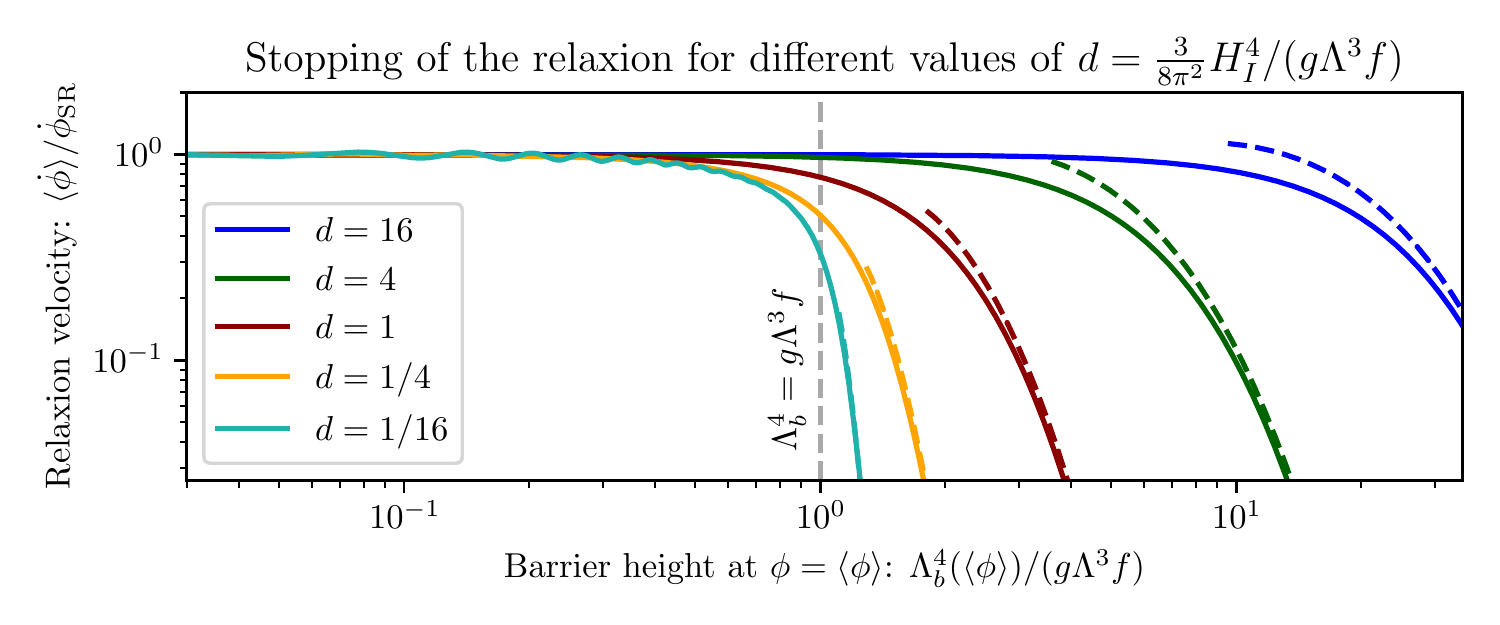}
	\caption{The relaxion velocity from numerical simulation of the Fokker-Planck equation~(\ref{FP_eq}). On the horizontal axis is the rescaled height of the barriers at $\phi = \langle \phi \rangle(t)$, $\Lambda_b^4(\langle \phi \rangle) / (g\Lambda^3 f)$. This variable is equal to one near the first local minima of the potential. On the vertical axis is the mean field velocity $\langle \dot \phi \rangle(t)$ in units of the slow-roll velocity in the absence of wiggles~(\ref{SR_define}). Different colors correspond to different values of the parameter $d$ defined in~(\ref{parameter_d}), demonstrating the modified stopping for large values of this parameter. The dashed lines show the analytical solution  (\ref{velo_minima}) in the region where the argument in the exponent is larger than one. As can be seen the analytical and the numerical approaches agree well.}
	\label{slow_down}
\end{figure}

\begin{figure}[!t]
	\centering
	\includegraphics[width=\textwidth]{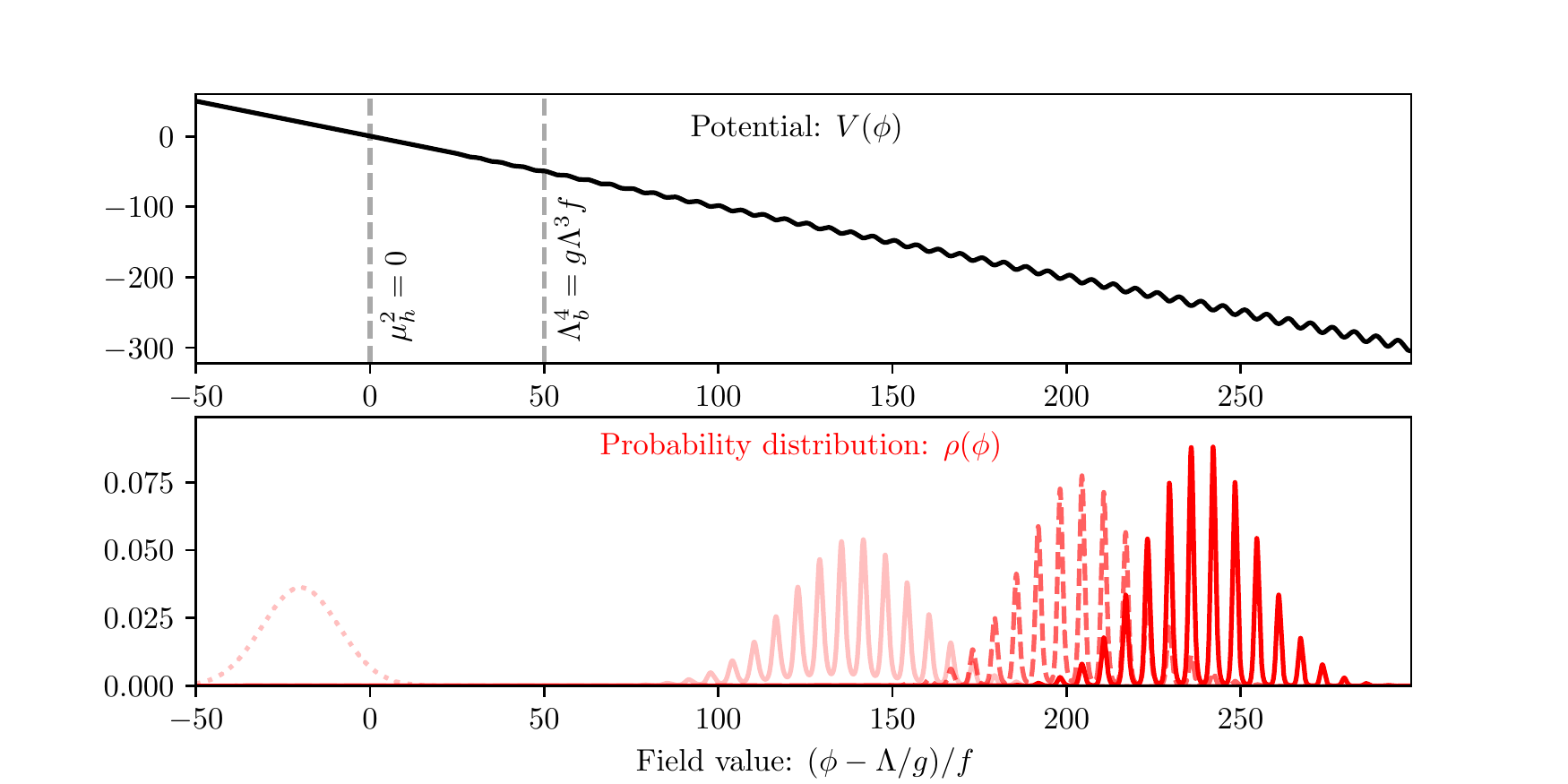}
	\caption{Several snapshots of the relaxion distribution $\rho(\phi)$ (bottom) from numerical simulation of the Fokker-Planck equation~(\ref{FP_eq}), together with the relaxion potential $V(\phi)$ (top). The time interval between the snapshots is $\Delta t = 200({f}/{\dot \phi_{\mathrm{SR}}} )$. Due to relatively strong diffusion effects, $d=2$, the distribution slows down significantly later compared to the first local minimum, at $\Lambda_b^4\sim g\Lambda^3f$.}
	\label{snapshot}
\end{figure}

We first set $b=100$ and perform simulations for different values of $d$. In the figure~\ref{slow_down} the mean velocity of the field $\langle \dot \phi \rangle$ is plotted as a function of the height of the barrier at which the relaxion distribution is centered at that time, $\Lambda_b(\langle \phi \rangle)$. As can be seen, initially $\langle \dot \phi \rangle = \dot \phi_{SR}$, while at late times the velocity shrinks which, however, can happen much later after the field reaches the first local minimum. In this late stage of the dynamics we observe good agreement between the numerical simulation and the analytical estimate based on Eq.~(\ref{velo_minima}), which is shown using dashed lines. We have also checked that the results are insensitive to the value of the parameter $b$, as long as it is large.

In Fig.~\ref{snapshot} we plot several snapshots of the field distribution for $d= 2$ and $b=50$ illustrating the relaxion dynamics. The bottom panel shows the distribution function while in the upper panel is the potential. The snapshots are shown at times separated by intervals of $\Delta t = 200({f}/{\dot \phi_{\mathrm{SR}}} )$. Note that as the distribution reaches larger wiggles, its modulation gets more pronounced. The last stage of the dynamics, when the evolution of the mean field slows down, involves a slow decay of the metastable vacua.

We conclude this section by stressing that, especially for large values of $H_I$, the distribution becomes broad and several minima get populated. One can thus predict the mean field value and the variance of the distribution at some late time, but not the specific value of the field. Moreover, the final value of the mean field and the variance depend on the duration of inflation, even though at late times the velocity is extremely suppressed, thus, the displacement at late times is small. For simplicity, in most of our computations in the following sections we use the expression from Eq.~(\ref{late_stop}) based on $B\sim 1$, as the stopping condition.

\section{Implications for the classical-beats-quantum regime}
\label{sec_implication_CbQ}

The stochastic formalism from the previous section is not restricted to the QbC regime. The later will be discussed in Sec.~\ref{sec:QbC}, while here we revisit the original classical model including stochastic effects. Importantly, even in the CbQ regime the relaxion does not necessarily get trapped in the first local minimum due to these effects. We determine the final minimum of the relaxion in section~\ref{ssec:whichmin}. Knowing in which minimum the relaxion ends up allows one to study more carefully the stability of that local minimum after inflation. In section~\ref{ssec:finitedensity} we discuss the implications for the stability constraints in dense astrophysical environments studied in \cite{Balkin:2021wea}.

Due to the large spread of its probability distribution $\rho(\phi)$, the relaxion can populate several local minima of its potential by the end of inflation. Each of these minima corresponds to a value of the Higgs VEV. An upper bound $H_I<v_{h}$ ensures that the spread in the Higgs VEV over these vacua, according to Eq.~(\ref{variance_at_zero_mass}), $\Delta \mu_h^2 \sim H_I^2$, does not exceed the weak scale and that all vacua correspond to a small Higgs VEV. Throughout this work we impose $H_I<100\mathrm{GeV}$, as it was done in~\cite{Nelson:2017cfv}.

\bigskip

\textbf{No domain walls:} As explained in the previous section, the relaxion slows down around local minima for which the transition rate to the next minimum is very small. These minima have long lifetimes, comparable to the total duration of inflation $N_I$, see Eq.~(\ref{lifetime_of_final_min}). Meanwhile, our observable universe has originated from inflation of a single Hubble patch by around $50$ to $60$ e-folds. For $N_I\gg 60$, which is usually required for a successful implementation of the relaxion mechanism, it is thus extremely unlikely to have domains with different vacua in our surrounding~\cite{Graham:2015cka}.
In other words, inside our observable universe, the relaxion is trapped in one of those local minima that got significantly populated during inflation. This happens to correspond to $v_h \approx 246 \mathrm{GeV}$. The field (and the Higgs mass) can thus be treated as homogeneous, with the usual isocurvature perturbations $\delta \phi \propto H_I$ on top.

\subsection{The final minimum of the relaxion}
\label{ssec:whichmin}

We now estimate the final local minimum of the relaxion in the CbQ regime, both in the QCD as well as in the non-QCD models.

Let us first confirm that $\delta \ll 1$ always holds in the CbQ regime. For this we insert the upper bound on the inflationary Hubble scale $H_I<g^{1/3}\Lambda$ into the stopping condition from Eq.~(\ref{late_stop}) and use $g\Lambda^3f<\Lambda_b^4$ that any local minimum of the relaxion potential satisfies,
\beq
\sin \delta - \delta \cos \delta  = \frac{3H_I^4}{16\pi^2\Lambda_b^4} < \frac{3\Lambda_b^{4/3}}{16\pi^2 f^{4/3}} \: \: \: \: \: \: \: \: \: \: \: \: \: (\text{CbQ}).
\eeq
We observe that the last term is always small both in the QCD and the non-QCD models, implying that $\delta \ll 1$. Recalling that $\cos \delta = {g\Lambda^3f}/{\Lambda_b^4}$ (eq.~\ref{exp_delta}), we observe that the final relaxion minimum indeed satisfies the usual condition from Eq.~(\ref{equal_slopes}), $\Lambda_b^4 \approx g\Lambda^3f$, in the CbQ regime.

We denote by $l$ the index of the final minimum of the relaxion. In  appendix~\ref{app:potential}, it was found that 
$\delta_{l} = \sqrt{l} \delta_{1}$ and $\delta_1^2 = \Lambda_b^4/(\Lambda^2 (-\mu_h^2))$.
In the small $\delta$ limit one can approximate $\sin \delta - \delta \cos \delta \approx \delta^3/3$, and use the above relations 
to express the typical value of $l$ as
\beq
\label{value_of_l_CbQ}
l \sim \frac{1}{\delta_1^2} \Bigl( \frac{9 H_I^4}{16\pi^2\Lambda_b^4} \Bigr)^{2/3}\approx  \frac{(-\mu_h^2)}{g'\Lambda f}\Bigl( \frac{9 H_I^4}{16\pi^2\Lambda_b^4} \Bigr)^{2/3}.
\eeq
The maximal value of $l$ that the relaxion can have in the CbQ regime is then given by
\beq
\label{l_CbQ}
l_{\mathrm{max}} = \frac{(-\mu_h^2)}{g'\Lambda f}\Bigl( \frac{9}{16\pi^2} \frac{\Lambda_b^{4/3}}{f^{4/3}} \Bigr)^{2/3} \: \: \: \: \: \: \: \: \: \: \: \: \: (\text{CbQ}).
\eeq
Similarly, the minimum value of $l$ can be determined from the lower bound on the Hubble scale.

\begin{figure}[!t]
\centering
\includegraphics[width=0.49\textwidth]{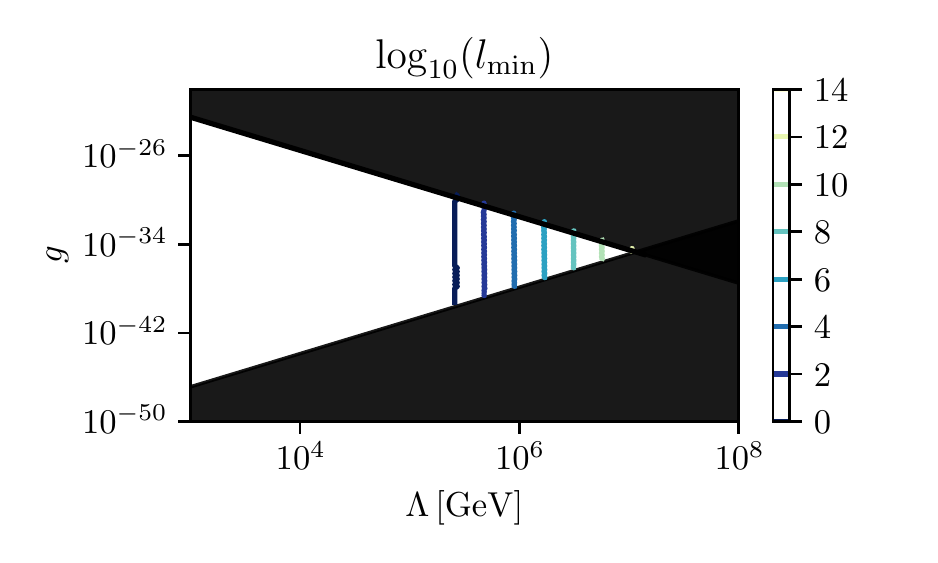}
\includegraphics[width=0.49\textwidth]{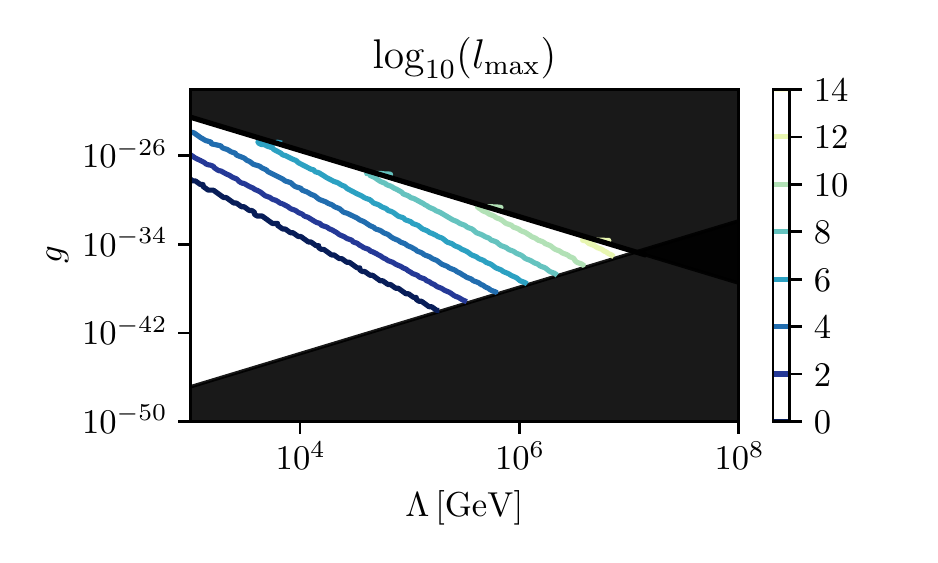}
\caption{In which ($l$-th) minimum does the QCD relaxion stop in the CbQ regime? The contours of the minimal (left panel) and maximal (right panel) values of $l$ (depending on $H_I$) for each point in the $g$ vs $\Lambda$ plane are shown. Even though $l$ can be as large as $10^{13}$, $\theta_{QCD}$ is still of order one in these minima and a modification to this minimal setup (e.g.~the change of the slope of the relaxion potential at the end of inflation) is required to solve the strong CP problem.}
\label{l_QCD}
\end{figure}

\begin{figure}[!b]
\centering
\includegraphics[width=0.32\textwidth]{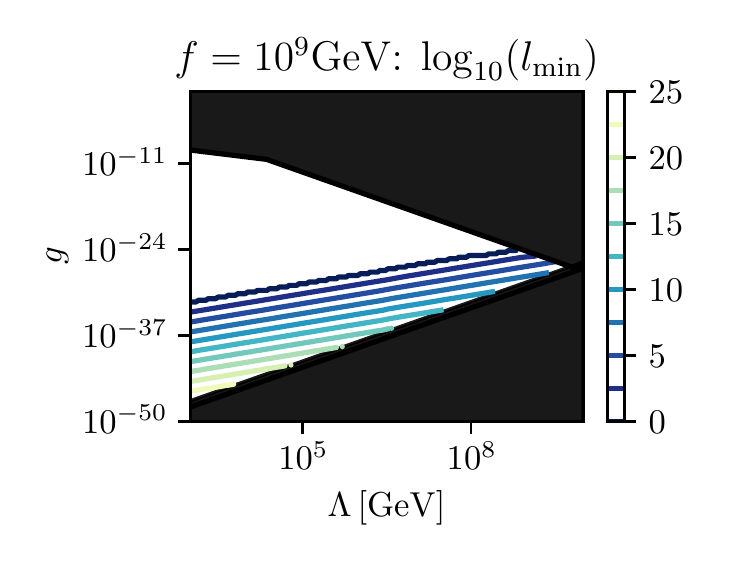}
\includegraphics[width=0.32\textwidth]{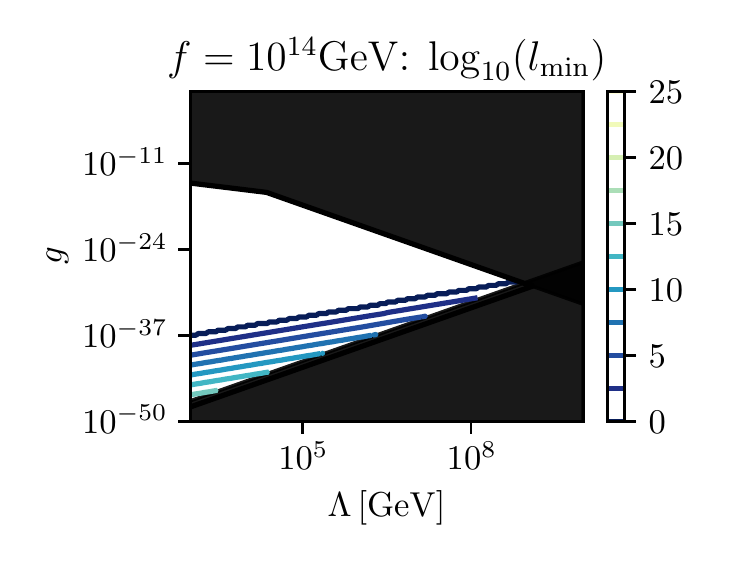}
\includegraphics[width=0.32\textwidth]{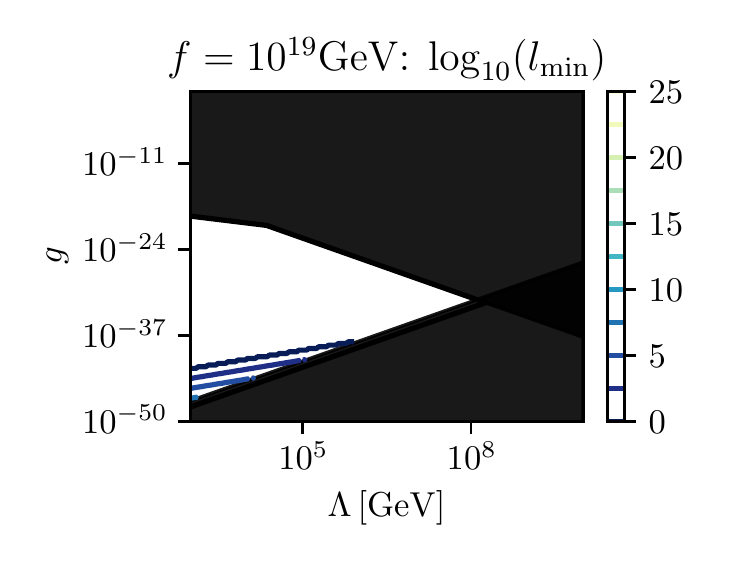}
\includegraphics[width=0.32\textwidth]{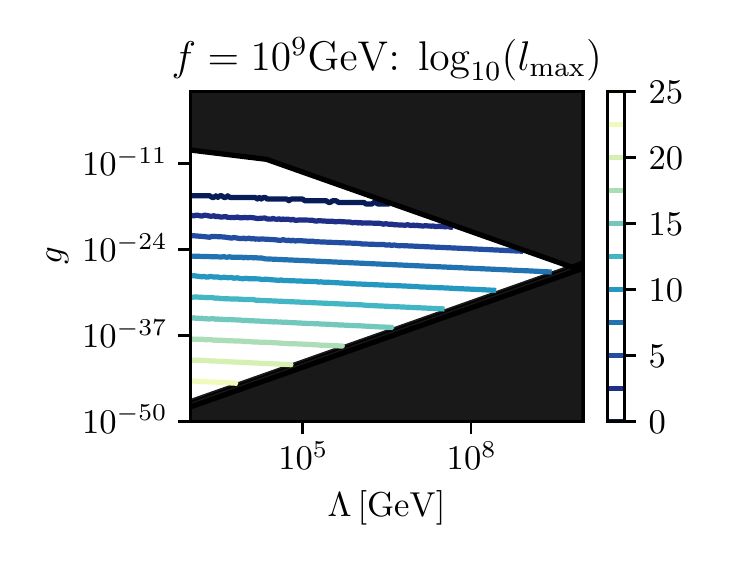}
\includegraphics[width=0.32\textwidth]{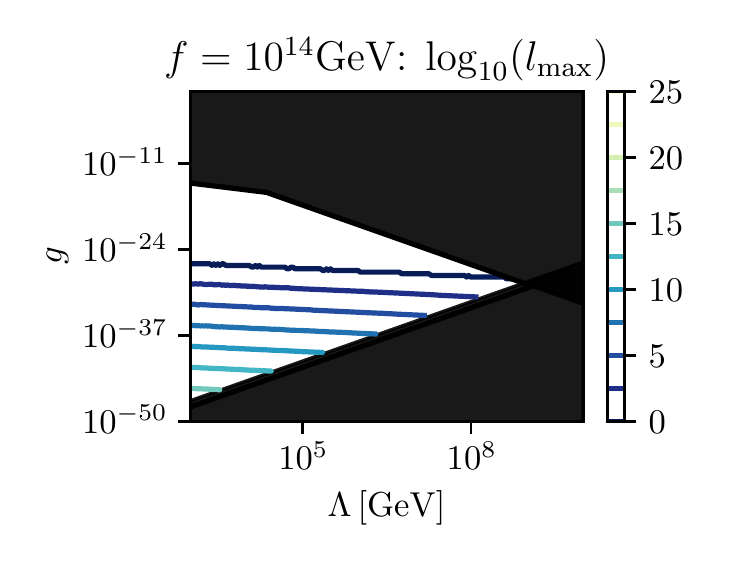}
\includegraphics[width=0.32\textwidth]{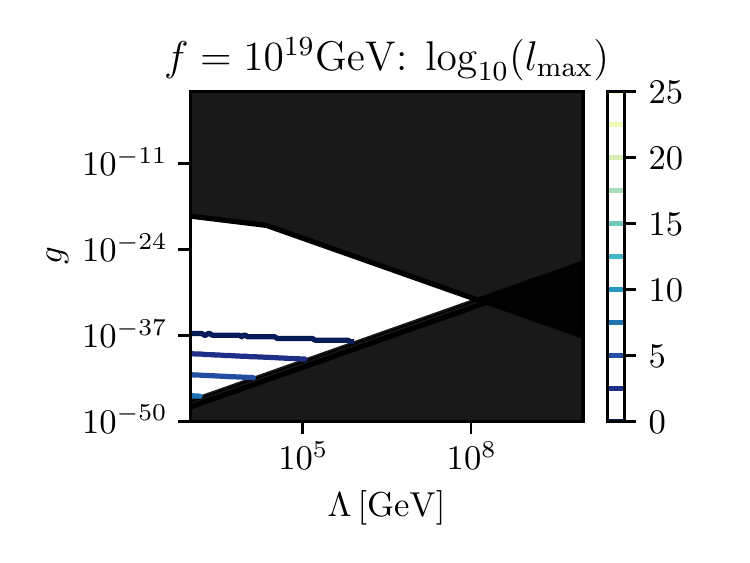}
\caption{In which ($l$-th) minimum does the non-QCD relaxion stop in the CbQ regime? For three values of the decay constant $f=10^{9}\rm GeV$, $f=10^{14}\rm GeV$ and $f=10^{19}\rm GeV$ (corresponding to the first, second and the third columns), the minimal (upper row) and maximal (lower row) values of $l$ (depending on $H_I$) for each point in the $g$ vs $\Lambda$ plane are shown. As can be seen, $l$ can be as large as $10^{25}$ for small values of $g$, while $l=1$ in the white regions for large $g$.}
\label{l}
\end{figure}

\bigskip

\textbf{The QCD model:} Let us consider the QCD relaxion, without any additional change of slope after inflation and without requiring $\theta_{\mathrm{QCD}}$ to be small. In Fig.~\ref{l_QCD} we plot the minimum and maximum values of $l$ as a function of $g$ and $\Lambda$. The value of the decay constant $f$ is fixed from the stopping condition, after which $l$ is a function of the Hubble scale $H_I$. As can be seen, in some regions of the parameter space $l=1$ (the large white regions), while in other regions $l$ can be as large as $10^{13}$.
Even though $l$ can take large values, $\theta_{\mathrm{QCD}}$ is still of order one in all those local minima. This follows from Eq.~(\ref{theta_angle}) and the fact that $\Lambda_b^4 \approx g\Lambda^3f$ holds in the CbQ regime.

We now  determine the relation between $l$ and $\theta_{\mathrm{QCD}}$. We note that the final relaxion field value can be determined from the expression for the Higgs mass in Eq.~(\ref{eq_Higgs_mass}), $\theta = -\mu_h^2/(g' \Lambda f)$. Here the rescaled field variable is defined as $\theta = (\phi - \Lambda/g')/f$ (see also appendix~\ref{app:potential}). Furthermore, the field value at the first minimum can be expressed as
\beq
\theta_{l=1} = \theta  \Bigl( \frac{g\Lambda^3f}{\Lambda_b^4} \Bigr)^2 = \theta \sin^2{\theta_{\mathrm{QCD}}}.
\eeq
where we used the fact that in the QCD model the barriers scale approximately as $\Lambda_b^4(\theta) \propto \sqrt{\theta}$. The value of $l$ can then be estimated as $2\pi l \approx \theta - \theta_{\mathrm{l=1}}$, which leads to
\beq
\label{l_thetaQCD}
l \approx  \frac{\mu_h^2}{g'\Lambda (2\pi f)} \Bigl[ 1- \Bigl( \frac{g\Lambda^3f}{\Lambda_b^4} \Bigr)^2 \Bigr] = \frac{\mu_h^2}{g'\Lambda (2\pi f)}(1-\sin^2\theta_{QCD}).
\eeq
Since $\theta_{QCD} \ll 1$ the last term in the brackets can be neglected. Therefore, the value of $l$ needed to accomodate a small $\theta_{QCD}$ is much larger compared to the actual value from Eq.~(\ref{l_CbQ}) in the CbQ regime. We thus conclude that in the CbQ regime, the QCD relaxion necessarily spoils the solution to the strong CP problem, even when including stochastic effects. As we will see, this conclusion changes in the QbC regime.

\bigskip

\textbf{The non-QCD model:} In Fig.~\ref{l} we plot the minimum and maximum values of $l$ as a function of $g$ and $\Lambda$ for the non-QCD relaxion model from section~\ref{nonQCD_CbQ}. We fix several benchmark values for the decay constant $f$, after which $l$ is again a function of the Hubble scale $H_I$. As can be seen, in some regions of the parameter space $l$ is one (the large white regions in the upper parts), while in other regions $l$ can be as large as $10^{25}$. This determination of $l$ is necessary to assess the fate of the relaxion potential in dense astrophysical environments, which we do next.

\subsection{Implications for the finite-density effects}
\label{ssec:finitedensity}

In~\cite{Balkin:2021wea}, the authors considered the behavior of the relaxion in dense environments, such as stars. Finite density effects can modify the effective relaxion potential and suppress the height of its barriers. For the non-QCD relaxion this effect arises because the Higgs VEV, which determines the height of the barriers, depends on the density of the fields coupled to the Higgs, including baryons. Based on this, the authors derived constraints on the relaxion parameter space by requiring that no relaxion bubbles of lower local minima can form in neutron stars, white dwarfs and sun-like stars. We use the relevant formulas from~\cite{Balkin:2021wea}. One condition is to  require that the barrier disappears inside the core of the star if the average baryonic density $n$ for the objects under consideration satisfies
\beq
n > 3\times 10^{-3} \mathrm{MeV}^3 \Bigl( \frac{\mathrm{TeV}}{\Lambda/\sqrt{l}} \Bigr)^2 \Bigl( \frac{\Lambda_b}{\mathrm{MeV}} \Bigr)^4,
\eeq
while the second condition is that the bubble that emerges from this can overcome the pressure and expand outwards, leading to a bound on the typical radius $r$ of the object
\beq
r > \frac{\Lambda_b^2}{g\Lambda^3}.
\eeq
We use the values of $n$ and $r$ given in ~\cite{Balkin:2021wea}. 

	\begin{figure}[!t]
	\centering
	\includegraphics[width=0.32\textwidth]{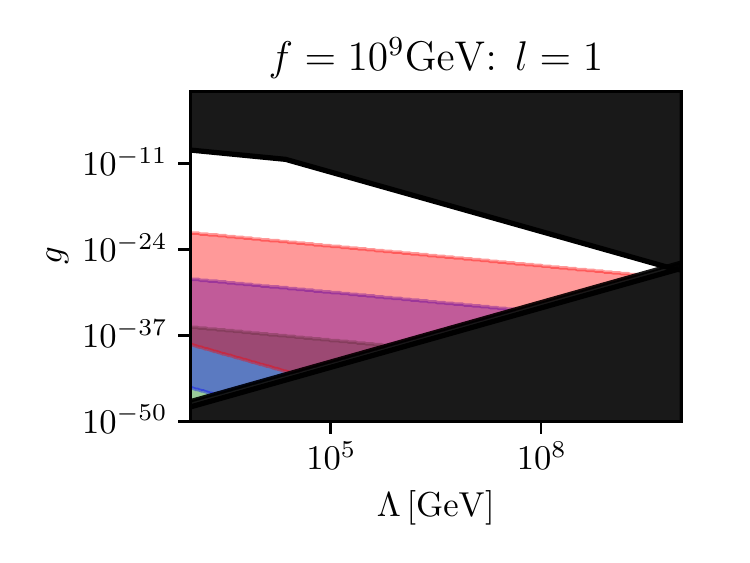}
	\includegraphics[width=0.32\textwidth]{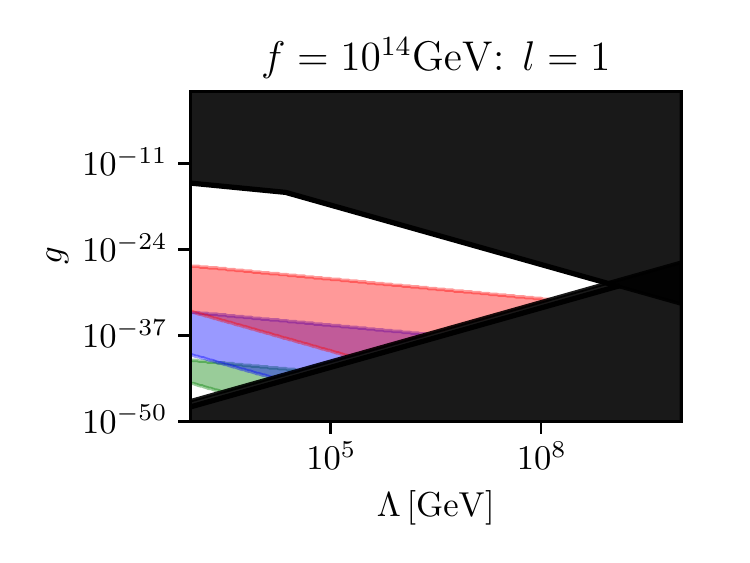}
	\includegraphics[width=0.32\textwidth]{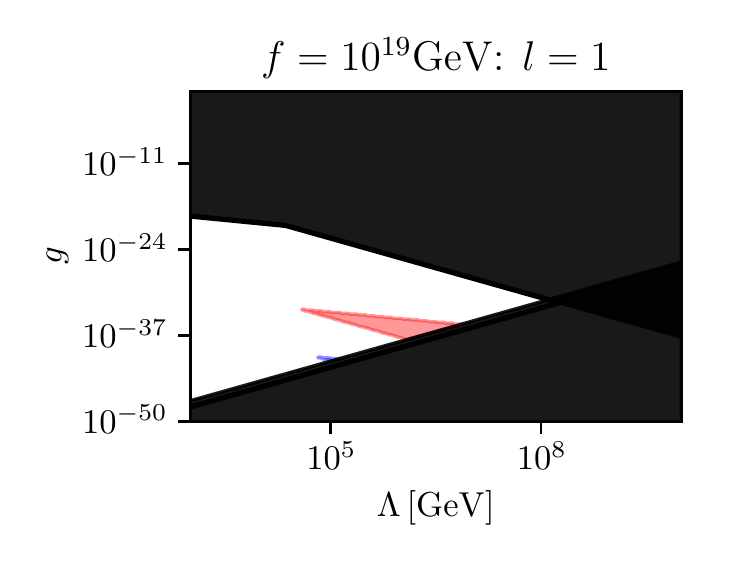}
	\includegraphics[width=0.32\textwidth]{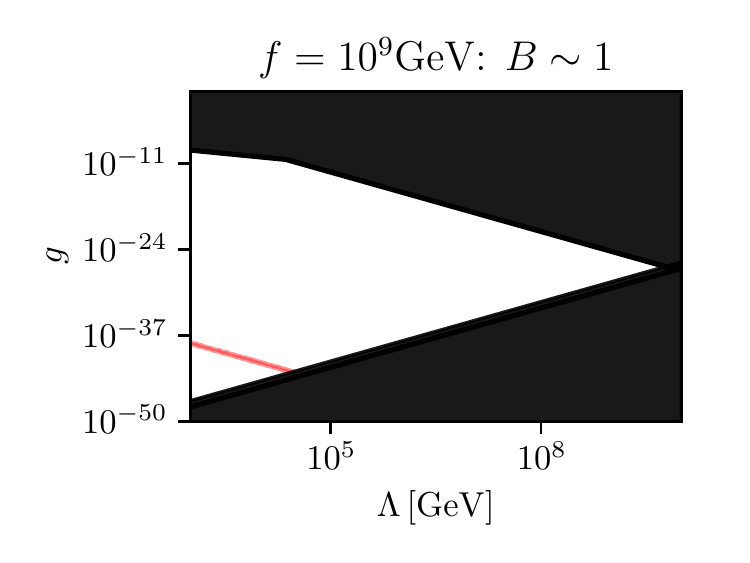}
	\includegraphics[width=0.32\textwidth]{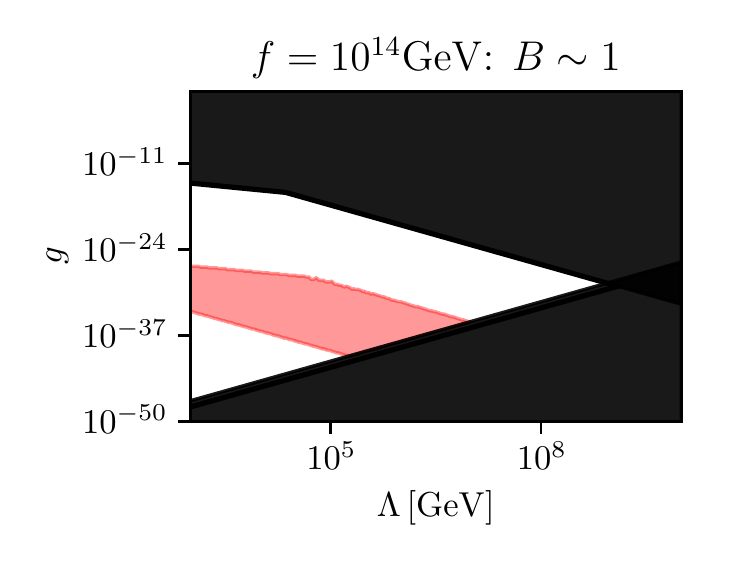}
	\includegraphics[width=0.32\textwidth]{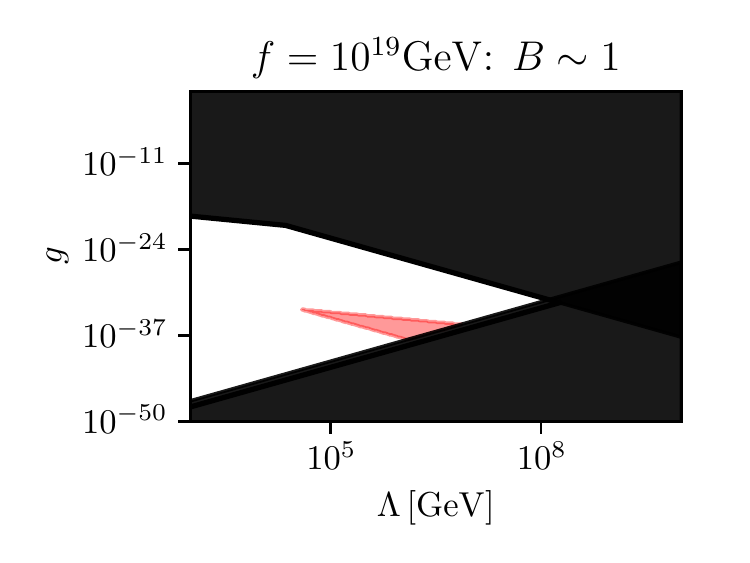}
	\caption{The constraints from the formation and escape of a relaxion bubble induced by neutron stars (red), white dwarfs (blue), and the Sun (green), in the case of the non-QCD relaxion in the CbQ regime. In the upper row $l$ is set to one, while in the lower row it is determined from Eq.~(\ref{value_of_l_CbQ}) corresponding to a small escape rate from the local minimum.}
	\label{stars}
\end{figure}

The first condition above depends strongly on the value of the final local minimum $l$. In the upper panel of figure~\ref{stars}, we present the resulting exclusion regions assuming $l=1$ as was shown in ~\cite{Balkin:2021wea}. They are particularly severe at low $f$ values. On the other hand, we show in the lower panel the correct constraints when using the proper value of $l$ as determined in our analysis from the condition that $B\sim1$ (and displayed in figure \ref{l}). We see that, as expected,  the constraints become much weaker as at the large $l$ values, the potential is less shallow and less sensitive to the finite density effects. In particular, only the neutron stars remain relevant. We will see in section \ref{sec:QbC} and figure \ref{even_more_g_L} that these constraints are even weaker in the QbC regime due to the larger values of $l$.

For the QCD relaxion, the barrier height depends on $\Lambda_{\mathrm{QCD}}$ which in turn changes with the baryon density. The QCD relaxion has to be trapped in a deep, almost $CP$-conserving minimum. In the CbQ regime, this cannot be achieved without a further modification of the potential (e.g.~the change of the slope). Hence,  the value of $l$ determined above is not relevant in this case.

We conclude that the correct determination of the final minimum for the relaxion is crucial for the derivation of the late-time constraints on the relaxion potential from neutron stars and other compact objects and largely opens up the viable relaxion parameter space.

\section{Implications of dropping the classical-beats-quantum condition}
\label{sec:QbC}

We now discuss the implications of dropping the QbC condition for the relaxion. Depending on the value of the Hubble scale during inflation we identify the following three regimes:
\begin{itemize}
	\item The CbQ regime,
	$$\frac{\Lambda^2}{M_{Pl}} < H_I < (g\Lambda^3)^{1/3}.$$
	This case was discussed in sections~\ref{sec:review} and~\ref{sec_implication_CbQ}. The stopping of the relaxion is determined by the relation from Eq.~(\ref{equal_slopes}).
	
	\item The QbC I regime,
	$$(g\Lambda^3)^{1/3} < H_I < \Bigl(\frac{16\pi^2}{3}g\Lambda^3f\Bigr)^{1/4}.$$
	Diffusion effects are now stronger compared to the previous case. However, they are not too strong (the parameter $d$ from Eq.~(\ref{parameter_d}) is small) and the barrier height at the final minimum still satisfies the relation from Eq.~(\ref{equal_slopes}), $\Lambda_b^4 \sim g\Lambda^3f$. One can check that the expression on the left side of the above inequality is always smaller compared to the one on the right side.
	
	\item The QbC II regime with
	$$\Bigl(\frac{16\pi^2}{3}g\Lambda^3f\Bigr)^{1/4} < H_I < 2\pi T_b,$$
	where $T_b$ is the temperature at which the relaxion barriers shrink, i.e.~the QCD scale $\Lambda_{\mathrm{QCD}}$ for the QCD axion model and, at most, the weak scale $v_h$ for the non-QCD model. In this case the diffusion effects are strong ($d>1$) and modify the stopping condition. The final minimum at which the field can get trapped satisfies $\Lambda_b^4 \sim 3H_I^4/16\pi^2$ or, in other words, $\Lambda_b \sim  H_I$.
\end{itemize}

As in the previous section, we impose the constraint $H_I<100 \mathrm{GeV}$ to ensure that the final spread in the Higgs VEV over the relaxion minima that get populated is small. The arguments from the previous section about the absence of different local minima populated inside our observable universe and, correspondingly, about the absence of domain walls~\cite{Graham:2015cka}, apply also in the QbC regime.

In sections~\ref{QCD_stochastic} and~\ref{nonQCD_stochastic} we consider the QCD and the non-QCD models, respectively.

\subsection{The QCD model}
\label{QCD_stochastic}

The main concern when assuming that the relaxion is the QCD axion is the reappearance of the strong CP problem as $\theta_{\mathrm{QCD}}\sim {\cal O}(1)$ is predicted. In the original GKR paper \cite{Graham:2015cka}, this is solved by assuming the existence of some dynamics at the end of inflation that modifies the slope of the relaxion potential, in which case the relaxion can address the hierarchy problem up to a cutoff of order ${\cal O}(30)$ TeV.
In this section, we exploit the new stopping condition from the previous section to reconcile the QCD relaxion mechanism with a small $\theta_{\mathrm{QCD}}$. Another attempt to address this issue was presented by Nelson and Prescod-Weinstein (NP) in ~\cite{Nelson:2017cfv}.
Following~\cite{Gupta:2018wif}, we refer to the mechanisms from~\cite{Graham:2015cka} and~\cite{Nelson:2017cfv} as GKR and NP, respectively.

The $\theta$-angle for QCD, given by (\ref{theta_angle}), is of order one if $\Lambda_b^4 \sim g \Lambda^3 f$. From the modified stopping condition it follows that the relaxion can stop at a much deeper minimum with a smaller $\theta_{\mathrm{QCD}}$ if the inflationary Hubble scale is large enough. We thus require the QbC II regime, where
\beq
H_I \sim \Lambda_b,\: \: \: \: \: \: \: \: \: \: \: \: \: \: \: \: \: \: \: \: g\Lambda^3 f <\Lambda_b^4.
\eeq 
In other words, the mechanism predicts the inflationary Hubble parameter to be close to $75 \mathrm{MeV}$. This however does not guarantee that $CP$ violation is sufficiently small. From Eq.~(\ref{theta_angle}) it follows that the additional constraint
\beq
g\Lambda^3 f < 10^{-10} \Lambda_b^4
\eeq
reconciles $\theta_{\mathrm{QCD}}$ with the experimental bound $10^{-10}$.

As in the original model, we require the relaxion energy density to be subdominant during inflation, see the condition (\ref{subdominant}). The combination of this inequality with the above relation on the Hubble parameter from the stopping condition imposes an upper bound on the cut-off,
\beq
\label{lb_cutoff}
\Lambda< \sqrt{\Lambda_b M_{\mathrm{Pl}} } = 10^{9} \rm GeV.
\eeq
We observe that in the QbC regime, the QCD relaxion can solve the strong CP problem without any further modification, in contrast with the GKR model, where an additional trick with the rolling potential is mandatory.

\begin{figure}[!t]
	\centering
	\includegraphics[width=0.85\textwidth]{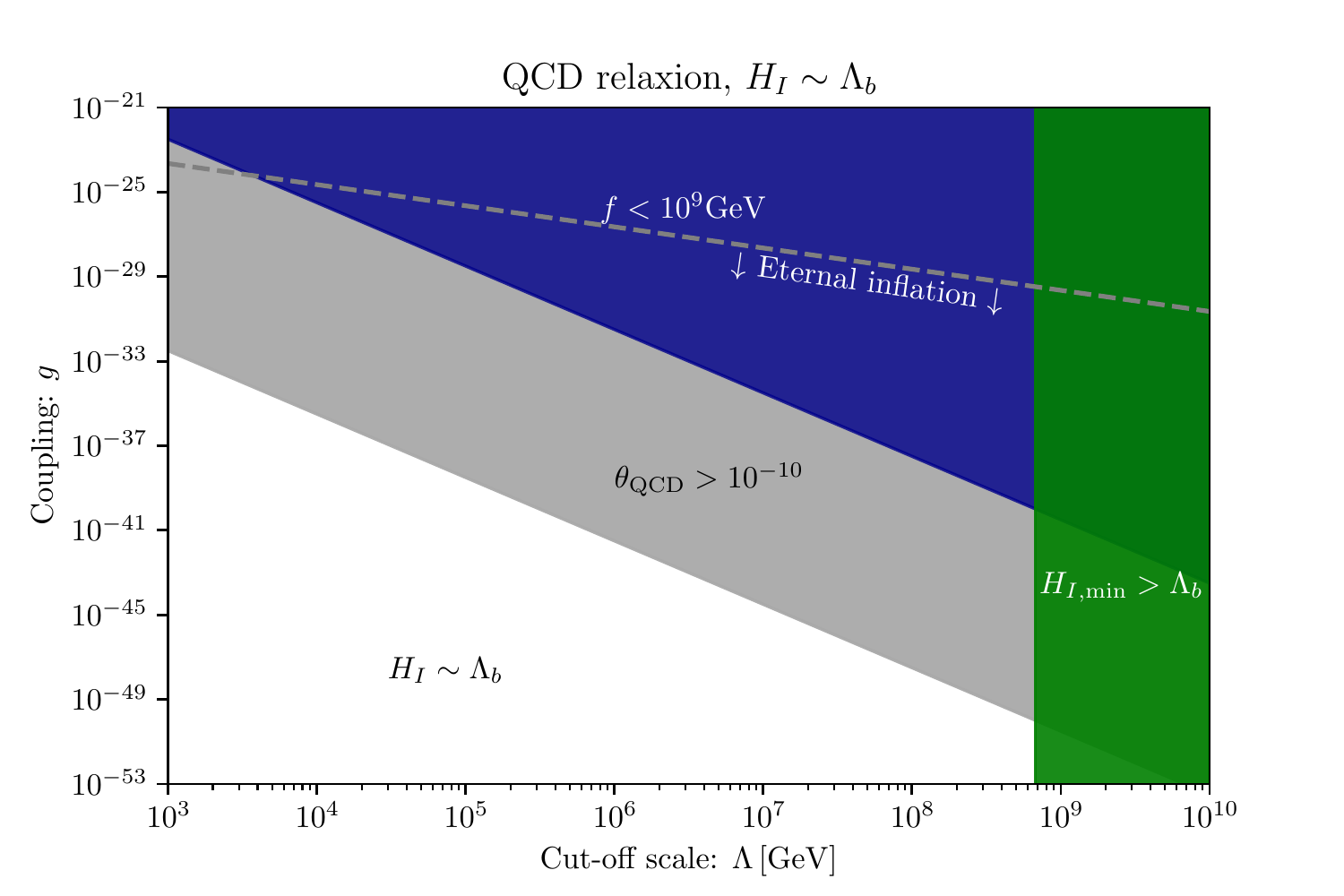}
	\caption{ The QCD relaxion parameter space in the QbC regime in the $g$ vs $\Lambda$ plane. In this scenario $H_I \sim \Lambda_b = 75 \mathrm{MeV}$ and the allowed parameter region is shown in white. The cut-off scale can now be pushed to almost $10^{9} \mathrm{GeV}$ with the maxmial value following from the lower bound on $H_I$ from (\ref{lb_cutoff}). The mechanism requires eternal inflation as the allowed region falls below the dashed line, which corresponds to (\ref{noneternal}).  }
	\label{fig_QCD_QbC}
\end{figure}

We can construct the $g$ vs $\Lambda$ parameter space for the relaxion as it was done in section~\ref{ssec:GKR_QCD}. This is shown in Fig.~\ref{fig_QCD_QbC}. The upper bound in the figure is the same as in the Fig.~\ref{QCD_CbQ} and arises from the combination of the lower bounds on $f$ with the inequality $g\Lambda^3 f <\Lambda_b^4$. The lower bound in the figure, which was due to the CbQ constraint, is no longer present. 

\bigskip

\begin{figure}[!b]
\centering
\includegraphics[width=0.75\textwidth]{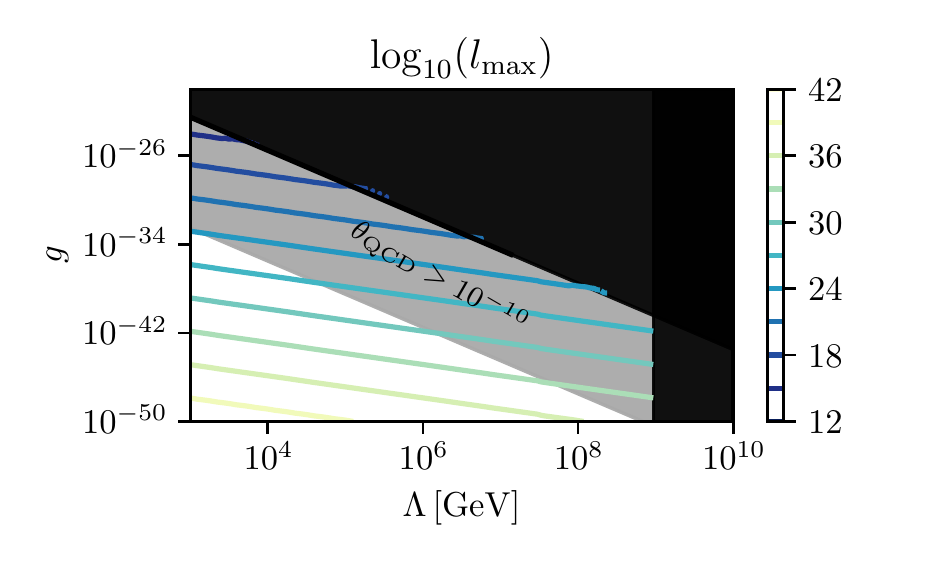}
\caption{ The contours of the maximal values of $l$ for each point in the $g$ vs $\Lambda$ plane for the QCD relaxion in the QbC II regime, with eternal inflation. $l$ can take much larger values compared to the CbQ regime (see Fig.~\ref{l_QCD}) and below the grey shaded region also $\theta_{\mathrm{QCD}}<10^{-10}$ holds. }
\label{l_QCDQbC}
\end{figure}

\textbf{The final minimum of the relaxion:} For completeness, in Fig.~\ref{l_QCDQbC} we show in the $g$ vs $\Lambda$ plane the contours for the maximal values of $l$ that the final relaxion minimum can have in the QbC regime. The region where $\theta_{\mathrm{QCD}}>10^{-10}$ is shaded in grey. The figure can be compared to Fig.~\ref{l_QCD} for the CbQ regime for which $\theta_{\mathrm{QCD}}\sim 1$. Not surprisingly, $l$ can now take much larger values. It can be determined from Eq.~(\ref{l_thetaQCD}) where for the maximal value one can set $f=10^{9}\mathrm{GeV}$.
We can conclude that the final minimum is deep enough such that it is safe from ``runaway" constraints discussed in Section \ref{ssec:finitedensity}.

\bigskip

\textbf{Comparison with the NP model:} Let us at this point compare our predictions with those in the NP model. In~\cite{Nelson:2017cfv} the relaxion stops near the first minimum of the potential, relying on the usual stopping condition
\beq
\Lambda_b^4(\phi; T_I) \sim g\Lambda^3 f.
\eeq
As a justification, the authors explain that in this region of the potential, the backreaction due to the wiggles will add order one corrections to the (initially gaussian) distribution function. The authors then assume that this backreaction stops $\phi$ from evolving. The backreaction time is estimated treating the wiggles as small perturbative corrections.

While  backreaction effects set in quite quickly, we find that they do not necessarily prevent the one-point function from evolving further. Even if the relaxion is trapped inside a local minimum, it can have a large probability to escape to a lower minimum due to diffusion effects, which generates a drift motion for the relaxion. Our analysis predicts that for the values of the inflationary Hubble parameter that were mentioned in the NP model, $H_I>3\mathrm{GeV}$, the relaxion would slow down around a very large barrier, $\Lambda_b(\phi) >3\mathrm{GeV}$. This would still increase after inflation by another factor of at least $10^{2.5}$ due to temperature dependence of $\Lambda_b$, which obviously corresponds to a larger value of the Higgs VEV. The only way to ensure that the relaxion stops near the correct minimum is to have $H_I \sim \Lambda_b \sim 75 \mathrm{MeV}$. At such low temperatures the QCD potential has already reached its zero temperature limit and, hence, is approximately constant.

\begin{figure}[!t]
	\centering
	\includegraphics[width=0.327\textwidth]{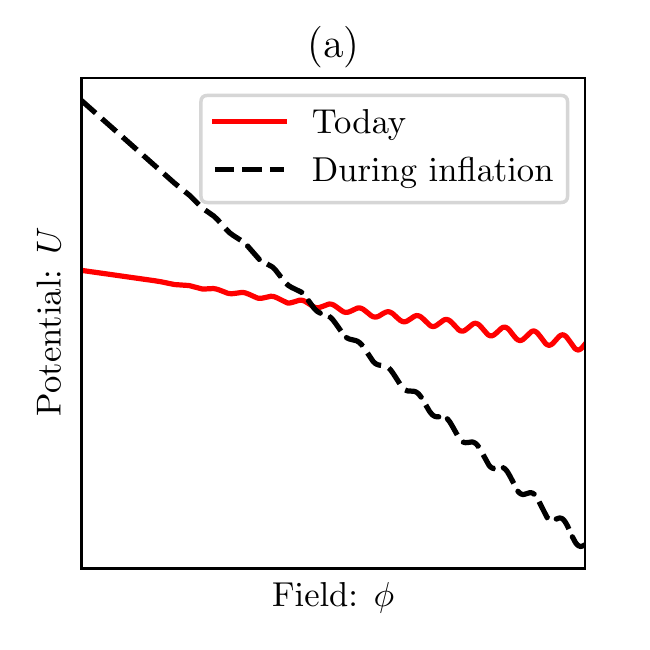}
	\includegraphics[width=0.327\textwidth]{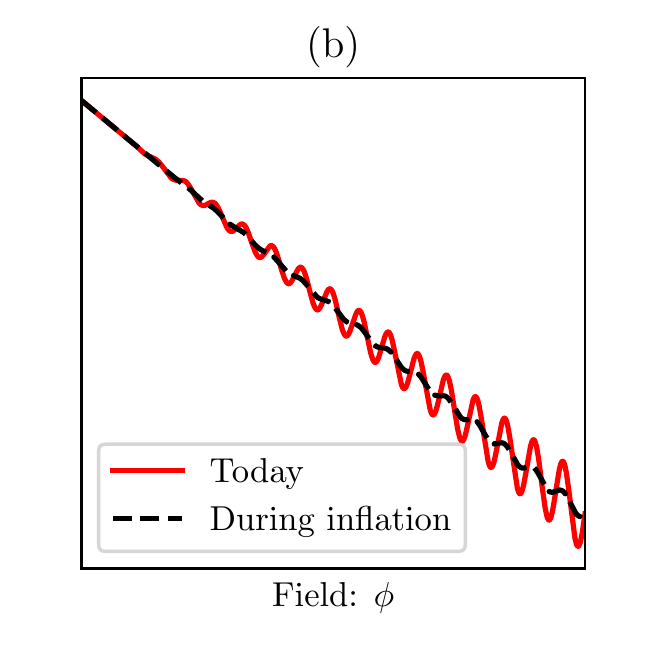}
	\includegraphics[width=0.327\textwidth]{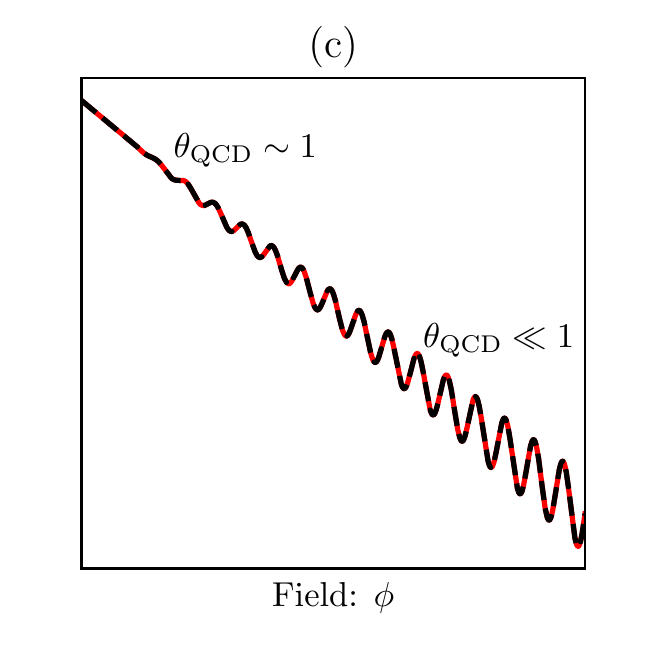}
	\caption{Illustrative comparison of different approaches to the strong CP problem in the QCD relaxion model. In the GKR approach (a) the slope of the rolling potential decreases after inflation, reducing $\theta_{\mathrm{QCD}}$ to a small value. In the NP model (b) the relaxion barriers are suppressed during inflation due to thermal effects at $T_I \sim H_I > \Lambda_{\mathrm{QCD}}$, and increase after inflation, reducing $\theta_{\mathrm{QCD}}$. However, NP assumed that the relaxion stops near the first local minimum, which our analysis contradicts, we therefore find that (b) is not viable. Based on the modified stopping condition explained in this work (c) the relaxion stops at a late minimum due to strong diffusion effects for $H_I \sim \Lambda_b$, resulting in a small $\theta_{\mathrm{QCD}}$. Note that in the third approach, there is no need for the potential to change after inflation.}
	\label{different_QCD}
\end{figure}

As a summary, in Fig.~\ref{different_QCD} we illustrate all three approaches to solving the CP problem.

\bigskip

\textbf{Eternal inflation:} As can be seen in Fig.~\ref{fig_QCD_QbC}, a new parameter region with small values of $g$ opens up if the CbQ constraint is dropped. This region corresponds to a small slope of the rolling potential and, therefore, a larger number of inflationary e-folds that are required to scan the Higgs mass. It turns out that there is a critical number of e-folds $N_{{c}}$, given by
\beq
\label{crit_num_efolds}
N_{{c}} \sim \frac{2\pi^2}{3} \frac{M_{\mathrm{Pl}}^2}{H_I^2}, 
\eeq
such that if inflation lasts longer compared to $N_c$, the universe is in a regime of eternal inflation~\cite{Graham:2018jyp, Dubovsky:2011uy}. In this case, inflation never ends globally, as in the non-eternal scenario. Instead, even though each Hubble patch eventually reheats, most of the universe is always inflating as inflating patches replicate themselves very efficiently due to the large Hubble expansion rate. For the simple case of a single-field slow-roll inflation, the condition of having eternal inflation can be shown to result from requiring the quantum-beats-classical regime for the inflaton dynamics.

Eternal inflation is not well-understood  when it comes to computing probabilities, which can lead to contradicting predictions. To avoid this scenario, we mostly focus on the non-eternal scenario in this work.

To have the relaxion mechanism without eternal inflation, the required number of inflationary e-folds (\ref{req_num_efolds}) should be smaller compared to $N_c$, which leads to a constraint,
\beq
\label{noneternal}
g\Lambda> \frac{3H^2}{\sqrt{2}\pi M_{Pl}}.
\eeq
If (\ref{noneternal}) holds, one can have a finite slow-roll inflation with $N_I$ e-folds such that $N_{\mathrm{req}} < N_I < N_{c}$. One can check that if the relaxion is in the CbQ regime, the required number of e-folds is guarantied not to exceed the critical bound and, therefore, eternal inflation can always be avoided. In contrast, for the QbC relaxion one has to check this explicitly. For the case of the QCD relaxion, it turns out that imposing the bound (\ref{noneternal}) puts very strong constraints on the parameter space and excludes the entire allowed region. In other words, the scenario described in this subsection \textit{requires eternal inflation}. The situation is different in the non-QCD model, which is discussed in the next subsection.

\subsection{The non-QCD model}
\label{nonQCD_stochastic}

\begin{figure}[!t]
	\centering
	\includegraphics[width=0.245\textwidth]{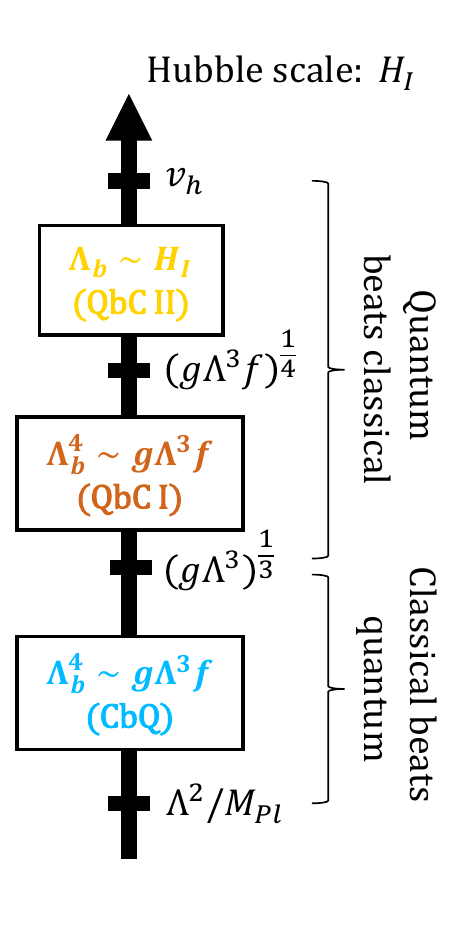}
	\includegraphics[width=0.73\textwidth]{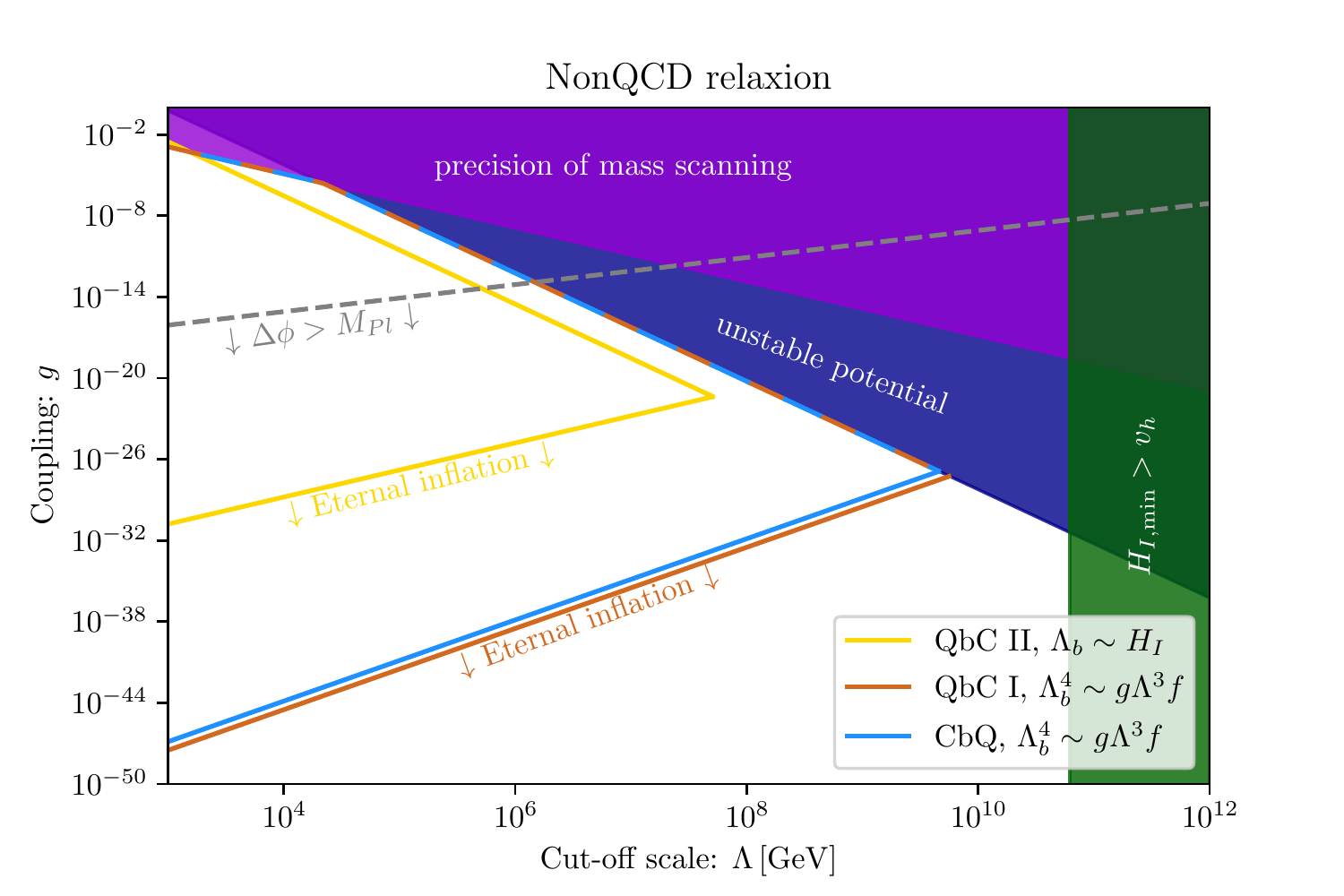}
	\caption{ The non-QCD relaxion parameter space. Depending on the value of $H_I$, the relaxion can be in one of the CbQ, QbC I or QbC II regimes, as illustrated in the left panel. In the right panel, the parameter region is shown in the $g$ vs $\Lambda$ plane. The contours for the QbC I and the QbC II regimes denote the regions where eternal inflation is not required. If one drops this requirement, also the lower part of the white region becomes available. Below the dashed line the relaxion requires a superplanckian displacement.  }
	\label{fig_nonQCD_QbC}
\end{figure}

We now study the implications of dropping the CbQ constraint for the non-QCD relaxion. 
As in the previous case, the $g$ vs  $\Lambda$ parameter space gets extended to small value of $g$ when the CbQ constraint is no longer imposed. In the non-QCD model, the $\theta$ angle is unconstrained, therefore, the new region includes both types of stopping conditions in the QbC regime. We summarize this in the left panel of Fig.~\ref{fig_nonQCD_QbC}. Throughout this work we use the yellow, brown and blue colors for the QbC II, QbC I and CbQ regimes of the non-QCD relaxion, respectively.

In the non-QCD model, $\Lambda_b$ is no longer fixed and can take values as large as (\ref{Lambda_b_upper}). The upper bound on the cut-off follows from the combination of (\ref{subdominant}) with the upper bound on the inflationary Hubble parameter $H_I<100 \mathrm{GeV}$, that is required to have a relatively small spread in the relaxion distribution at the end of relaxation, according to Eq.~(\ref{variance_at_zero_mass}),
\beq
\label{upper_bound_L_nonQCD}
\Lambda < \Bigl( \frac{3}{8\pi}\Bigr)^{1/4} \sqrt{M_{\mathrm{Pl}} H_I} \approx 2\times 10^{10} \mathrm{GeV} \Bigl(\frac{H_I}{100 \mathrm{GeV}}\Bigr)^{1/2},
\eeq

We show the allowed parameter region for this model in white in the right part of Fig.~\ref{fig_nonQCD_QbC}. Compared to Fig.~\ref{nonQCD_CbQ} for the CbQ relaxion, the lower bound on the parameter region is replaced by the upper bound on the cut-off from (\ref{upper_bound_L_nonQCD}).  The dashed line in the plot corresponds to $\Delta \phi \sim \Lambda/g' = M_{\mathrm{Pl}}$, below which the required displacement of the relaxion is superplanckian.

\bigskip

\textbf{Eternal versus non-eternal inflation:} As in the QCD model, it is important to check whether the relaxion necessarily requires eternal inflation or not, i.e.~the condition~(\ref{noneternal}). Combining this inequality with (\ref{subdominant}) and eliminating $H_I$ one arrives at the lower bound on $g$ for non-eternal inflation,
\beq
\label{noneternal_g_L}
g> \frac{8}{\sqrt{2}} \frac{\Lambda^3}{M_{\mathrm{Pl}}^3}.
\eeq
The resulting parameter region is shown with the brown line in Fig.~\ref{fig_nonQCD_QbC}. Below the lower bound of this region eternal inflation is required. Importantly, the lower bound approximately coincides with the one in the CbQ relaxion, arising from eliminating $H_I$ in (\ref{eq:range_of_H}). The allowed region for the CbQ relaxion is marked with the blue line in Fig.~\ref{fig_nonQCD_QbC}. The same applies for the upper bound on the cut-off, which can thus be comparable to the one in the CbQ case, given by Eq.~(\ref{upper_bound_L}).

If one restricts to the late stopping condition $\Lambda_b\sim H_I$, the allowed parameter region shrinks further. In this case, we have
$H_I^4 > ({16\pi^2}/{3}) g \Lambda^3 f >({16\pi^2}/{3}) g \Lambda^4$. Combining this with (\ref{noneternal}) and eliminating $H_I$, one arrives at the lower bound, whereas combining with $H_I<100 \mathrm{GeV}$ and eliminating $H_I$ gives the upper bound,
\beq
\frac{24 \Lambda^2}{M_{\mathrm{Pl}}} < g <  \frac{3}{16\pi^2} \frac{(100 \mathrm{GeV})^4}{\Lambda^4}.
\eeq
The corresponding region is encompassed with the yellow line in Fig.~\ref{fig_nonQCD_QbC}. The cut-off can be pushed to quite high values even in this case,
\beq
\Lambda<1.5 \times 10^{7} \mathrm{GeV} \Bigl( \frac{H_I}{100\mathrm{GeV}}\Bigr)^{2/3}.
\eeq
We therefore observe that, in contrast to the QCD model, eternal inflation is not required. 

\bigskip

\textbf{The volume-weighted distribution:} The requirement of non-eternal inflation, $N_I<N_{\mathrm{req}}$ with $N_{\mathrm{req}}$ given by Eq.~(\ref{noneternal}), plays a crucial role from the point of view of volume-weighting effects. Recall that  Eq.~(\ref{FP_eq}) describes the relaxion evolution inside a single Hubble patch and does not take into account the dependence of the Hubble expansion rate on the relaxion energy $V(\phi)$. This dependence is expected to be small since the relaxion is assumed to constitute a tiny fraction of the total energy density of the universe from~(\ref{subdominant}). The small difference in the energy density can however add up over long periods of time and become significant in the end. Furthermore, once the slow-roll dynamics of the inflaton completes after $N_I$ e-folds,~(\ref{subdominant}) is clearly no longer satisfied. It turns out that the condition~(\ref{noneternal}) ensures that the mechanism is not spoiled if volume-weighting is included in the computation of the probabilities, as we clarify below:

\begin{figure}[!t]
	\centering
	\includegraphics[width=0.49\textwidth]{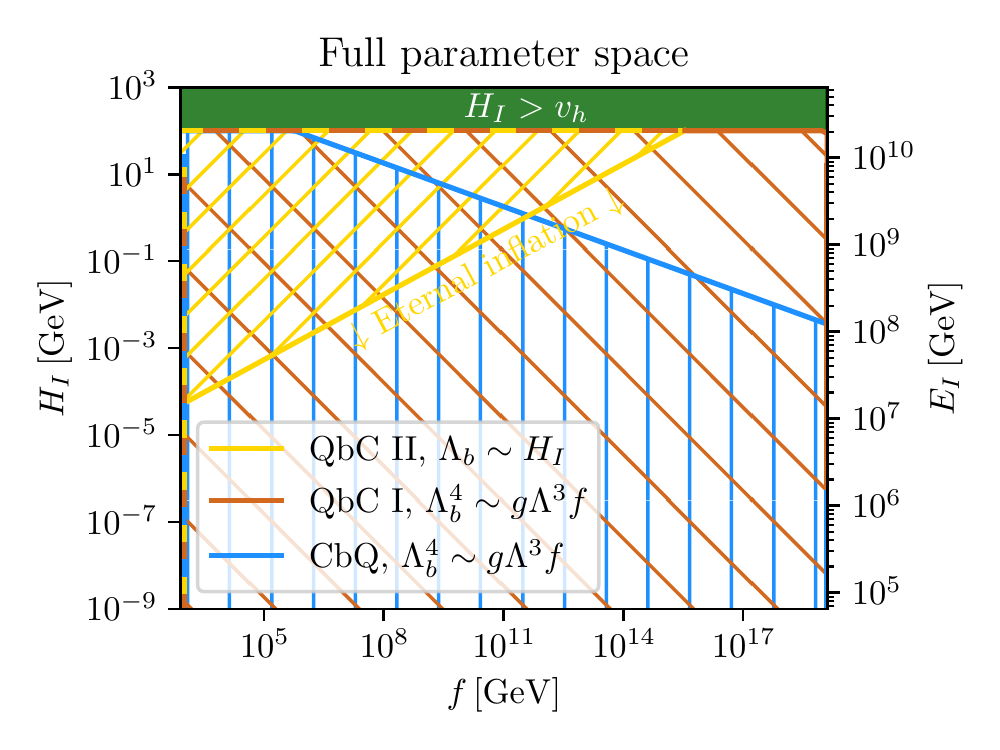}
	\includegraphics[width=0.49\textwidth]{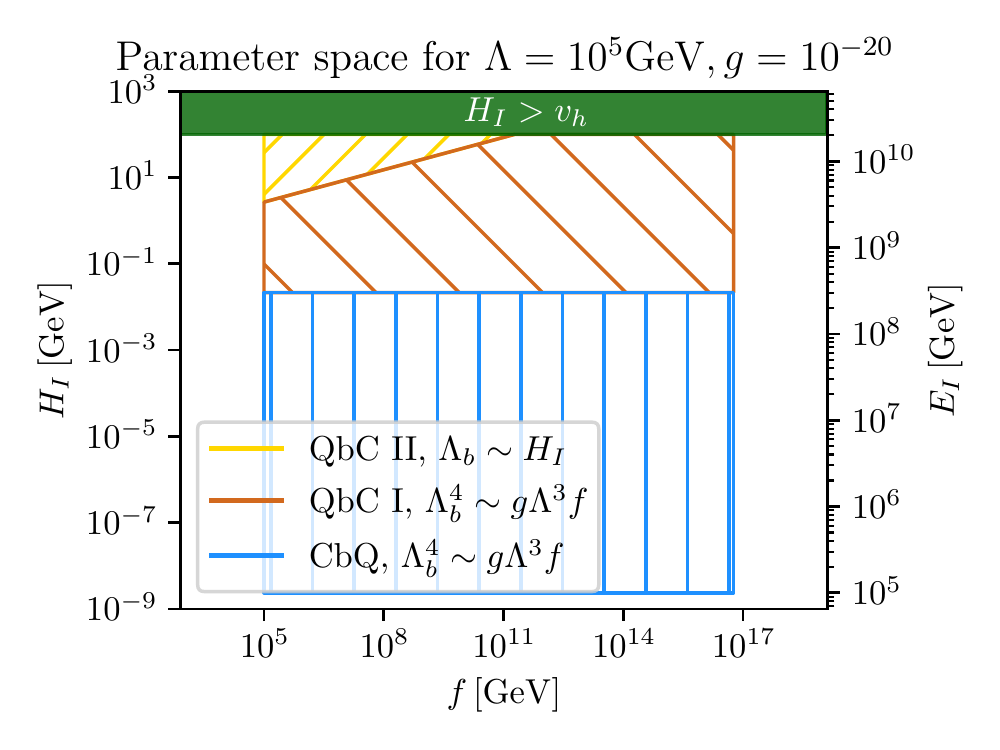}
    \caption{The extended parameter region for the non-QCD relaxion model in the QbC regime, in the $H$ vs $f$ plane. The blue line encloses the available region in the CbQ regime (GKR) whereas the yellow and brown lines correspond to QbC I and QbC II regimes, where eternal inflation is not required. In left panel, the remaining free parameters $g$ and $\Lambda$ are not fixed, while in the right panel we set $g=10^{-20}$ and $\Lambda = 10^{5}\mathrm{GeV}.$}
    \label{H_vs_f}
\end{figure}

\begin{itemize}
	\item Incorporating the volume-weighting in the FP equation is straightforward. As explained in appendix~\ref{app_vol} (see also~\cite{Graham:2018jyp}) this results in an additional term in the equation,
	\beq
	\label{FPV_eq}
	\frac{dP}{dt} = \frac{1}{3H_I}\frac{\partial(P \:  \partial_\varphi V)}{\partial\phi} + \frac{H_I^3}{8\pi^2} \frac{\partial^2P}{\partial\phi^2} + \frac{4\pi}{M_{\mathrm{Pl}}^2} \frac{V}{H_I  } P.
	\eeq
	Here $P(\phi, t) = e^{3(H(\phi)-H_I)t} \rho(\phi, t)$, which can be understood as the rescaled total volume occupied by patches with relaxion field value $\phi$. It is still assumed that the relaxion is subdominant compared to the inflaton. As we demonstrate in appendix~\ref{app_vol}, as long as the inequality (\ref{noneternal}) is satisfied, the difference in the potential energy of the relaxion can be neglected and the volume-weighted probability distribution $P$ ends up in the correct local minimum at the end of inflation.
	
	\item  After $N_I$ e-folds of (slow-roll) inflation the universe may still keep inflating in some parts. In particular, there will be a tiny fraction of patches with a wrong Higgs/relaxion VEV, including the ones where the Higgs mass is large, $\mu_h^2\sim \Lambda^2$. The small probability of such patches may potentially be compensated by a faster expansion rate, associated to the larger potential energy of the relaxion in those patches (the difference in potential energies for $\Delta \phi \sim \Lambda/g$ is $\Delta U \sim \Lambda^4$). One might wonder whether this can spoil the solution to the hierarchy problem. The fate of the ``wrong'' patches is also discussed in the appendix~\ref{app_vol}. We follow the more detailed analysis, considering the measure problem, from~\cite{Gupta:2018wif}. The conclusion is that if the condition (~\ref{noneternal}) for non-eternal inflation is imposed, patches with a small Higgs VEV are still the most likely ones.
\end{itemize}

So far we did not make any statement about the value of cosmological constant (CC) in the correct patches and, in particular, did not require that we end up in a value with a small CC. As in \cite{Graham:2015cka}, we assume that the CC in our observable universe is tuned to the correct value independently of the relaxion mechanism.

\begin{figure}[!t]
	\centering
	\includegraphics[width=0.32\textwidth]{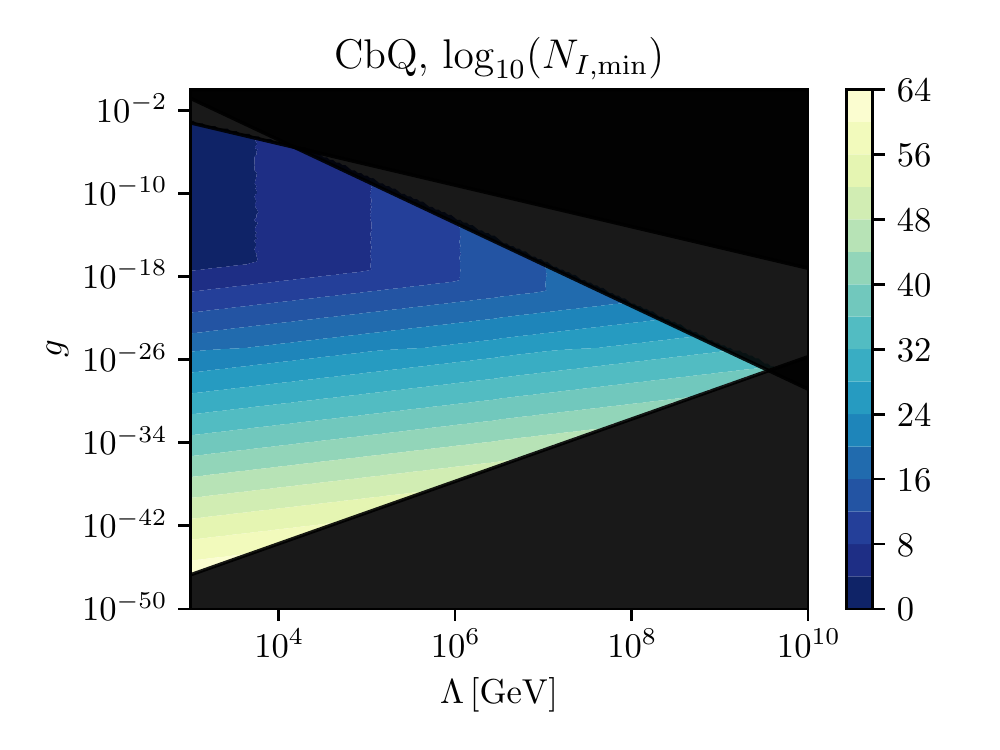}
	\includegraphics[width=0.32\textwidth]{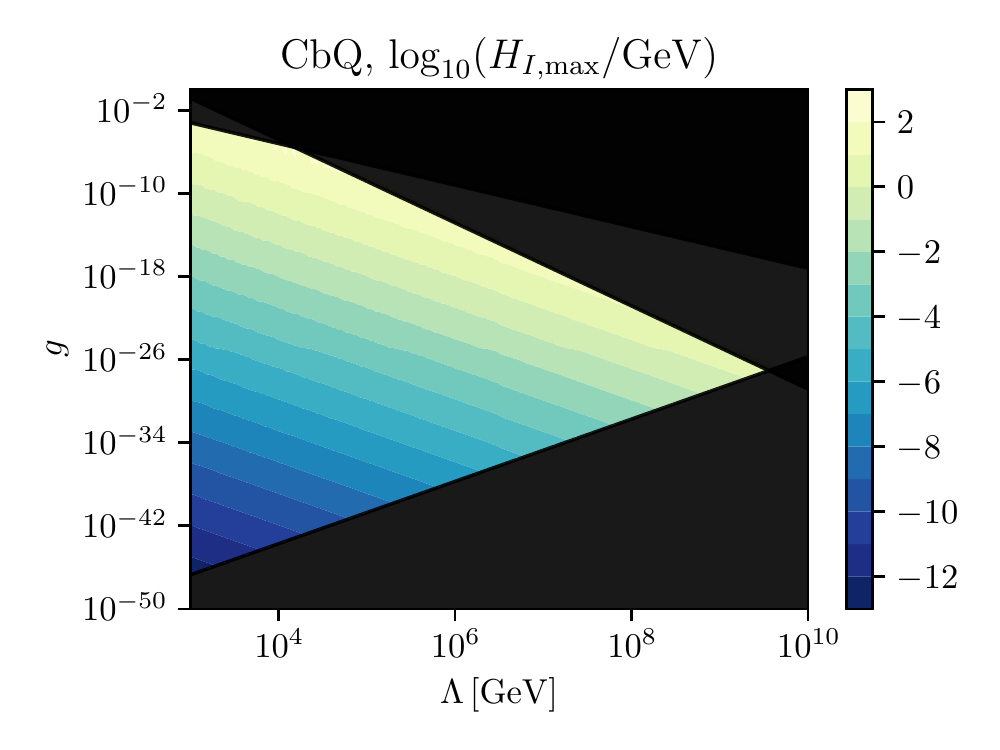}
 	\includegraphics[width=0.32\textwidth]{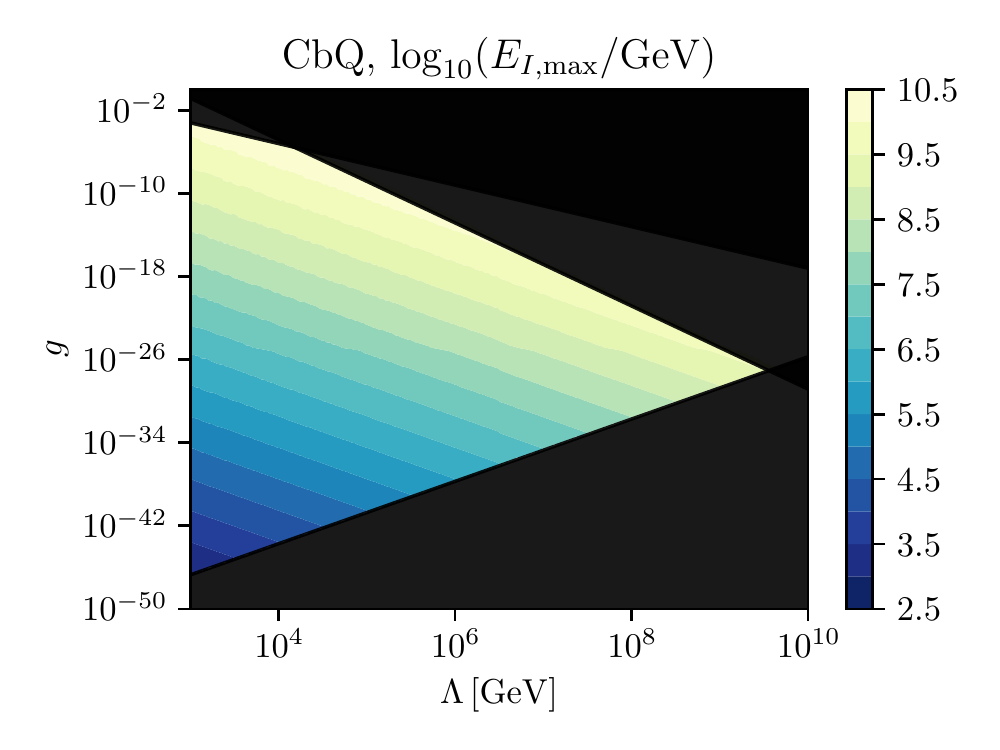}
  	\includegraphics[width=0.32\textwidth]{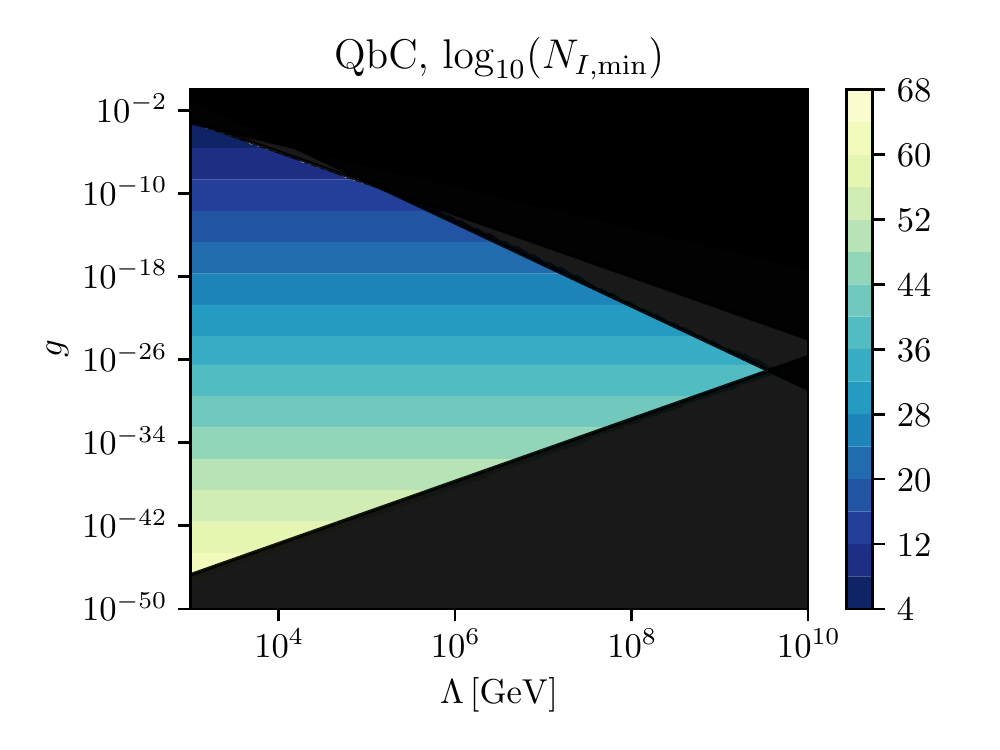}
	\includegraphics[width=0.32\textwidth]{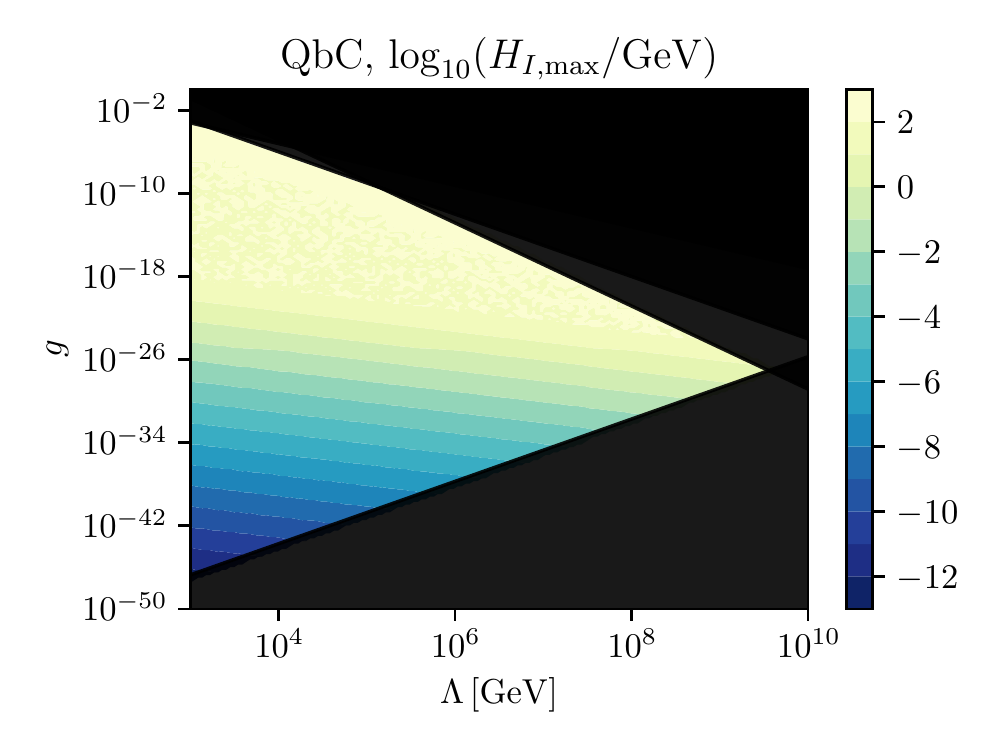}
 	\includegraphics[width=0.32\textwidth]{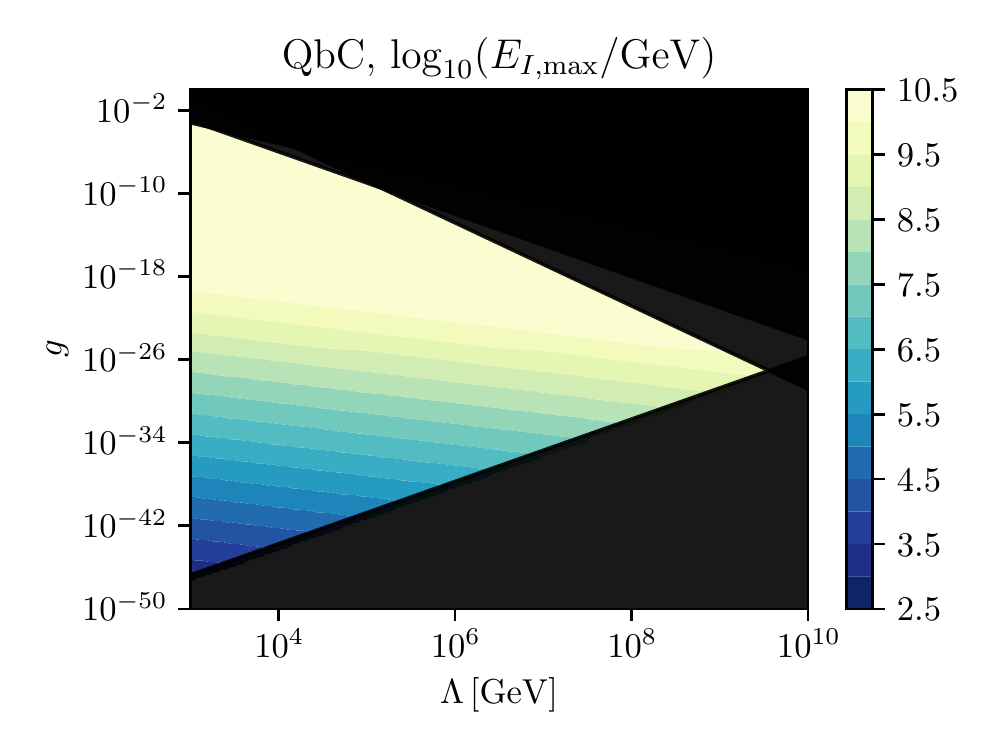}
    \includegraphics[width=0.32\textwidth]{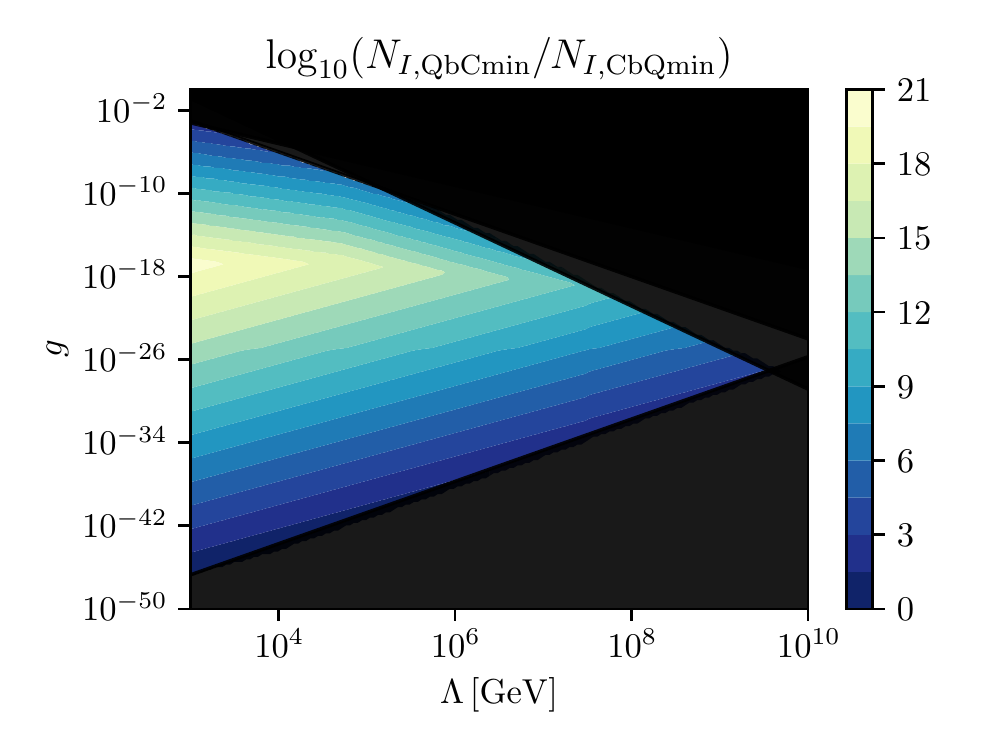}
	\includegraphics[width=0.32\textwidth]{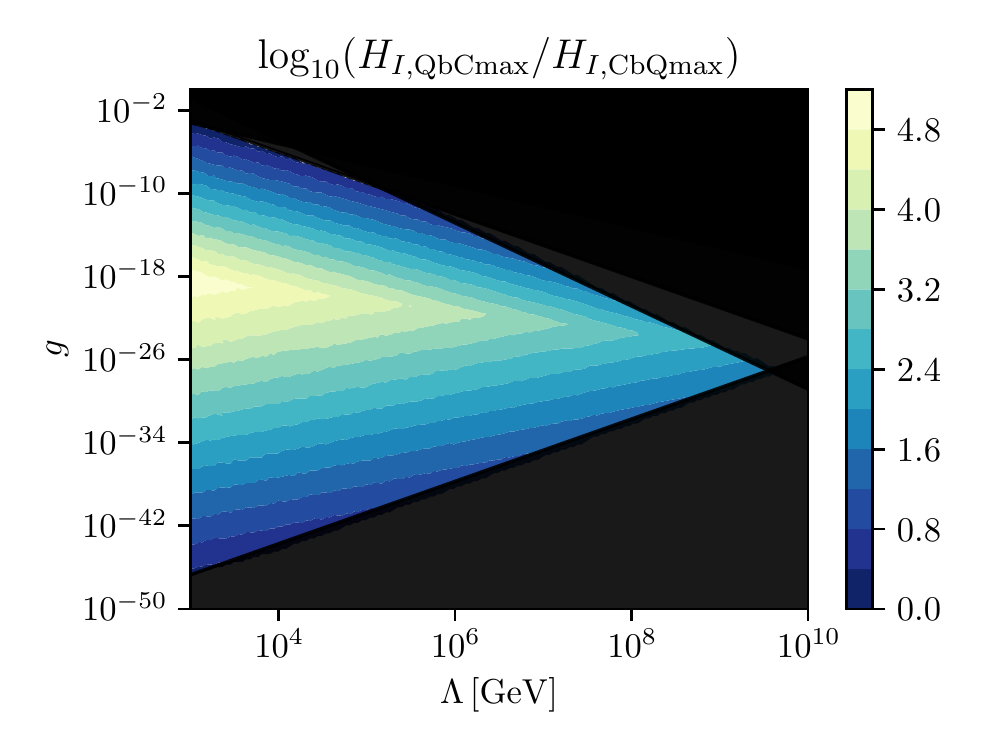}
 	\includegraphics[width=0.32\textwidth]{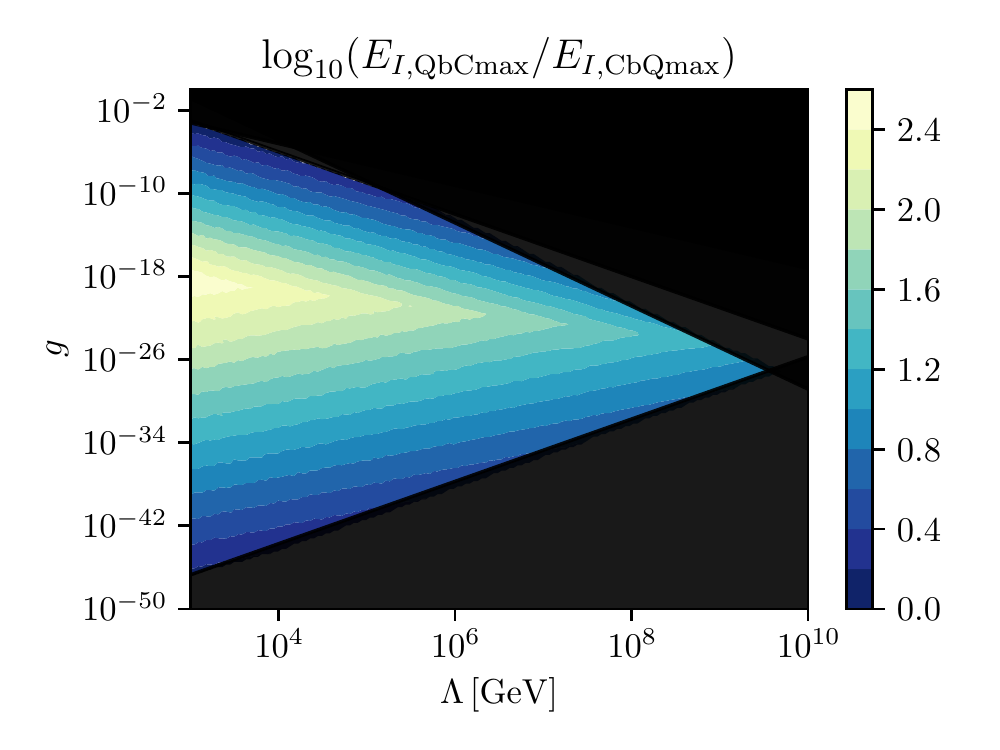}
	\caption{Comparison of predictions between the CbQ regime (upper panel) and the QbC regime without eternal inflation (middle panel) for  the non-QCD relaxion model: Values of the required minimal number of e-folds for relaxation (left), maximal possible value for the inflationary Hubble scale (center), maximal possible value of the inflation scale $E_I$ (right). The ratio of values is shown in the lower panel.}
	\label{more_g_L_test}
\end{figure}

\begin{figure}[!t]
	\centering
    \includegraphics[width=0.32\textwidth]{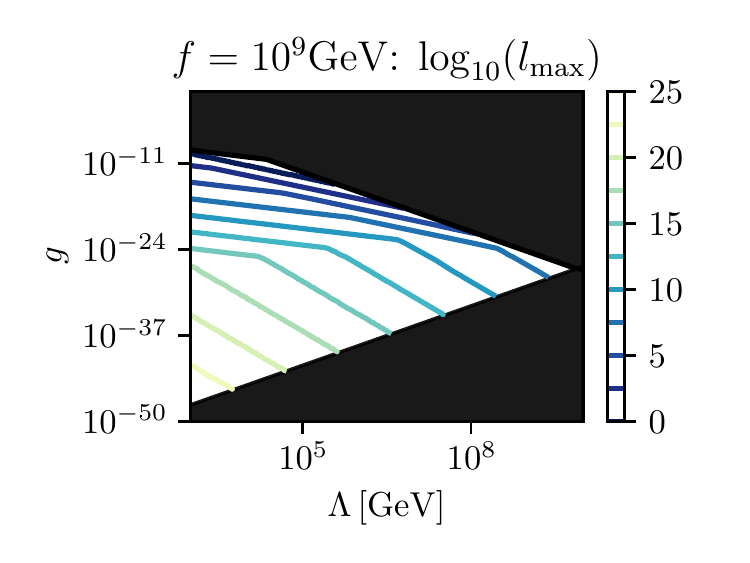}
    \includegraphics[width=0.32\textwidth]{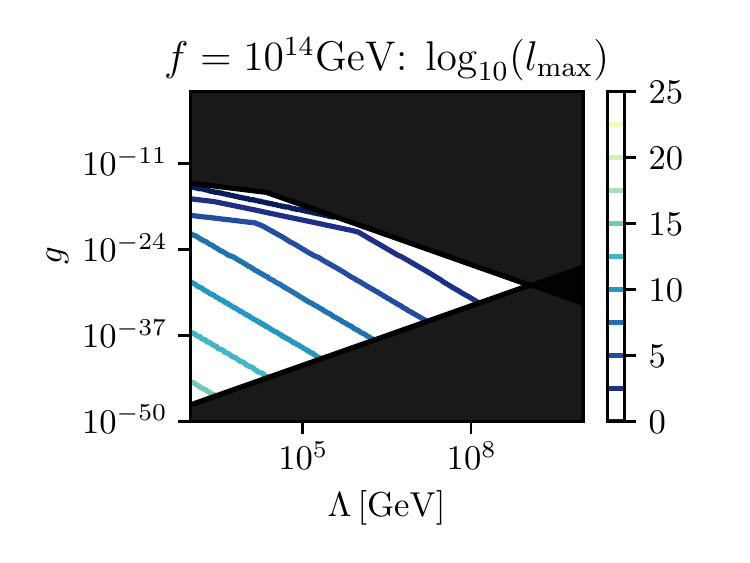}
    \includegraphics[width=0.32\textwidth]{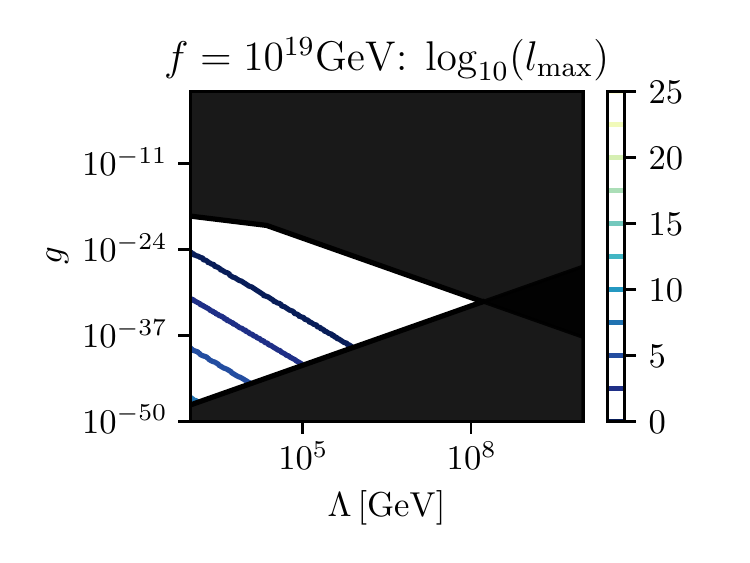}
	\includegraphics[width=0.32\textwidth]{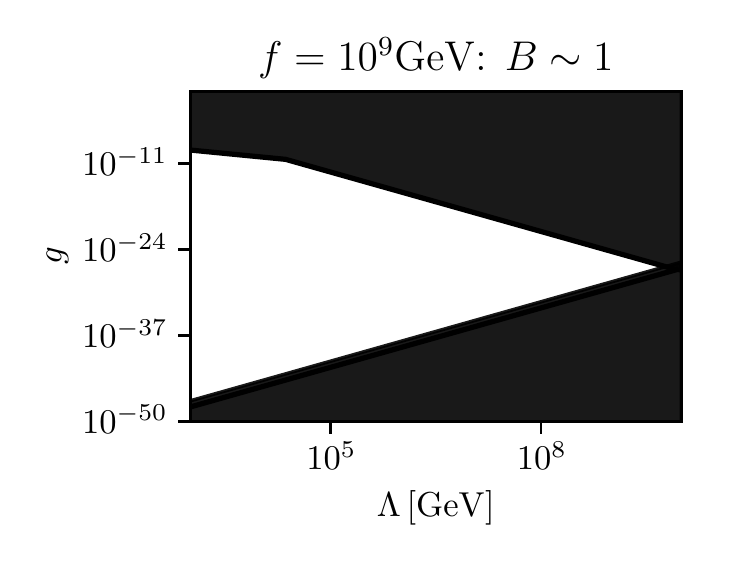}
	\includegraphics[width=0.32\textwidth]{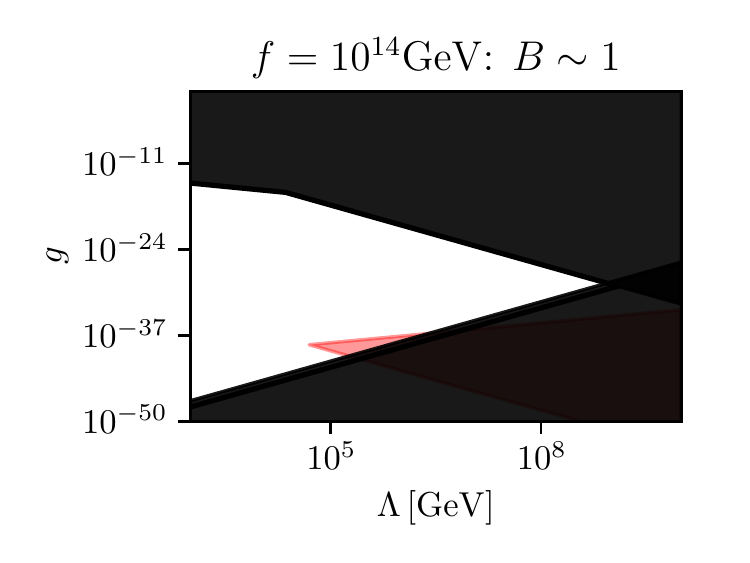}
	\includegraphics[width=0.32\textwidth]{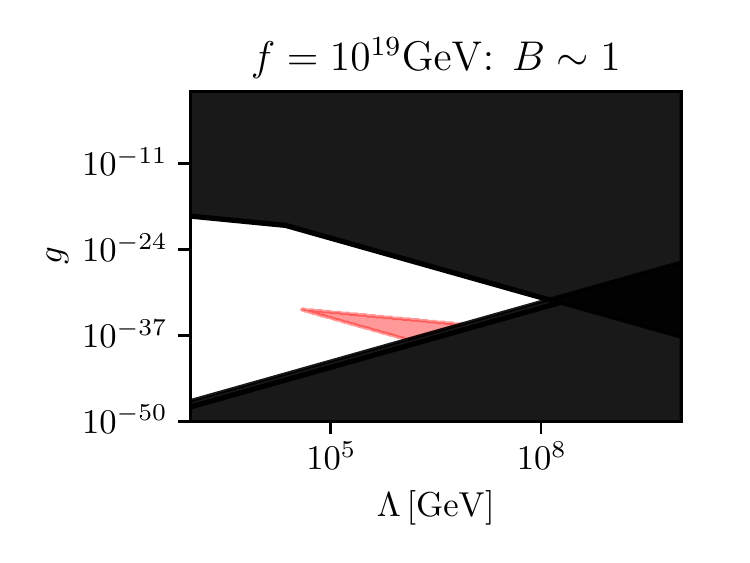}
	\caption{The parameter space for the non-QCD relaxion model in the QbC regime without eternal inflation, in the $g$ vs $\Lambda$ plane, demonstrating the maximal values of the number of the local minimum $l$ (upper panel), and constraints from the formation and escape of a relaxion bubble induced by neutron stars (red) (ower panel). Compared to the CbQ regime, the values of $l$ are larger and the `runaway relaxion' constraints are weaker.}
	\label{even_more_g_L}
\end{figure}

For the non-eternal inflation window, which we mostly focus on in this work, the parameter region in the $g$ vs $\Lambda$ plane is more or less the same as in the CbQ regime. However, for each point in this allowed region, now larger values of inflationary Hubble parameter $H_I>g^{1/3}\Lambda$ are allowed. To demonstrate this, figure~\ref{H_vs_f} shows the $H$ vs $f$ parameter space for the CbQ and QbC models. Again, in the QbC case, the brown color denotes the region where the early stopping takes place whereas the yellow line corresponds to the stopping determined by the Hubble scale. As can be seen in the left panel, a new region has opened up compared to the CbQ scenario (blue line). We impose $H_I<100 \mathrm{GeV}$ to ensure that the spread in the Higgs masses after relaxation is not too large. The other upper bound for the CbQ scenario arises from combining (\ref{equal_slopes}), (\ref{cbq}) and (\ref{Lambda_b_upper}). The lower bound for the QbC regime with $\Lambda_b\sim H_I$ follows from (\ref{noneternal}) combined with the stopping condition and $\Lambda>\mathrm{TeV}$. In each point of the left panel, the values of $g$ and $\Lambda$ are not fixed, which is why the three different regions overlap. In the right panel, we fix the values  $g=10^{-20}$ and $\Lambda = 10^{6}\mathrm{GeV}$, to illustrate how the QbC parameter region opens up for larger values of $H_I$.

In Fig.~\ref{more_g_L_test} we show the minimum number of inflationary e-folds and the largest possible Hubble scale of inflation (as well as inflationary scale $E_I$) for a successful relaxation for each point in the $g$ vs $\Lambda$ plane in the QbC model, in comparison with the predictions obtained in the CbQ regime.

\bigskip

\textbf{The final minimum in the QbC regime and finite-density constraints:} We close the section with figure~\ref{even_more_g_L}. The upper column shows the maximal values of the final minimum $l$ in the QbC regime. Here, only the non-eternal inflation scenario is considered. We use the same benchmark values for the decay constant as in Fig.~\ref{l}. We do not show the minimal values of $l$ since $l_{\mathrm{max}}^{CbQ} = l_{\mathrm{min}}^{QbC}$. The lower row shows the excluded regions of the parameter space, where relaxion bubble formation is possible in dense environments. The plots were constructed using the relations from section~\ref{ssec:finitedensity}, taken from~\cite{Balkin:2021wea}, inserting the value of $l$ determined from the stopping condition $B\sim 1$.  Constraints from neutron stars are the strongest for large decay constants.

\begin{comment}
In figure~~\ref{ps_f_m} we compare the $f$ vs $m$ parameter spaces. As can be seen, the two are not very different from each other, although in the quantum regime the relaxion can have somewhat larger masses. \AC{This is a bit strange.} More specifically, in the classical beats quantum regime the mass can be in the range,
\beq
10^{-16}\mathrm{eV} < m_{\phi} < 10 \mathrm{GeV},
\eeq
whereas, if we assume quantum stopping, the masses are typically larger,
\beq
10^{-3}\mathrm{eV} < m_{\phi} < 10^2 \mathrm{GeV}.
\eeq

Finally, we construct projected plots of $H$ vs $f$ and $f$ vs $m_{\phi}$ parameter spaces, by fixing $\Lambda$ and $g$ to some values. The results are shown in Figure~\ref{nonqcdproj}. The dark orange  color denotes the region allowed in the regime of classical beats quantum. The light orange and the yellow regions are obtained when quantum beats classical. In the yellow region the stopping condition is modified. We set $\Lambda = 10^5 \rm GeV$ and $g = 10^{-20}$. 

\end{comment}

\section{Properties of the relaxion}
\label{sec:properties}

In this section we discuss some properties of the relaxion, including its mass, interactions and the lifetime. We focus on the main interaction channel with the standard model particles, via the mixing with the Higgs. In addition, the relaxion can have pseudoscalar couplings e.g.~to photons, as the usual axion. These couplings are however model-dependent and, for simplicity, we do not discuss them here.
All these properties were already derived in \cite{Espinosa:2015eda,Flacke:2016szy, Banerjee:2020kww}. The formula are unchanged. We review them for completeness.  The main goal of this section is to determine the new phenomenological region in the mass-versus-mixing-angle plane, opened by the QbC regime.

\bigskip

\textbf{Mass and mixing with the Higgs:} Let us consider the relaxion-Higgs potential
\beq
V(h,\varphi) = -g\Lambda^3\varphi + \frac{1}{2}[\Lambda^2 - g'\Lambda \varphi]h^2 + \frac{\lambda}{4!}h^4 + \Lambda_{b}^4 \Bigl( \frac{\langle h \rangle }{v_{h}} \Bigr)^2\Bigl[1-\cos\Bigl(\frac{\varphi}{f}\Bigr)\Bigr].
\eeq
At the end of relaxation, the fields find themselves in the vicinity of a local minimum with $\langle h \rangle = H=v_{h}$ and $\langle \varphi \rangle = \phi = \phi_{\mathrm{stop}}$. The mixing of the relaxion with the Higgs, as well as its physical mass, can be computed by considering excitations around this minimum. Performing a Taylor expansion, $\varphi \rightarrow \varphi+\phi$ and $h \rightarrow h+H$ we arrive at
\beq
V(h, \varphi) = V(H, \phi) + \frac{\partial^2 V}{\partial H^2} \frac{h^2}{2} + \frac{\partial^2 V}{\partial \phi^2} \frac{\varphi^2}{2} + \frac{\partial^2 V}{\partial H \partial \phi} h \varphi +...
\eeq
where all the derivatives are evaluated at $H=v_{h}$ and $\phi=\phi_{\mathrm{stop}}$. The three terms above are identified with Higgs and relaxion masses $m_h^2$ and $m_{\phi}^2$, and the third term $m_{h \phi}^2$ describes the mixing. Computing the derivatives we arrive at,
\beq
m_h^2 = \frac{\partial^2 V}{\partial H^2}\Big|_{v_h, \phi_{\mathrm{stop}}} = \Lambda^2 - g'\Lambda \phi_{\mathrm{stop}} + \frac{\lambda_h}{2}v_h^2 + 2\frac{\Lambda_b^4}{v_h^2}\cos\Bigl(\frac{\phi_{\mathrm{stop}}}{f}\Bigr) = 2\lambda_hv_{h}^2,
\eeq
\beq
\label{relaxion_mass}
m_{\phi}^2 = \frac{\partial^2 V}{\partial \phi^2}\Big|_{v_h, \phi_{\mathrm{stop}}} = -\frac{\Lambda_b^4}{f^2}\cos\Bigl(\frac{\phi_{\mathrm{stop}}}{f}\Bigr) = \frac{\Lambda_b^4}{f^2}\sin\delta,
\eeq
\beq
m_{h \phi}^2 = \frac{\partial^2 V}{\partial H \partial \phi}\Big|_{v_h, \phi_{\mathrm{stop}}} = -g'\Lambda v_h - \frac{\Lambda_b^4}{f} \frac{2}{v_h}\sin\Bigl(\frac{\phi_{\mathrm{stop}}}{f}\Bigr).
\eeq
In the first line we used the fact that the first derivative of the potential should vanish.

The mixing angle is defined as the angle which rotates the mixing term away by means of a rotation in the $h, \varphi$ space to new fields $h'$ and $\varphi'$,
\beq
h = h'\cos\theta_{h\phi} + \varphi' \sin\theta_{h\phi}, \: \: \: \: \: \: \: \: \varphi = \varphi'\cos\theta_{h\phi} - h' \sin\theta_{h\phi}.
\eeq
Inserting this into the expression for the potential above, we arrive at the following expression for the mixing angle
\beq
\label{mixing_angle_general}
\sin(2\theta_{h\phi}) = - \frac{2m_{h \phi}^2}{\sqrt{(2m_{h \phi}^2)^2 + (m_h^2 - m_{\phi}^2)^2}}.
\eeq
By construction the mixing angle is between one (in the case of equal masses) and zero (if there is no mixing at all).

\begin{figure}[!t]
	\centering
	\includegraphics[width=0.49\textwidth]{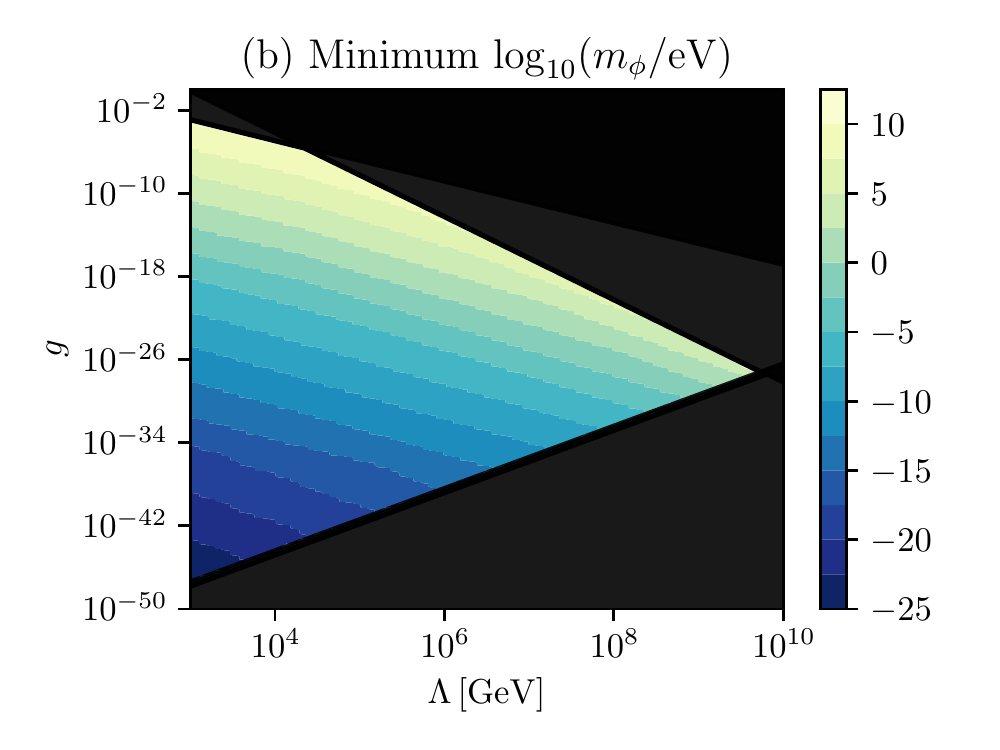}
	\includegraphics[width=0.49\textwidth]{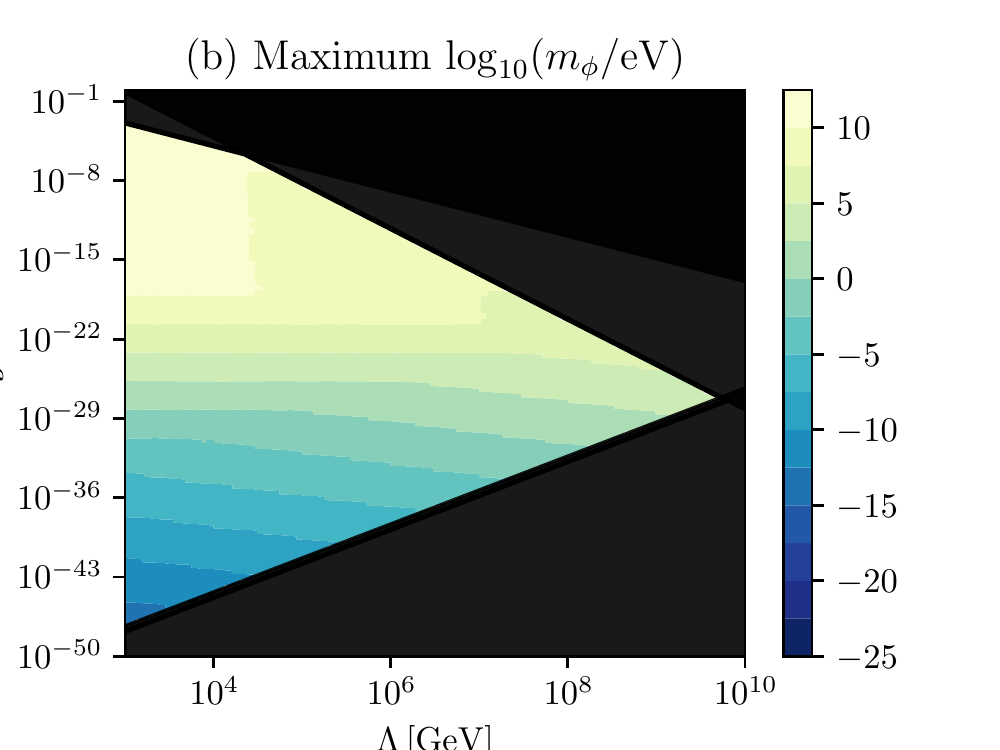}
	\caption{The parameter space for the non-QCD relaxion model in the QbC regime without eternal inflation, in the $g$ vs $\Lambda$ plane, featuring the values of the minimum (a) and maximum (b) possible value for the relaxion mass for each point of the allowed parameter region.}
	\label{more_g_L_2}
\end{figure}

There is one important caveat that one has to keep in mind. When the relaxion mass is close the Higgs mass, the physical mass of the relaxion and the Higgs is the one obtained in the diagonal basis~\cite{Flacke:2016szy}. To find $m_h$ and $m_{\phi}$, one diagonalizes the matrix and equates the largest of the two diagonal entries to $m_h$. Moreover, in this case, the adiabatic approximation that the Higgs tracks its minimum might not hold anymore and one would have to evolve the coupled system of two fields. We thus mostly focus on the regime where $m_{\phi}< \mathrm{few}\times \mathrm{GeV}$. The maximal and minimal relaxion masses are shown in Fig~\ref{more_g_L_2} for each point in the $g$ vs $\Lambda$ plane.

If one approximates $\sin(\phi_{\mathrm{stop}}/f)=0$, the mixing angle can be expressed as 
\beq
\sin(2\theta_{h\phi}) =  \frac{2g\Lambda v_h}{\sqrt{(2g\Lambda v_h)^2 + (m_h^2 - m_{\phi}^2)^2}},
\eeq
However, in the many cases when the field stops not too far from the first minimum, $\sin(\phi_{\mathrm{stop}}/f)=0$ is not satisfied and the second contribution to the mixing angle dominates over the first one.

In the small angle limit, where $\sin(\theta_{h\phi}) \approx \theta_{h\phi}$, one can write
\beq
\sin\theta_{h\phi} \approx - \frac{m_{h\phi}^2}{(m_h^2 - m_{\phi}^2)} \approx - \frac{m_{h\phi}^2}{m_h^2} = - \frac{1}{m_h^2} \frac{\partial^2 V}{\partial H \partial \phi}\Big|_{v_h, \phi_{\mathrm{stop}}}. 
\eeq

\bigskip

\textbf{The lifetime:} We consider the decay of the relaxion through its mixing with the Higgs, and estimate its lifetime, depending on its mass and the mixing angle. The decay rate is proportional to the mixing angle squared and can be written as
\beq
\Gamma_{\phi} = \sin^2\theta_{h\phi} \times  \Gamma_{h}(m_{\phi}).
\eeq
The decay of the Higgs can proceed through different channels~\cite{Bezrukov:2009yw}. For masses of the Higgs below $0.2\rm GeV$, the main decay channels are the decay into leptons (tree-level, via the Yukawa couplings) and, more importantly, the decay into a pair of photons. The later process is one-loop, involving $W$-bosons, charged leptons or quarks in the loop. The analytical expressions for this perturbative processes are
\beq
\Gamma_{h\rightarrow \bar{l}l}(m_h) = (1-y)^{3/2}\frac{m_l^2}{4\pi}\frac{m_h}{2v_h^2},
\eeq
and
\beq
\Gamma_{h\rightarrow \gamma \gamma}(m_h) = \Bigl( \frac{\alpha_{QED}}{4\pi}\Bigr)^2 \frac{m_h^3}{16\pi v_h^2}\Bigl|F_{W}(m_W) + \sum_l e_l^2 F_f(m_l) + 3\sum_q e_q^2 F_f(m_q)\Bigr|^2,
\eeq
Here $$y=\frac{4m_i^2}{m_h^2},$$and, in the second expression we defined,
\beq
F_W = 2 + 3y(1+(2-y)x^2), \: \: \: \: \: \: \: F_f = -2y(1+(1-y)x^2),
\eeq
where $x = \mathrm{atan}(1/\sqrt{y-1})$ if $y>1$ and $x = (\pi + i\mathrm{ln}[1+\sqrt{1-y}] - \mathrm{ln}[1-\sqrt{1-y}])/2$ if $y<1$. When the mass of the Higgs is much smaller than the mass of the particle participating in the loop, the above functions approach the limits $F_W\rightarrow 7$ and $F_{f}\rightarrow -4/3$.

\bigskip

\begin{figure}[!t]
	\centering
	\includegraphics[width=0.98\textwidth]{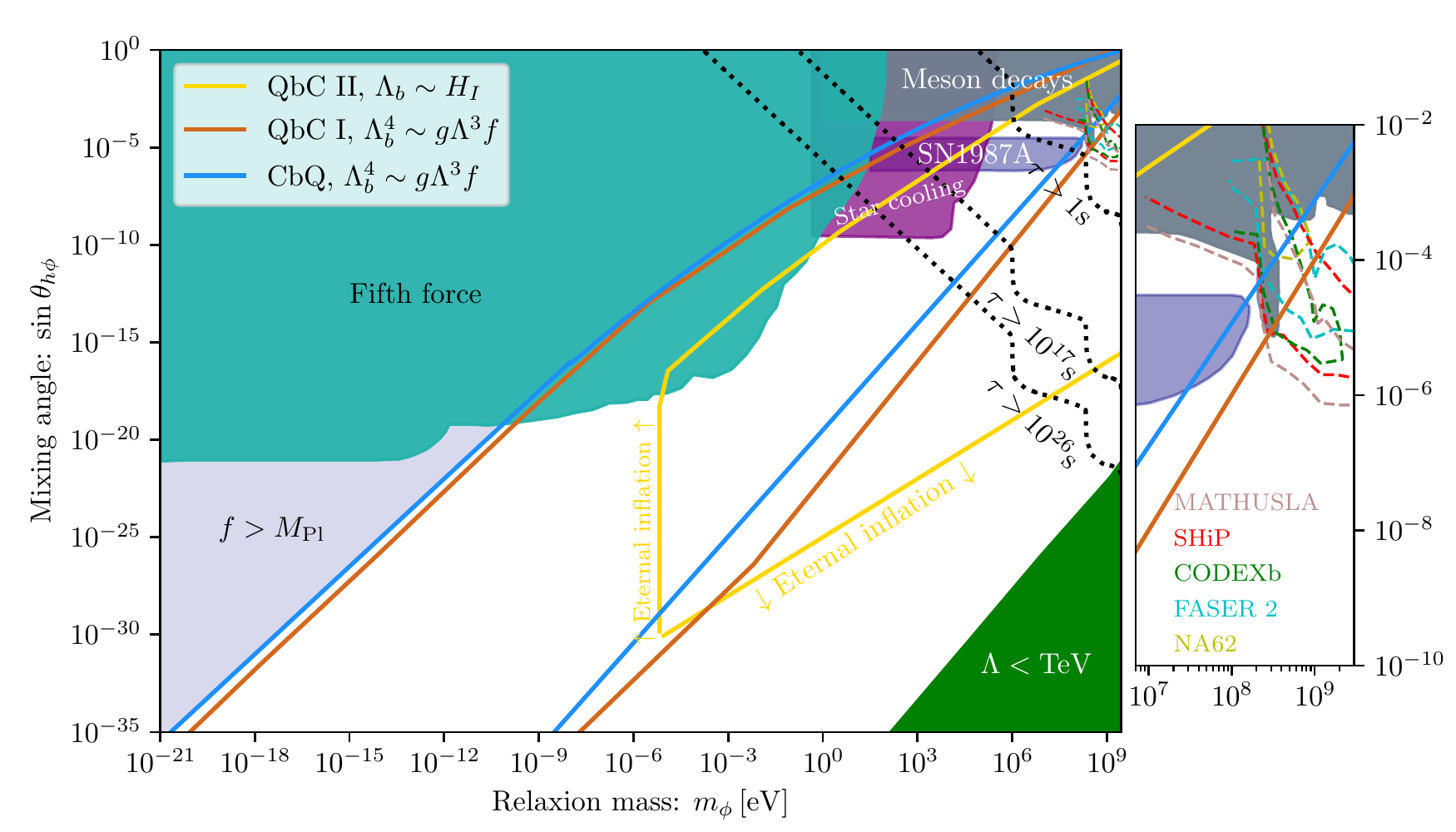}
	\caption{The relaxion parameter space in the $m_{\phi}$ versus $\sin\theta_{h \phi}$ plane, showing the regions associated with each regime (CbQ, QbCI, QbCII) inside each of the respective colored contours, and the experimental and astrophysical bounds. Black dotted lines indicate the contours for the relaxion lifetime. The high mass region is zoomed in on the right for a better illustration of the projected sensitivities of future experiments shown with dashed lines.}
	\label{m_angle}
\end{figure}

\textbf{Laboratory and astrophysical probes of the relaxion:} Various experiments can probe the relaxion through its mixing with the Higgs. A detailed study of the different constraints can be found in~\cite{Flacke:2016szy, Banerjee:2020kww} and here we present a brief summary. The constraints are reported in Fig.~\ref{m_angle} which shows the parameter space for the relaxion in the $m_{\phi}$ vs $\sin\theta_{h\phi}$ plane. The blue line shows the available region in the CbQ regime (see also~\cite{Banerjee:2020kww}). The yellow and brown lines enclose the parameter region for the QbC regime, with the  stopping near $\Lambda_b\sim H_I$ and $\Lambda_b^4 \sim g\Lambda^3f$, respectively. The QbC II region can be further extended to smaller masses and smaller mixing angles if one allows for eternal inflation. We used the general expression (\ref{mixing_angle_general}) for the mixing angle for constructing the parameter regions. Contours of fixed lifetime for the relaxion are also shown in the figure as dotted lines. Below we briefly outline the constraints from~\cite{Flacke:2016szy, Banerjee:2020kww} arising from fifth force experiments, astrophysical observations and experiments studying meson decays.

\begin{itemize}
	\item \textbf{Fifth force experiments:} The strongest constraints on the relaxion parameter space arise from fifth force experiments, which measure the deviations from the gravitational interactions between two neutral particles and thus can detect the existence of a new interaction. These include Eot-Wash experiments, InvSqL experiments and tests of the Casimir force~\cite{Hoskins:1985tn, Bordag:2001qi,Kapner:2006si,Schlamminger:2007ht, Bordag:2009zz, Berge:2017ovy}. The excluded region is shaded in light green.

 	\item \textbf{Astrophysical constraints:} Powerful probes on the relaxion parameter space come from astrophysics. This includes the bounds from stellar evolution, where the relaxion can be produced via its coupling to electrons and photons~\cite{Grifols:1988fv,Cadamuro:2011fd,Raffelt:2012sp,Hardy:2016kme}. The relaxion can also be produced inside the supernova cores via its coupling to nucleons~\cite{Turner:1987by,Frieman:1987ui,Burrows:1988ah,Krnjaic:2015mbs}. These put limits on the couplings and, thus, on the mixing angle. The excluded regions are shaded in purple and navy, where the bound from the SN1987A supernova is subject to some uncertainties~\cite{Bar:2019ifz}.
 
	\item \textbf{Meson decays:} In the $\mathrm{MeV}-5\mathrm{GeV}$ range of relaxion masses, the most sensitive laboratory probes are the proton beam dump and accelerator experiments. These test $K$- and $B$-meson decays and put constraints on the production of the relaxion. Strong constraints come from the CHARM experiment, as well as LHCb, Belle, BABAR, E787 and E949 experiments~\cite{BNL-E949:2009dza, Belle:2009zue,BaBar:2013npw,LHCb:2015nkv, LHCb:2016awg, CHARM:1985anb}. The constraints from these experiments are shaded in grey.  The sensitivity of future experiments, such as MATHUSLA~\cite{Evans:2017lvd,Curtin:2018mvb}, SHiP~\cite{lanfranchi2017sensitivity}, CODEX-b~\cite{Gligorov:2017nwh, Beacham:2019nyx} and FASER-2~\cite{Feng:2017vli, Beacham:2019nyx} are also shown as dashed lines. In the right part of Fig.~\ref{m_angle} the parameter region relevant for these experiments is zoomed in for a better vizualization.
 
\end{itemize}

 The QbC regime extends the available parameter region for the relaxion to smaller mixing angles. As can be seen in the zoomed plot of figure~\ref{m_angle}, future experiments have the potential of probing parts of this new region.

\section{Summary and outlook}
\label{sec:conclusion}

We provided a new perspective into the mechanism of cosmological relaxation of the electroweak scale, based on a stochastic formalism and the Fokker-Planck equation, which takes into account the effect of fluctuations during inflation on the dynamics of the relaxion field. This formalism enabled us to determine the minima that get populated at the end of relaxation, the spread in the field values/Higgs masses, the typical relaxion mass in the minimum and the height of its barrier. Such an analysis allowed us to identify the different regimes of relaxation, the corresponding properties of the relaxion, and the constraints on its parameter space. We summarise the main findings below.

\begin{itemize}

\item In addition to revisiting the relaxion in the conventional classical-beats-quantum (CbQ) regime, we also extended the discussion to the quantum-beats-classical (QbC) regime. Here, two regimes were identified, QbC I and QbC II. In the first one, the stopping of the relaxion is determined by the same condition as in the CbQ regime on the barrier height, $$\Lambda_b^4\sim g\Lambda^3 f .$$ In the second regime, we derived a new stopping relation,
$$\Lambda_b \sim H_I .$$

\item We determined precisely in which minimum the relaxion stops in the CbQ regime (see Fig.~\ref{l} for the non-QCD relaxion) as well as in the QbC regime, see Fig.~\ref{even_more_g_L}.

\item We derived in turn the constraints from the late-time formation and escape of a relaxion bubble induced by neutron stars, white dwarfs, and the Sun,  which crucially depends on the stopping minimum of the relaxion as (baryonic) high-density effects on the relaxion potential are relatively suppressed when the minimum is deeper.  These constraints are shown for the non-QCD relaxion in figure \ref{stars} in the CbQ regime, as well as in Fig.~\ref{even_more_g_L} for the QbC regime. We conclude that most of the relaxion parameter space is safe from the ``runaway" threat discussed in \cite{Balkin:2021wea}.

\item We examined the QCD relaxion including the stochastic dynamics. Because of our modified stopping condition in the QbC II regime, the relaxion can stop at a much deeper minimum associated with a smaller $\theta_{\mathrm{QCD}}$. This solves the strong CP problem without the need for extra model building at the end of inflation, in contrast with \cite{Graham:2015cka}. Besides, the cutoff scale $\Lambda$ can be much larger than ${\cal O} (30)$ TeV, up to $6\times 10^8$ GeV. Nevertheless, this happens necessarily in the eternal inflation regime. The QCD relaxion parameter space in the QbC regime is shown in fig.~\ref{fig_QCD_QbC}, to be compared with 
the one for the CbQ regime in Fig.~\ref{QCD_CbQ}. 
We also conclude that the final minimum is deep enough such that it is safe from ``runaway" constraints in dense astrophysical environments.

\item Another attempt to solve the strong CP problem for the QCD relaxion was proposed in
~\cite{Nelson:2017cfv}. The idea was to consider large Hubble scales for which the relaxion barrier height is suppressed during inflation due to de-Sitter-temperature effects, shifting the position of the first minimum, which is now much deeper, thus removing the need to cancel the slope of the potential at the end of inflation.  However, the full FP equation was not solved   and there was no discussion about tunneling probabilities. Instead, a perturbative approach around gaussian distributions was followed. The relaxion was  assumed to stop at the first minimum, with the usual CbQ  stopping condition $ \Lambda_b^4\sim g\Lambda^3 f$. Our analysis contradicts these results.
We predict a different Hubble scale for the QCD relaxion than ~\cite{Nelson:2017cfv}.  The value they request is incompatible with the stopping condition  used in ~\cite{Nelson:2017cfv}. The comparison between different approaches to the strong CP problem for the QCD relaxion is illustrated  in figure \ref{different_QCD}.

\item We showed that relaxation in the QbC regime is possible with a non-QCD relaxion and that eternal inflation is not required in that scenario. The distinct parameter regions  associated with each regime are shown in the [$\Lambda,g$] plane in Fig.~\ref{fig_nonQCD_QbC} and in the  [$f,H_I$] plane in Fig.~\ref{H_vs_f}. The values of the cut-off scale $\Lambda$ that can be successfully relaxed are summarized in the table below.
The preditions for the required number of efolds of inflation as well as the inflation scale are shown in Fig.~\ref{more_g_L_test}.

\item  The different regimes are also shown in the phenomenological plane  [mixing angle, mass] that is relevant for experimental searches of the relaxion (see Fig.~\ref{more_g_L_2} and \ref{m_angle}). This complements the results presented in \cite{Banerjee:2020kww} which focused on the CbQ regime. The main effect of the QbC regime is to open up a large new region, in particular with larger relaxion mass, not subject to eternal inflation. Part of it will be covered by future experiments such as NA62, MATHUSLA, SHiP, FASER, CODEXb.
It will be interesting to investigate further cosmological probes.
Section \ref{sec:properties} serves the basis for the promising analysis of relaxion dark matter that is presented in a companion paper \cite{Chatrchyan:2022asg}.

\end{itemize}

 We also summarise the main results of the technical derivations  that are deferred to the appendices:

\begin{itemize}

\item In appendix~\ref{app:escape}, we presented the flux-over-population method~\cite{Kramers:1940zz} for finding the escape rate from a false minimum in our framework. 
Such method has been considered in cosmology for studying phase transitions in de Sitter spacetime \cite{Camargo-Molina:2022ord}, with potential application the study of vacuum survival during inflation \cite{Mantziris:2022fuu}. The main result of this appendix is the analytical formula for the decay rate given in equation (\ref{eq_k}) (and reproduced in (\ref{decay_rate_anal}), that is compared with the decay rate extracted from the numerical solution to the FP equation. Figure \ref{slow_down} is one of our main results showing that the relaxion can stop much further away from the first minimum, and with a very good agreement between the analytical and numerical estimates.

\item In appendix~\ref{app:potential} the 
calculation needed for the determination of the final minimum of the relaxion using the decay rate formula  is presented. This depends on $\delta$ the field distance between a maximum of the potential and its next consecutive minimum.
The relaxion potential is analyzed, including the location of the minima/maxima, the mass and the barrier heights, generalizing the analysis from~\cite{Banerjee:2018xmn} to the case of late stopping and beyond the small $\delta$ regime (given by \ref{eq:smalldelta}) relevant in the QbC regime.
The important formulas are (\ref{eq:delta1},\ref{eq:deltal} and \ref{eq:generalformulaDeltaV}) that are then used to calculate $B$ in (\ref{define_B}).

\item 
 In appendix~\ref{app:stop} we discuss the dynamics and the stopping condition for the relaxion in the Fokker-Planck formalism at a somewhat more technical level. 
This appendix provides a more careful alternative derivation for the relaxion velocity  given in (\ref{velo_minima}), in terms of equation (\ref{eq_velocity}). As the velocity never vanishes, it cannot be used as a stopping condition and instead $B\sim 1$ is used as stopping condition, with $B$ given in (\ref{define_B}).

\item 
The aim of appendix~\ref{app_vol} is to address the question ``Does the relaxion climb up?”. We demonstrate, following~\cite{Gupta:2018wif}, that including the volume-weighting for the probability distribution does not spoil the mechanism as long as we are in the regime of non-eternal inflation. Main condition for this is (\ref{eq:noclimbing}).

\end{itemize}

The next question is whether these findings open any new observable opportunities for the relaxion which is notoriously difficult to test. The answer is positive. One important implication of the QbC regime is the possibility the explain dark matter with the relaxion. The relic abundance of the relaxion from the misalignment mechanism depends on the initial displacement of the field which originates from fluctuations during inflation. It thus depends on the Hubble scale during inflation $H_I$ as well as the relaxion mass $m_{\phi}$. In the CbQ regime, the relaxion is always underabundant. However, in the QbC regime, the values of $H_I$ and $m_{\phi}$  are different and allow for the relaxion to be dark matter. This is the topic of a separate article \cite{Chatrchyan:2022asg}.

 The QCD relaxion also predicts a larger value for  $\theta_{\mathrm{QCD}}$ than the one from the `standard' QCD axion and could be probed by future neutron EDM searches \cite{Abel:2018yeo, Filippone:2018vxf}. 
 Further development of atomic precision physics experiments have also the ability  to push forward the searches for the relaxion~\cite{Kim:2022ype,Banerjee:2019epw}.
 Generally, there are also potential additional signatures in the non-QCD models associated with the degrees of freedom responsible for the barrier, as  investigated in 
\cite{Beauchesne:2017ukw,Barducci:2020axp} (although \cite{Espinosa:2015eda} showed that the relaxion mechanism can work
 without the need for any new weak-scale fermions, but instead with an additional light axion-like scalar as a companion to 
the relaxion). 
Finally, we conclude by stressing that relaxion models require a long period of relatively low-scale inflation, $H_I<v_h$. A key for testing the relaxion mechanism will be to investigate low-scale inflationary models consistent with Planck data and their associated signatures (e.g.~\cite{Evans:2017bjs,Takahashi:2018tdu,Daido:2017wwb,Daido:2017tbr,Matsui:2020wfx}).

\bigskip

\begin{table}[h!]
	\begin{center}
		\label{tab:table1}
		\begin{tabular}{l|c|c|c|c|c} 
			\textbf{} & \textbf{CbQ} & \textbf{QbC I}  & \textbf{QbC I} & \textbf{QbC II}  & \textbf{QbC II} \\
			\textbf{} &  &  &\small{eternal infl.}&  &\small{eternal infl.}\\	\hline
		    & & & & & \\
			\textbf{QCD relaxion} & - & - & - & - & $10^9 \mathrm{GeV}$ \\
		(with constant slope)	& & & & & \\
			& & & & & \\
			\textbf{Non-QCD relaxion} & $4 \times 10^9 \mathrm{GeV}$ & $4\times 10^9 \mathrm{GeV}$ & $2\times 10^{10} \mathrm{GeV}$ &  $10^7 \mathrm{GeV}$ & $2\times 10^{10} \mathrm{GeV}$ \\  \hline
		\end{tabular}
	\end{center}
 \caption{The maximal cut-off scales that the relaxion mechanism can reach.}
\end{table}

\section*{Acknowledgements}

We are grateful to Hyungjin Kim for many insights as well as collaboration on related work. We also thank Andreas Ekstedt, Oleksii Matsedonskyi, Enrico Morgante, Konstantin Springmann and Stefan Stelzl for useful discussions, as well as Bachelor student Kyrill Michaelsen.
This work is supported by the Deutsche Forschungsgemeinschaft under Germany Excellence
Strategy - EXC 2121 ``Quantum Universe'' - 390833306.

\appendix
 
\section{Escaping from a local minimum}
\label{app:escape}

In this appendix we compute the escape probability from a local minimum for the Fokker-Planck equation. The computation is based on the flux-over-population method~\cite{Kramers:1940zz, Berera:2019uyp}, and applies the method to an overdamped system. As a reminder, the Fokker-Planck equation has the form,
$$
\frac{\partial \rho}{\partial t} = \frac{1}{3H}\frac{\partial(\rho \:  \partial_\phi V)}{\partial\phi} + \frac{H^3}{8\pi^2} \frac{\partial^2\rho}{\partial\phi^2}.
$$

We consider a situation where the field distribution is centered in a local minimum of the potential at $\phi=0$, with $V=V_0$. There is a local maximum at $\phi=\phi_b$ with $V(\phi_b)=V_b$, followed by another local minimum at some $\phi>\phi_b$. 

We introduce the total probability to be in the local minimum, $n_f$, given by
\beq
n_f = \int_{-\infty}^{\phi_b}\rho(\phi)d\phi,
\eeq
the flux of probability from the local minimum along the barrier $j_b$ (per unit time), 
\beq
j_b = \frac{dn_f}{dt} =  \int_{-\infty}^{\phi_b}\Bigl[  \frac{1}{3H}\frac{\partial(\rho \:  \partial_\phi V)}{\partial\phi} + \frac{H^3}{8\pi^2} \frac{\partial^2\rho}{\partial\phi^2} \Bigr] d\phi =   \frac{\rho \:  \partial_\phi V}{3H} + \frac{H^3}{8\pi^2} \frac{\partial\rho}{\partial\phi} = \frac{H^3}{8\pi^2} \frac{\partial\rho}{\partial\phi}\Big|_{\phi_b},
\eeq
and the escape rate from the local minimum $k$ defined as the ratio of these two quanitites,
\beq
k=\frac{j_b}{n_f}.
\eeq
The escape rate gives the flow rate of the probability from the local minimum through the barrier.

The equilibrium configuration is given by
\beq
\rho_{\mathrm{eq}}(\phi) = \frac{1}{N} \exp{ \Bigl(-\frac{V(\phi) - V(\phi_0)}{D}\Bigr) },
\eeq
where we defined
\beq
D = \frac{3H^4}{8\pi^2}
\eeq
and $N$ is the constant which normalizes the total probability. Note that the parameter $D$ used in this chapter has a dimension $4$ and should not be confused with the dimensionless parameter $d$ used in the main body of the paper.

To study the escape from a local minimum we  assume that the field is in equilibrium only partially, inside the local minimum. Due to the large height of the barrier it has not yet equilibrated globally, beyond the barrier, as the field has not yet properly probed that region of the potential. It is also assumed that the distribution is in a steady state,
\beq
\rho(\phi) = \rho_{\mathrm{eq}}(\phi) \xi(\phi),
\eeq
where $\xi\rightarrow 1$ for $\phi \rightarrow -\infty$ and $\xi \rightarrow 0$ for $\phi \rightarrow \infty$. This ansatz is known as the Kramers ansatz~\cite{Kramers:1940zz}. It can be obtained by adding a source and a sink to the system to maintain the steady flow. In reality there will be no sources or sinks, and the population in the local minimum will gradually decrease. The function $\xi$ parametrizes the thermal activation in the vicinity of the saddle point.

We can then evaluate $j_b$ to be
\beq
j_b = \frac{H^3}{8\pi^2} \Bigl[ \frac{\partial\rho_{\mathrm{eq}}}{\partial\phi}\xi + \rho_{\mathrm{eq}}\frac{\partial\xi}{\partial\phi} \Bigr]\Big|_{\phi_b} = \frac{H^3}{8\pi^2} \rho_{\mathrm{eq}} \frac{\partial\xi}{\partial\phi} \Big|_{\phi_b}.
\eeq

If we insert the Kramers ansatz into the Fokker-Planck equation and separate the $\partial_t\rho_{\mathrm{eq}}=0$ part we arrive at,
\beq
0 = \frac{1}{3H}\frac{\partial(\rho_{\mathrm{eq}}\xi \:  \partial_\phi V)}{\partial\phi} + \frac{H^3}{8\pi^2} \frac{\partial^2(\rho_{\mathrm{eq}}\xi)}{\partial\phi^2},
\eeq
or,
\beq
0 = (\partial_{\phi}\xi) (\partial_{\phi} V) \rho_{\mathrm{eq}} + D [ 2 (\partial_{\phi}\rho_{\mathrm{eq}})(\partial_{\phi}\xi) + \rho_{\mathrm{eq}}(\partial_{\phi}^2\xi) ].
\eeq

Next, we assume that the function $\xi$ changes from $1$ to $0$ in the vicinity of the barrier and we expand the potential near the local maximum up to the quadratic order,
\beq
V(\phi) = V_b + \frac{(\phi-\phi_b)^2}{2}V_b'' + ...
\eeq
Inserting this into the above equation, as well as into the equilibrium distribution, we arrive at the following simple differential equation
\beq
0 = -(\phi-\phi_b) \partial_{\phi}\xi V_b'' +  D \: \partial^2_{\phi}\xi 
\eeq
plus terms higher order in $(\phi-\phi_b)$. Together with the boundary conditions for $\xi$ this has the solution
\beq
\xi(\phi) = \sqrt{ \frac{|V_b''|}{2\pi D} } \int_{\phi}^{\infty} d \widetilde \phi \exp{\Bigl(-\frac{(\widetilde \phi - \phi_b)^2V_b''}{2D}\Bigr)}.
\eeq

Having found the form of $\xi$ we can now insert it (evaluated at $\phi_b$) into the expression for the probability flux to arrive at 
\beq
j_b = - \frac{H^3}{8\pi^2} \rho_{\mathrm{eq}}(\phi_b) \sqrt{ \frac{|V_b''|}{2\pi D} }.
\eeq

The only thing left is to express $n_f$. The simplest way to estimate it is to assume a small diffusion parameter $D\ll\Delta V_b$, such that most of the distribution is around the quadratic minimum of the potential, which then allows one to approximate the integral of $n_f$ as that of a gaussian function with infinite boundaries,
\beq
n_f \approx \int_{-\infty}^{\phi_b}\rho_{\mathrm{eq}}(\phi)d\phi \approx \frac{1}{N} \int_{-\infty}^{\infty} \exp{ \Bigl( -\frac{(\phi-\phi_0)^2V_0''}{2D} \Bigr)} d\phi = \frac{1}{N} \sqrt{ \frac{V_0''}{2\pi D} }.
\eeq
Inserting everything back into the expression for the absolute value of the flux rate $k$ we arrive at
\beq
\label{eq_k}
\boxed{
	k = \frac{\sqrt{V_{0}'' |V_b''|}}{6\pi H} \exp{ \Bigl( - \frac{8\pi^2 (V_b - V_0)}{3H^4} \Bigr)} =  \frac{\sqrt{V_{0}'' |V_b''|}}{6\pi H} e^{  - \frac{8\pi^2 \Delta V_b}{3H^4} }. 
}
\eeq
The flux rate is indeed of the form of the Hawking-Moss instanton~\cite{Hawking:1981fz} in the exponent.

\bigskip

\textbf{Analytics vs numerics:} To test the validity range of the above analytical formula, we perform numerical simulation of the Fokker-Planck equation. As an example we consider the potential $V(\phi) = \phi^2/2 - \phi^3/3 + \phi^4/24$. For this potential $V(\phi_0)=0$ and $V''(\phi_0) = 1$. The local maximum is at $\phi_b = 1.267$, where $V(\phi_b)=0.232$ and $V''(\phi_b) = -0.732$ (or, ignoring the minus sign, $|V''_b|=0.732$).

\begin{figure}[!h]
	\centering
	\includegraphics[width=0.55\textwidth]{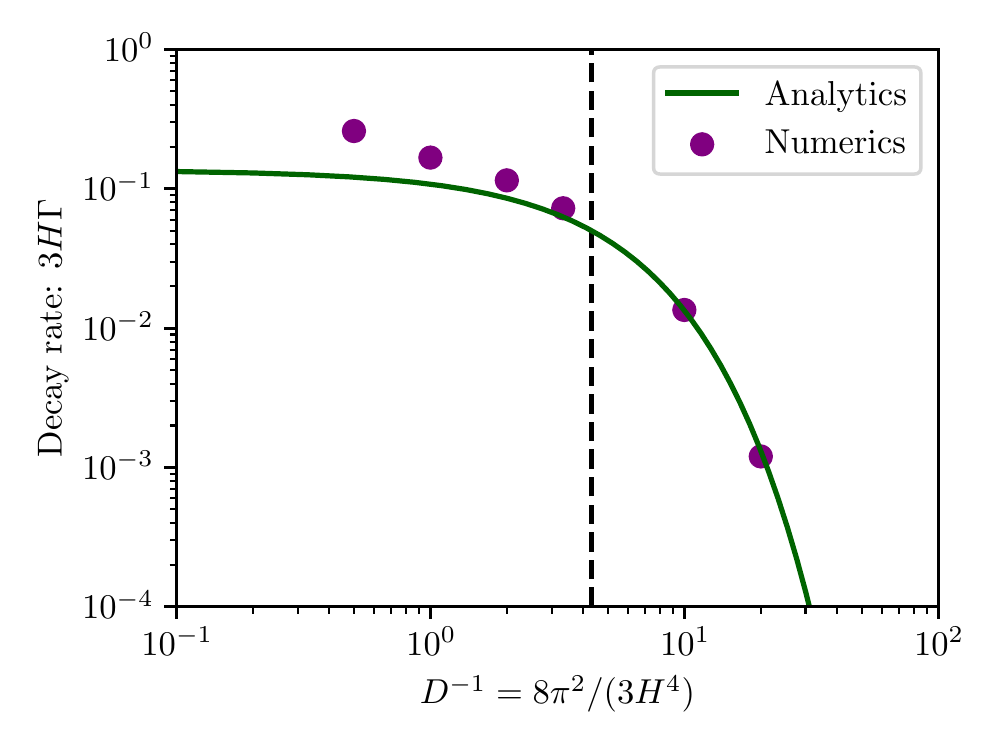}
	\caption{The decay rate for different values of the Hubble parameter extracted from the numerical simulation of the Fokker-Planck equation (purple data points) and using the analytical formula (\ref{decay_rate_anal}) (the green line). As can be seen, the two agree very well for $D<\Delta V_b = 0.232$.}
	\label{decay_rate_compare}
\end{figure}

The flux rate $k$ can be related to the decay rate of the false minimum in this potential. Indeed, including the backwards flux from the true minimum we can write
\beq
\frac{dn_f}{dt} = -k_{f\rightarrow t} n_f + k_{t\rightarrow f} n_t.
\eeq
In our numerical simulations we take the probability distribution $\rho(\phi)$ to be initially concentrated in the false minimum around $\phi=0$. In this case, $n_f \approx 1$ and $n_t \ll n_f$ at early times. We can write
\beq
\frac{dn_f}{dt} \approx -k_{f\rightarrow t} n_f = -\Gamma n_f
\eeq
and, therefore, expect an exponential decay of $n_f$,
\beq
\label{eq_Gamma}
n_f(t) \propto e^{-\Gamma t}.
\eeq
We extract the decay rate $\Gamma$ numerically, by evolving the system, tracking $n_f(t)$ and performing a fit to it according to Eq.~(\ref{eq_Gamma}). We then compare the extracted value with the analytical formula from Eq.~(\ref{eq_k}) with $\Gamma = k$. The comparison is shown in figure~\ref{decay_rate_compare} for several values of $D$. The two curves agree quite well, especially when $\Delta V_b \gtrsim D$, indicated by the black dashed line, after which the decay is very slow. We observe that the decay rate is insensitive to the precise form of the initial distribution of the field.

\section{The shape of the relaxion potential}
\label{app:potential}

In this appendix we discuss some properties of the relaxion potential, including the curvature near the local minima/maxima and the height of potential barriers.

\bigskip

\textbf{The rescaled potential:} Let us again write the equations that describe the relaxion dynamics, including the explicit dependence of the barrier height on the Higgs VEV,
\beq
\dot \phi =  - \frac{V'}{3H_I}  \approx \frac{1}{3H_I} \Bigl[ g\Lambda^3 - \frac{\Lambda_b^4}{f} \Bigl( \frac{\langle h \rangle}{v_h} \Bigr)^m  \sin\Bigl( \frac{\phi}{f}\Bigr)\Bigr],
\eeq
where we have used the relaxion potential
\beq
V(\phi) =  - g\Lambda^3\phi + \Lambda_{b}^4 \Bigl( \frac{\langle h \rangle}{v_h} \Bigr)^{m}\Bigl[1-\cos\Bigl(\frac{\phi}{f}\Bigr)\Bigr].
\eeq
The Higgs mass is given by $\mu_h^2 = \Lambda^2 - g'\Lambda\phi$. Note that the above parametrization includes both the QCD and the non-QCD axion models. In the first case $m=1$ whereas in the second case usually $m = 2$. The axion field has to stop at $\langle h \rangle = v_h$. By definition, $\Lambda_b$ determines the barrier height at the correct Higgs VEV and, therefore, for the QCD axion $\Lambda_b \approx 75 \mathrm{MeV}$.

For simplicity, it is convenient to introduce new variables, following Appendix B of~\cite{Fonseca:2019lmc}. First of all, noting that the wiggles turn on at $\phi = \Lambda/g'$, we define a new dimensionless field variable
\beq
\theta = \frac{1}{f}\Bigl( \phi - \frac{\Lambda}{g'}\Bigr).
\eeq
In this way $\theta=0$ corresponds to the Higgs mass being zero. For positive values of $\theta$, the Higgs mass/VEV can be expressed in terms of this variable as
\beq
\label{field_trans}
\mu_h^2 = -\theta (g'\Lambda f).
\eeq
Moreover, the field by definition stops at $\theta = -\mu_h^2/(g'\Lambda f) = (88\mathrm{GeV})^2/(g'\Lambda f)$.

Furthermore, we introduce a dimensionless time variable,
\beq
\tau = \frac{\dot \phi_{\mathrm{SR}}}{f} t =\frac{g\Lambda^3}{3H_If} t.
\eeq
In terms of this variable, the rescaled field $\theta$ slow-rolls along the linear potential with a velocity equal to one.

In contrast with \cite{Fonseca:2019lmc}, we work in the regime with large Hubble that implies $\dot \phi_{\mathrm{SR}}/(3Hf)\ll 1$. In this case,  
inserting everything into the original field equation one arrives at the simpler equation,
\beq
\frac{d\theta}{d\tau} = 1-\Bigl( \frac{\theta}{b} \Bigr)^{m/2} \sin\theta = 1-f(\theta)\sin\theta,
\eeq
where the dimensionless parameter $b$ characterizes the shape of the potential and is given by
\beq
\label{eq:stopping_relation_b}
b = \Bigl( \frac{-\mu_h^2}{g\Lambda f} \Bigr) \Bigl( \frac{g\Lambda^3 f}{\Lambda_b^4} \Bigr)^{2/m}.
\eeq
Note that to make the discussion even more general we will consider potentials of the form $$V(\theta) = -\theta + f(\theta)[1-\cos\theta],$$ with some slowly and monotonically increasing, positive function $f$. If one assumes the stopping condition for classical slow-rolling, given by (\ref{equal_slopes}), then the term in the second brackets is equal to one, and one obtains that the field stops around $\theta =  b$, which should come with no surprise, since only from this point the derivative of the potential,
\beq
\label{eq_rescaled_potential}
V(\theta) = -\theta +\Bigl( \frac{\theta}{b} \Bigr)^{m/2}[1 - \cos\theta]
\eeq
can be zero. Note that the classical dynamics of $\theta$ is determined only via the parameter $b$. This parameter, which characterizes how quick the barriers of the potential grow in height, is usually large. The potential is related to the physical potential via $V_{\mathrm{orig}}(\phi) = (g\Lambda^3f)V(\theta)$, with $\phi$ given by (\ref{field_trans}). In what follows we set $m=2$ in our expressions unless stated otherwise.

\bigskip

\textbf{The location of local minima:} Let us count the periods of the cosine term of the potential by $n$, i.e.~$\theta = 2\pi n$. We want to find the first local minimum, i.e.~the point where
\beq
V'(\theta) = -1 + \frac{\theta}{b}\sin(\theta) = 0.
\label{vdot}
\eeq
Note that we neglect the other derivative term from the second term, because for large $b$-s it is much smaller. The above condition can be satisfied only once $\theta>b$, within one period. Let us denote the corresponding index of that period by $n_1$. 

A minimum of $V'(\theta)$ lies between the first local minimum and the first local maximum of the potential, and is approximately located at $$\theta_{\star} = 2 \pi n + \pi/2,$$ such that the sine term is maximized, $\sin\theta_{\star}=1$. Let us denote the separation of the local minimum (the local maximum) from this central point by $\delta$, i.e.~$\theta_{min} = \theta_{\star} - \delta$ and $\theta_{max} = \theta_{\star} + \delta$. The parameter $\delta$ can be found from the condition (\ref{vdot})
\beq
1 = \frac{\theta_{min}}{b}\sin(\theta_{min}) = \frac{(\theta_{\star} - \delta)}{b}\sin(\theta_{\star} - \delta) \approx  \frac{\theta_{\star} }{b}\cos(\delta)
\eeq
or, in other words, $$\cos(\delta) = \frac{b}{\theta_{\star}}  ,$$In the more general case one would have $\cos(\delta) ={1}/{f(\theta_{\star})}$.

Importantly, the above expression is valid for any minimum/maximum of the potential with index $i$ (note that $n_i = n_1 + (i-1)$ for $i\geq 1$) and, thus, carries an additional index $i$,
\beq
\cos(\delta_i) = \frac{b}{\theta_{\star, i}}, \: \: \: \: \:  \: \: \: \: \: \theta_{\star, i} \approx 2 \pi n_i.
\eeq

Let us mention some basic properties of the parameter $\delta$. This parameter determines the location of the relaxion minimum, $\sin(\theta_{min}) = \cos(\delta)$. If $\theta/b\approx 1$ (or $f(\theta)\approx 1$) it is very small, whereas in the limit of large $\theta$ it approaches $\pi/2$. In the case of the QCD axion $\delta$ also determines the $\theta$-angle of QCD. As we will see, the value of $\delta$ determines how the relaxion mass is suppressed compared to the naive expectation. It is thus important to realize what is the minimal value that $\delta$ can take, near the first local minimum. In the small-$\delta$ limit
\beq
\label{eq:delta1}
\delta_1 \approx \sin(\delta_1) = \sqrt{1 - \Bigr(\frac{b}{\theta_{\star, 1}}\Bigr)^2} \approx \sqrt{\frac{\theta_{\star, 1}^2 - b^2}{b^2}}\approx \sqrt{\frac{4\pi b}{b^2}}\sim b^{-1/2}.
\eeq
Similarly, assuming that the $i$-th local minimum satisfies $\theta_{\star, i} - b \ll b$, one can write
\beq
\delta_i \approx \sqrt{\frac{4\pi i b}{b^2}} \sim \sqrt{i} \delta_1.
\label{eq:deltal}
\eeq

\bigskip

\textbf{The mass at the minima:} Let us evaluate the mass squared near the local minimum,
\beq
V_{0, i}'' = V''(\theta_{min, i}) =  \frac{\theta_{min, i}}{b}\cos(\theta_{min, i}) \approx  \frac{\theta_{\star, i}}{b} \sin(\delta_i) = \tan(\delta_i).
\label{vdot}
\eeq
As can be seen from the above expression, the mass squared is approximately equal to $\theta_{min}/{b}$ when $\theta_{min}\gg{b}$, which is the naive expectation for the mass. In contrast, if $\theta_{min} \approx {b}$, as it is the case near the first local minimum $n_1$, the mass squared is very small compared to the naive expectation. Taylor expanding the sine term one arrives at
\beq
V_{0, i}'' = \frac{\theta_{\star, i}}{b} \sin(\delta_i) \approx \frac{\theta_{\star, i}}{b} \delta_i.
\eeq

In the more general case with a function $f(\theta)$, one would simply replace $\theta_{\star}/b$ by $f(\theta_{\star})$ in the above expressions.

The same computation reveals that the mass near the minimum is equal (with a minus sign) to the curvature in the neighboring local maximum, $V_{b, i}'' = V''(\theta_{max, i})  = -V_{0, i}''$.

\bigskip

\textbf{The potential barriers:} Let us now evaluate the energy difference between the local minimum and the neighboring local maximum. This difference determines the transition rate to the next minimum.
\beq
\Delta V^{\rightarrow}_{b, i} = V(\theta_{max, i}) - V(\theta_{min, i}) \approx -\theta_{max, i} - \frac{\theta_{max, i}}{b}\cos(\theta_{max, i}) - \theta_{min, i} + \frac{\theta_{min, i}}{b}\cos(\theta_{min, i})
\eeq
Inserting the expressions for $\theta_{min, i}$ and $\theta_{max, i}$ we arrive at
\beq
\label{eq:generalformulaDeltaV}
\Delta V^{\rightarrow}_{b, i}= - 2 \delta_i + 2\frac{\theta_{\star, i}}{b}\sin(\delta_i)  = 2\frac{\theta_{\star, i}}{b}[\sin(\delta_i)- \delta_i\cos(\delta_i)] = 2[\tan(\delta_i) - \delta_i].
\eeq
Let us again consider two limits. If  $\theta_{\star, i}\gg{b}$, $\delta$ becomes approximately $\pi/2$ and the above expression coincides with the naive expectation, $\Delta V^{\rightarrow}_{b, i} = 2\theta_{\star, i}/b$. In contrast, when $\theta_{\star, i} \approx {b}$, the barrier is suppressed. Performing a Taylor expansion one arrives at
\beq
\Delta V^{\rightarrow}_{b, i} \approx 2\frac{\theta_{\star, i}}{b} \frac{\delta_i^3}{3}.
\label{eq:smalldelta}
\eeq

Again, generalizing to $f(\theta)$ is trivial. Finally, it is easy to check that potential barrier in the opposite direction is larger and is given by
\beq
\Delta V^{\leftarrow}_{b, i} = 2[\tan(\delta_i)- \delta_i+\pi ].
\eeq

\bigskip

\section{Analysis of the new stopping condition}
\label{app:stop}

In this appendix, we perform a more careful quantitative analysis of the new stopping condition for the stochastic relaxion dynamics.

Transforming the Fokker-Planck equation to the rescaled variables, mentioned in the previous appendix, one arrives at
\beq
\label{FP_rescaled}
\partial_{\tau} \rho = \Bigr[\rho\Bigr( -1 - \Bigl( \frac{\theta}{b} \Bigr)^{m/2}\sin\theta\Bigl)\Bigl]' + d\rho'',
\eeq
where $d  = 3H_I^4/(8\pi^2 g\Lambda^3 f)$ is the dimensionless parameter defined in Eq.~(\ref{parameter_d}) characterizing the strength of diffusion effects. The stochastic dynamics depends not only on $b$, but also on $d$.

\bigskip
\textbf{Effective Fokker-Planck equation:} Below, we derive the evolution of the field in the late-time regime. The field typically spreads over several local minima, and in the vicinity of each of them it is in a meta-equilibrium state (a state which is slowly decaying via tunneling). We can thus write
\beq
\rho(t, \theta) = \sum_{i}\rho_{\mathrm{eq}}(\theta, i)N_i(t),
\eeq
where both $\rho$ and $\rho_{\mathrm{eq}}$ are normalized to one and $N_i$ shows the fraction of the total probability in that local minimum. In this representation, $N_i$ are now the variables which evolve in time.

Taking into account the transitions between nearest local minima, the time evolution of $N_i$ can be written as
\beq
\frac{dN_i}{d\tau} = -k_{\rightarrow}(i)N_i - k_{\leftarrow}(i)N_i + k_{\rightarrow}(i-1)N_{i-1} + k_{\leftarrow}(i+1)N_{i+1},
\eeq
where $k_{\rightarrow}$ and $k_{\leftarrow}$ give the flux rates in the positive and negative direction from the given $i$-th minimum. It is convenient to introduce the following flux rates
\beq
k(i) = \frac{1}{2}(k_{\rightarrow}(i) + k_{\leftarrow}(i)), \: \: \: \: \: \: \: \: \: \: \delta k(i) = \frac{1}{2}(k_{\rightarrow}(i) - k_{\leftarrow}(i)),
\eeq
from which $k$ gives a mean flux rate (in both directions) and $\delta k$ is the asymmetry between the flux rates in both directions. It is this term that causes the field to continue rolling in the positive direction even in the tunneling dominated regime. In terms of these quantities the equation of motion can be rewritten as
\beq
\frac{dN_i}{d\tau} = (N_{i+1}+N_{i-1}-2N_i)k(i) + N_{i+1}(k_{\leftarrow}(i+1) - k_{\leftarrow}(i) - \delta k(i)) + N_{i-1}(k_{\rightarrow}(i-1) - k_{\rightarrow}(i) + \delta k(i)).
\eeq
One can check that in the last two brackets the term $\delta k$ usually is much larger compared the difference terms involving $k_{\rightarrow}$ and $k_{\leftarrow}$. Neglecting these terms one arrives at
\beq
\frac{dN_i}{d\tau} = (N_{i+1}+N_{i-1}-2N_i)k(i) + (N_{i+1}-N_{i-1})( - \delta k(i)).
\eeq
The above expression is similar to a Fokker-Planck equation. Indeed, if one takes the continuum limit, the first term reduces to a Laplacian, whereas the second term is a first order derivative. We thus arrive at
\beq
\frac{\partial N}{\partial \tau} = \partial_{\theta} [N(-4\pi \delta k(\theta)))] + (4\pi^2k(\theta))\,\partial_{\theta}^2N.
\eeq
Starting from the original Fokker-Planck equation with a wiggly potential, we have effectively ``integrated out" the fluctuations (except for those that account fro transitions between local minima) and arrived at an ``effective" Fokker-Planck equation, which however has quite different properties. In particular, there a no wiggles, the new drift term has the form $-4\pi \delta k(\theta)$, and decreases as $\theta$ grows. The second term contains the new diffusion term which also decreases with $\theta$ in contrast to the original equation where it was constant.

We can estimate the dynamics of the mean field, $\langle \theta \rangle = \int_{\theta} N(\theta) \theta d\theta$, as
\beq
\label{eq_velocity}
\langle \dot \theta \rangle  =  4\pi \delta k( \langle \theta \rangle ) = 2\pi (k_{\rightarrow} - k_{\leftarrow} ),
\eeq
which coincides with Eq.~(\ref{velo_minima}) derived in the main body of the paper. The forward and backward fluxes can be computed using Eq.~(\ref{eq_k}) from appendix~\ref{app:escape} for the escape rate and the results from appendix~\ref{app:potential} for the relaxion potential. Using the rescaled variables one arrives at
\beq
k_{\rightarrow} = \frac{\tan(\delta)}{2\pi} \exp\Bigl( -\frac{2}{d} \Bigl[\tan(\delta) - \delta \Bigr]\Bigr), \: \: \: \: \: \: \: \: \: k_{\leftarrow} = k_{\rightarrow} e^{-(2\pi)/d}.
\eeq
In the limit of $d\ll 1$, $k_{\leftarrow}$ is negligible compared to $k_{\rightarrow}$. In contrast, when $d\gg 1$, the backwards tunneling is important and $k_{\rightarrow} - k_{\leftarrow} \approx k_{\rightarrow} (2\pi/d)$.

\bigskip

\textbf{Modified stopping condition:} We now compare the stopping condition for the relaxion, extracted from solving numerically the real-time Fokker-Planck equation, and the one obtained by solving Eq.~(\ref{eq_velocity}). The relaxion does not strictly-speaking stop and, instead, has a nonzero velocity which decreases very rapidly with time. For simplicity, here we compare the field values at the moment when the field velocity has decreased by a factor of $10$ compared to its initial value, $ \langle \dot \theta \rangle = 0.1$ (where time derivative is taken with respect to dimensionless variable $\tau$).

\begin{figure}[!t]
	\centering
	\includegraphics[width=0.8\textwidth]{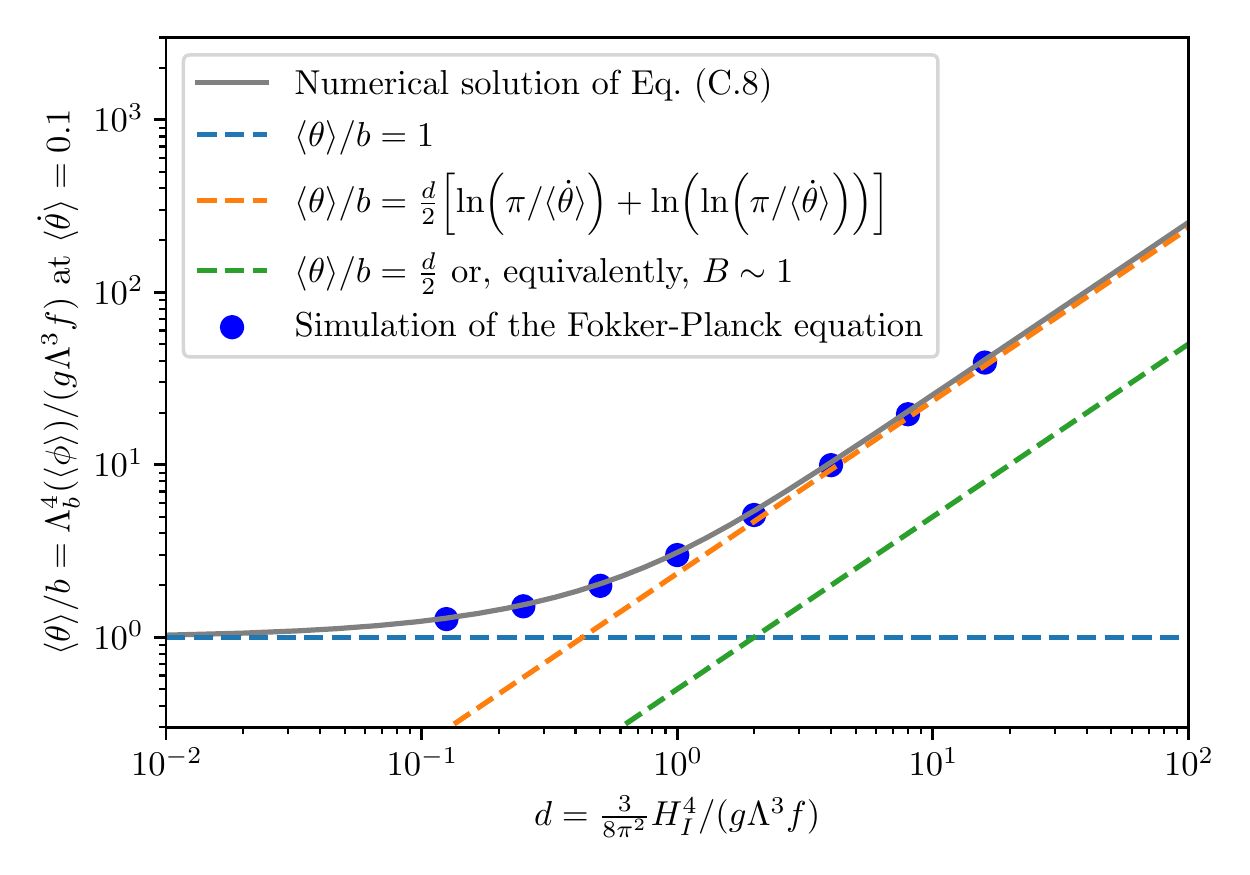}
	\caption{The modified stopping condition for the relaxion. On the horizontal axis is the diffusion parameter $d$, defined in (\ref{parameter_d}). On the vertical axis is the rescaled mean field $\langle \theta\rangle $ divided by the parameter $b$ from (\ref{eq:stopping_relation_b}). We use $m=2$ and $b=100$. The blue points correspond to the numerical simulation of the Fokker-Planck equation. The grey line is the numerical solution to ~(\ref{eq_velocity}). The dashed lines correspond to the approximate expressions in the small-$d$ limit (blue line), in the large-$d$ limit from Eq.~(\ref{analytical_large_d_stop}) (orange line) and the approximate solution corresponding to the stopping condition $B\sim 1$ (green line).}
	\label{stop_and_compare}
\end{figure}

The comparison is shown in figure~\ref{stop_and_compare}. On the horizontal axis is the diffusion parameter $d$, defined in (\ref{parameter_d}) and characterizing the strength of diffusion effects. On the vertical axis is the rescaled field $\theta$ divided by the parameter $b$ from (\ref{eq:stopping_relation_b}). We use $b=100$ and $m=2$. The parameter $b$ gives the approximate location of the first local minimum and, thus, $\theta\approx b$ in the case of the usual stopping condition, which is shown as the blue dashed line. The blue points in the plot are obtained by solving numerically the real-time Fokker-Planck equation and extracting the field expectation value and its derivative as a function of time. The simulation is then stopped when $\langle \dot \theta \rangle = 0.1$. The grey line instead corresponds to the numerical solution of Eq.~(\ref{eq_velocity}) for $\langle \theta \rangle$. This solution has to limiting cases:
\begin{itemize}
	\item If $d\ll1$, the solution is simply $\theta \approx b$. This is the standard stopping condition and, as can be seen in figure~\ref{stop_and_compare}, the numerical solution indeed approaches this limit.
	
	\item If $d\gg 1$, the solution can be expressed as
	\beq
	\label{analytical_large_d_stop}
	\frac{\langle \theta \rangle }{b} = \frac{d}{2} \Bigl[ \ln\Bigl( \frac{\pi}{\langle \dot \theta \rangle }\Bigr) + \ln\Bigl(\ln\Bigl( \frac{\pi}{ \langle \dot \theta \rangle }\Bigr) \Bigr) + ... \Bigr]
	\eeq
	where we used the formulas for the mass and the barrier height in the $\theta\gg b$ limit from appendix~\ref{app:potential}, as well as the asymptotic expansion of the Lambert-W function. The orange line in Fig.~\ref{stop_and_compare} is obtained by keeping the two terms in Eq.~(\ref{analytical_large_d_stop}) and already agrees very well with the numerical solution. For a general value of $m$ the term on the left-hand side of Eq.~(\ref{analytical_large_d_stop}) should be replaced by $({\langle \theta \rangle }/{b})^{m/2}$.
\end{itemize}

The last important question is what exactly can be used as the stopping condition for the field, given that its velocity never becomes zero. In most of the paper we used $B\sim 1$, where $B$ is the argument in the exponent of $k_{\rightarrow}$ which agrees with Eq.~(\ref{analytical_large_d_stop}) if one neglects all the logarithmic corrections. We show this solution, i.e.~$\langle \theta \rangle = bd/2$ as the green dashed line in the figure. As can be observed, for $d>2$ the difference in the values of $\Lambda_b$ is less than a factor of two, which justifies the use of this simpler form of the stopping condition.

\section{Volume-weighting and the measure problem}
\label{app_vol}

In this appendix we revisit the Fokker-Planck equation and discuss some effects due to volume weighting. Our main conclusion is that as long as we are in the regime of non-eternal inflation, these effects are unimportant.

\bigskip

\textbf{The volume-weighted Fokker-Planck equation:} We represent the energy budget of the universe during inflation as a sum of contributions from the inflaton and the relaxion sectors, $V(\phi, \sigma)=V_I(\sigma) + V(\phi)$, where $\sigma$ denotes the inflaton field. The FP equation for the relaxion distribution function $\rho(t, \phi)$ has the form~\cite{Gupta:2018wif}
\beq
\label{FP_general}
\frac{d\rho}{dt} = \frac{\partial}{\partial\phi} \Bigl[ \frac{H^{3/2}(\phi, \sigma)}{8\pi^2} \frac{\partial}{\partial\phi} \Bigl( H^{3/2}(\phi, \sigma) \rho \Bigr) + \frac{V'(\phi) \rho}{3H(\phi, \sigma)} \Bigr],
\eeq
with $H^2(\phi, \sigma) = (8\pi/3) V(\phi, \sigma)/M^2_{\mathrm{Pl}}$. Neglecting the relaxion energy density, $H(\phi, \sigma) \approx H_I$ (which is justified in the limit when (\ref{subdominant}) holds), and ignoring the time-dependence of the inflaton $\sigma$, the FP equation takes the simpler form given in Eq.~(\ref{FP_eq}), which was used in the previous sections.

We now introduce the function $\mathcal{V}(t, \phi)$ which gives the total volume of patches with a field value $\phi$,
\beq
\mathcal{V} = e^{3H t}\rho
\eeq
 This function satisfies the same FP equation with an additional term,
\beq
\frac{d\mathcal{V}}{dt} = \frac{\partial}{\partial\phi} \Bigl[ \frac{H^{3/2}(\phi, \sigma)}{8\pi^2} \frac{\partial}{\partial\phi} \Bigl( H^{3/2}(\phi, \sigma) \mathcal{V} \Bigr) + \frac{V'(\phi) \mathcal{V}}{3H(\phi, \sigma)} \Bigr] + 3H(\phi, \sigma) \mathcal{V}.
\eeq
In the limit of $H(\phi, \sigma) \approx H_I$ the solution to the above equation is related to the one for Eq.~(\ref{FP_eq}). Instead, we keep the leading-order corrections to $H_I$ from the relaxion sector,
\beq
H= \sqrt{  \frac{8\pi[V_I + V(\phi)]}{3M^2_{\mathrm{Pl}}} } \approx H_I + \frac{4\pi}{3} \frac{V(\phi)}{H_I M_{\mathrm{Pl}}^2 }.
\eeq
Inserting this expansion into Eq.~(\ref{FP_general}), and rescaling away the $\phi$-independent growth due to the inflaton, by defining 
\beq
P(\phi) = \mathcal{V}(\phi)e^{-3H_It}, 
\eeq
we arrive at the following FP equation for the function $P(t, \phi)$,
\beq
\label{P_eq}
\frac{dP}{dt} = \frac{1}{3H_I}\frac{\partial(P \:  \partial_\varphi V)}{\partial\phi} + \frac{H_I^3}{8\pi^2} \frac{\partial^2P}{\partial\phi^2} + \frac{4\pi}{M_{\mathrm{Pl}}^2} \frac{V(\phi)}{H_I  } P.
\eeq
Note that neither $\mathcal{V}(\phi)$ nor $P(\phi)$ are normalized distribution functions.

\bigskip

\textbf{Does the relaxion climb up?} The additional term in the last equation gives a preference to regions with a large potential energy or, in other words, the volume-weighted probability density $P$ grows faster in those regions. We would like to understand which field values will dominate the total volume in the end of relaxation.

For simplicity we can ignore the wiggles and consider $V(\phi) = -g\Lambda^3 \phi$. We then use the solution for $\rho$ from Eq.~(\ref{sol:nowiggles}), and multiply it  by the $\phi$-dependent growth factor, arising from the last term in Eq.~(\ref{P_eq}), to arrive at
\beq
P(\phi, t) = \sqrt{ \frac{2\pi}{H_I^3 t} } \exp{\Bigl\{- \frac{2\pi^2 }{H^3_It}\Bigl(\phi - \frac{g\Lambda^3t}{3H_I}\Bigr)^2 \Bigr\}} \times \exp{\Bigl\{\frac{4\pi(-  g\Lambda^3 \phi )t }{M^2_{\mathrm{Pl}}H_I} \Bigr\}}.
\eeq
We now check how the motion of the distribution peak is modified after the volume-weighting. Requiring $\partial_{\phi} P = 0$ one arrives at
\beq
\phi_{\mathrm{peak}}(t) = \frac{g\Lambda^3t}{3H_I} - \frac{g\Lambda^3 H_I^2 t^2 }{M_{\mathrm{Pl}}^2 \pi} = \dot \phi_{\mathrm{SR}} t - \frac{g\Lambda^3 H_I^2 t^2 }{M_{\mathrm{Pl}}^2 \pi} .
\eeq
The second term arises from volume-weighting and pushes the distribution in the opposite direction compared to the slow-roll. This term however grows as $t^2$ and, thus, will become comparable to the first term only after some time, corresponding to
\beq
N \sim \frac{\pi M_{\mathrm{Pl}}^2 }{3 H_I^2} \sim N_c
\eeq
e-folds, where $N_{c}$ is the critical value for eternal inflation, given by Eq.~(\ref{crit_num_efolds}). This implies that in the non-eternal inflation regime, which we mostly focus on in this work, volume-weighting does not affect the predictions for the relaxion dynamics. The situation is different in the eternal inflation regime, where the sensitivity to the choice of how to count probabilities can be attributed to the measure problem.

\bigskip

\textbf{The fate of the wrong Hubble patches:} Having clarified that at the end of inflation most Hubble patches will have a small Higgs VEV, we now discuss whether the relaxion may start climbing the potential after inflation. Indeed, in this regime the relaxion field can no longer be safely treated as subdominant in terms of its energy density. Here we follow~\cite{Gupta:2018wif}.

\bigskip

One of the constraints on the relaxion sector is that its energy density is much smaller compared to the inflationary energy scale, 
\beq 
H(\phi, \sigma)=H_I \gg \frac{\Lambda^2}{M_{Pl}},\: \: \: \: \: \: \: \text{ (during  inflation)}
\eeq 
which ensures that all Hubble patches expand roughly at the same rate during inflation. This condition is no longer satisfied after inflation and, after some time, the Hubble parameter is 
\beq
H \gtrsim H(\phi) = \frac{\sqrt{V(\phi)}}{M_{Pl}}.\: \: \: \: \: \: \: \text{ (after  inflation)}
\eeq
Here we assume that we are in the regime of non-eternal inflation, and define by $t_I =  N_I/H_I$ the proper time by which the $\sigma$-driven inflation ends globally. The ``wrong'' patches with a large Higgs mass have a much larger energy density for $t>t_I$ and experience a faster expansion.

\bigskip

The regime where relaxation is possible without eternal inflation satisfies the constraint 
\beq
\label{eq:noclimbing}
 (g\Lambda^3)^{1/3} > \frac{\Lambda^2}{M_{Pl}}. 
\eeq
This implies that $ (g\Lambda^3)^{1/3} > H(\phi)$ holds even for the largest relaxion field value, with $\mu_h^2 \sim \Lambda^2$. Hence the post-inflationary dynamics of the relaxion in the ``wrong'' patches is governed by classical slow-roll and not by quantum fluctuations. This prevents the relaxion from effectively ``climbing up'' the potential, as in the eternal inflation scenario.

\bigskip

We estimate the additional expansion of the ``wrong'' patches during the time when the relaxion in these patches slow-rolls down to the vicinity of the first minimum of the potential~\cite{Gupta:2018wif},
\beq
\exp\Bigl[ 3\int H(\phi_{SR}(t))dt\Bigr] = \exp\Bigl[ 9\int \frac{H^2(\phi_{SR})}{V'(\phi_{SR})}d\phi_{SR}\Bigr] \sim \exp\Bigl[ \frac{9}{M_{Pl}^2}\int \phi d\phi\Bigr] \sim \exp\Bigl[\frac{9(\Delta \phi)^2}{2 M_{Pl}^2} \Bigr].
\eeq
It turns out that if the non-eternal inflation constraint (\ref{noneternal}) is satisfied, this extra expansion does not compensate the suppressed probability of such patches. The latter can be estimated using the gaussian approximation for the distribution function,
\beq
\rho(\phi) \sim \exp\Bigl[-\frac{2\pi^2}{H_I^3t_I} (\Delta \phi)^2 \Bigr] \sim \exp\Bigl[-\frac{2\pi^2 g^2\Lambda^2 (\Delta \phi)^2 }{H_I^4} \Bigr].
\eeq
In the last part we used $H_It_I \sim N_I  \sim N_r \sim H_I^2/(g^2\Lambda^2)$. Comparing the above expression to  the previous one, one observes that if (\ref{noneternal}) is satisfied, the gain in volume is small and the final probability is large in patches with a small Higgs mass.

\bigskip

\textbf{The measure problem in eternal inflation:} To compare the probabilities/volume fractions of different patches in an eternally inflating universe one has to introduce a time regularization since otherwise quantities might diverge. We restrict our attention to finite spacetime regions $V^{(4)}(t_c)$ swept out by the geodesics at $t<t_c$, where $t_c$ is some cutoff which is taken to infinity at the end of the calculations. The relative probabilities can then be expressed as
\beq
\frac{p(A)}{p(B)} =\lim_{t_c \rightarrow \infty}\frac{V^{(4)}_A(t_c)}{V^{(4)}_B(t_c)}.
\eeq
As there is no unique way to set a global time direction across casually disconnected regions of spacetime, the regularization procedure is not unambiguous and leads to the measure problem~\cite{Gupta:2018wif}. Two common choices here are the proper time cut-off measure~\cite{Linde:1986fd} and the scale factor cut-off measure~\cite{Linde:1993xx, Linde:1993nz}. The first choice is more intuitive however gives a strong preference to patches with large energy and, in particular, results in the so-called ``youngness'' paradox~\cite{Linde:1994gy, Bousso:2007nd}. The second choice uses the scale factor to regularize the volumes, thus effectively avoiding the bias for large energies~\cite{DeSimone:2008bq}.

Using the proper time cut-off measure is similar to weighting the probabilities with the volume and, as it was already demonstrated, the wrong patches would dominate the total volume in the regime of eternal inflation. If one instead uses the scale factor cut-off measure, which is then similar to not weighting the probabilities with the volume, the ``wrong'' patches are not problematic, as it was explained in~\cite{Nelson:2017cfv, Gupta:2018wif}.

\bibliographystyle{h-physrev5}

\bibliography{masterbib-new}

\end{document}